\documentclass[traditabstract]{aa} 

\usepackage{graphicx}
\usepackage{txfonts}
\usepackage{natbib}
  \def\Td{T_{\hspace*{-0.2ex}\rm d}}
  \def\Tg{T_{\hspace*{-0.2ex}\rm g}}
  \def\Rin{R_{\rm in}}
  \def\Rout{R_{\rm out}}
  \def\HH{{\rm\langle H\rangle}}
  \def\nH{n_{\HH}}
  \def\etal{${\rm \hspace*{0.8ex}et\hspace*{0.7ex}al.\hspace*{0.7ex}}$}
  
  \def\ie{i.\,e.\ }
  \def\eg{e.\,g.\ }
  \def\amin{{a_{\rm min}}}
  \def\amax{{a_{\rm max}}}

\begin{document}

   \title{The unusual protoplanetary disk around the T\,Tauri star ET\,Cha}

   \author{P.~Woitke\inst{1,2,3}
      \and B.~Riaz\inst{4}
      \and G.~Duch{\^e}ne\inst{6,5}
      \and I.~Pascucci\inst{4}
      \and A.-Ran~Lyo\inst{8} 
      \and W.~R.~F.~Dent\inst{10}  
      \and N.~Phillips\inst{13}
      \and W.-F.~Thi\inst{5}
      \and F.~M{\'e}nard\inst{5}
      \and G.~J.~Herczeg\inst{14}
      \and E.~Bergin\inst{15} 
      \and A.~Brown\inst{16} 
      \and A.~Mora\inst{12}
     %\and C.~Pinte\inst{5}
      \and I.~Kamp\inst{7}
     %\and I.~Tilling\inst{13}
     %\and J.-C.~Augereau\inst{5}
     %\and J.~M.~Alacid\inst{19,20} 
     %\and S.~Andrews\inst{21}  
     %\and D.~R.~Ardila\inst{22}
      \and G.~Aresu\inst{7}
     %\and D.~Barrado\inst{11,18}
      \and S.~Brittain\inst{11}
     %\and D.~R. Ciardi\inst{24}
     %\and W.~Danchi\inst{25}
     %\and C.~Eiroa\inst{10} 
     %\and D.~Fedele\inst{25,26,27}
      \and I.~de~Gregorio-Monsalvo\inst{10}
     %\and C.~A.~Grady\inst{14}
     %\and A.~Heras\inst{28}
     %\and C.~D.~Howard\inst{12} 
     %\and N.~Huelamo\inst{11}
     %\and A.~Krivov\inst{29} 
     %\and J.~Lebreton\inst{5}
     %\and R.~Liseau\inst{30} 
     %\and G.~Mathews\inst{9} 
     %\and C.~Martin-Zaidi\inst{5}
     %\and G.~Meeus\inst{10}
     %\and I.~Mendigut\'ia\inst{11}
     %\and B.~Montesinos\inst{11}
     %\and M.~Morales-Calderon\inst{37}
     %\and H.~Nomura\inst{32}
     %\and E.~Pantin\inst{16}
     %\and L.~Podio\inst{7}
     %\and D.~R.~Poelman\inst{3}
     %\and S.~Ramsay\inst{33}
     %\and K.~Rice\inst{13}
     %\and P.~Riviere-Marichalar\inst{11}
     %\and A.~Roberge\inst{13} 
      \and G.~Sandell\inst{9}
     %\and E.~Solano\inst{19,20}
     %\and B.~Vandenbussche\inst{17}
     %\and H.~Walker\inst{34}
     %\and J.~P.~Williams\inst{9}
     %\and G.~J.~White\inst{35,36}
     %\and G.~Wright\inst{2}
   }

   \institute{ %new 1 Woitke
             University of Vienna, Dept. of Astronomy, 
             T{\"u}rkenschanzstr.~17, A-1180 Vienna, Austria
         \and %2 Woitke Wright
             UK Astronomy Technology Centre, Royal Observatory, Edinburgh,
             Blackford Hill, Edinburgh EH9 3HJ, UK
         \and %3 Woitke Poelman
             SUPA, School of Physics \& Astronomy, University of St.~Andrews,
             North Haugh, St.~Andrews KY16 9SS, UK;
         \and %4 Pascucci Riaz
	     Space Telescope Science Institute, 3700 San Martin Drive, 
             Baltimore, MD 21218, USA
         \and %5 Menard Augereau Duchene Lebreton Pinte Thi Martin-Zaidi
  	     UJF-Grenoble 1 / CNRS-INSU, Institut de Plan{\'e}tologie et
             d'Astrophysique (IPAG) UMR 5274, Grenoble, F-38041,
             France
         \and %6 Duchene
             Astronomy Department, University of California, Berkeley, 
             CA 94720-3411, USA
         \and %7 Kamp Podio Aresu
             Kapteyn Astronomical Institute, Postbus 800,
             9700 AV Groningen, The Netherlands
         \and %8 Lyo
             Korea Astronomy and Space Science Institute, 61-1 Hwaam-dong, 
             Yuseong-gu, Daejeon 305-348, Korea
	% \and %9 Williams Mathews
	%     Institute for Astronomy, University of Hawaii at Manoa, 
        %     Honolulu, HI 96822, USA
        % \and %10 Meeus Eiroa
        %     Dep. de F\'isica Te\'orica, Fac. de Ciencias, UAM Campus 
        %     Cantoblanco, 28049 Madrid, Spain
	% \and %11 Montesinos Riviere-Marichalar Huelamo Mendigutia Barrado
        %     LAEX, Depto. Astrof{\'i}sica, Centro de Astrobiolog{\'i}a 
        %    (INTA-CSIC), P.O. Box 78, E-28691 Villanueva de la Ca\~nada, Spain
          \and %12 %9 Howard Sandell
             SOFIA-USRA, NASA Ames Research Center, Mailstop 211-3 Moffett 
             Field CA 94035 USA
	% \and %13 Roberge Danchi
        %     NASA Goddard Space Flight Center, Exoplanets \& Stellar 
        %     Astrophysics, Code 667, Greenbelt, MD 20771, USA
	% \and %14 Grady
        %     Eureka Scientific and Exoplanets and Stellar Astrophysics Lab, 
        %     NASA Goddard Space Flight Center, Code 667, Greenbelt, MD, 
        %     20771, USA
	 \and %15 %10 Dent de~Gregorio
             ESO-ALMA, Avda Apoquindo 3846, Piso 19, Edificio Alsacia, 
             Las Condes, Santiago, Chile
	% \and %16 Pantin
        %     CEA/IRFU/SAp, AIM UMR 7158, 91191 Gif-sur-Yvette, France
        % \and %17 Vandenbussche
        %     Instituut voor Sterrenkunde, KU Leuven, Celestijnenlaan 200D, 
        %     3001 Leuven, Belgium 
        % \and %18 Barrado
        %     Calar Alto Observatory, Centro Astron\'omico Hispano-Alem\'an
        %     C/Jes\'us Durb\'an Rem\'on, 2-2, 04004 Almer\'{\i}a, Spain
        % \and %19 Alacid Solano
        %     Unidad de Archivo de Datos, Depto. Astrof{\'i}sica, Centro de
        %    Astrobiolog{\'i}a (INTA-CSIC), P.O. Box 78, E-28691 Villanueva de 
        %     la Ca\~nada, Spain 
        % \and %20 Alacid Solano
        %     Spanish Virtual Observatory 
        % \and %21 Andrews
        %     Harvard-Smithsonian Center for Astrophysics, 60 Garden St., 
        %     Cambridge, MA, USA 
        % \and %22 Ardila
        %     NASA Herschel Science Center, California Institute of Technology,
        %     Pasadena, CA, USA.  
          \and %23 %11 Brittain
             Clemson University, Clemson, SC, USA
        % \and %24 Ciardi
        %     NASA Exoplanet Science Institute/Caltech 770 South Wilson 
        %     Avenue, Mail Code: 100-22, Pasadena, CA USA 91125 
        % \and %25 Fedele
        %     Departamento de Fisica Teórica, Facultad de Ciencias, Universidad
        %     Autónomade Madrid, Cantoblanco, 28049 Madrid, Spain 
        % \and %26 Fedele
        %     Max Planck Institut f{\"u}r Astronomie, K{\"o}nigstuhl 17, 69117
	%     Heidelberg, Germany 
        % \and %27 Fedele
        %     Johns Hopkins University Dept. of Physics and Astronomy, 3701 San
	%     Martin drive Baltimore, MD 21210 USA 
        % \and %28 Heras
	%    Research and Scientific Support Department-ESA/ESTEC, PO Box 299, 
        %     2200 AG Noordwijk, The Netherlands 
        % \and %29 Krivov
	%     Astrophysikalisches Institut und Universit{\"a}tssternwarte,
	%     Friedrich-Schiller-Universit{\"a}t, Schillerg{\"a}{\ss}chen 2-3, 
        %     07745 Jena, Germany 
        % \and %30 Liseau
        %     Department of Radio and Space Science, Chalmers University of
	%     Technology, Onsala Space Observatory, 439 92 Onsala, Sweden
	  \and %-  %12 Mora 
              ESA-ESAC Gaia SOC, P.O. Box 78. E-28691 Villanueva de 
              la Ca\~{n}ada, Madrid, Spain 
        %\and %32 Nomura
	%     Department of Astronomy, Graduate School of Science, 
	%     Kyoto University, Kyoto 606-8502,Japan 
	% \and %33 Ramsay
	%     European Southern Observatory, Karl-Schwarzschild-Strasse 2, 
        %     85748 Garching bei M\"unchen, Germany.  
        % \and %34 Walker 
        %     The Rutherford Appleton Laboratory, Chilton, Didcot, OX11 OQL, UK
        % \and %35 White
	%     Department of Physics \& Astronomy, The Open University, 
        %     Milton Keynes MK7 6AA, UK 
        % \and %36 White
        %     The Rutherford Appleton Laboratory, Chilton, Didcot,
	%     OX11 OQL, UK 
        % \and %37 Morales-Calderon 
        %     Spitzer Science Center, California Institute of Technology, 
        %     1200\,E California Blvd, 91125 Pasadena, USA.  
          \and %new 13 Phillips
             SUPA, Institute for Astronomy, University of Edinburgh,
             Royal Observatory, Blackford Hill, Edinburgh EH9 3HJ, UK
        % \and %38 de~Gregorio
        %     European Southern Observatory, Alonso de C{\'o}rdova 3107, 
        %     Vitacura, Casilla 19001, Santiago 19, Chile
	  \and %39 %14 Herczeg
             Max-Planck-Institut f\"ur extraterrestriche Physik, 
             Giessenbachstrasse 1, 85748 Garching, Germany
	  \and %40 %15 Bergin
             Department of Astronomy, The University of Michigan,
             500 Church Street, Ann Arbor, MI 48109-1042, USA
	  \and %41 %16 Brown
	     Center for Astrophysics and Space Astronomy, University
	     of Colorado, Boulder, CO 80309-0389, USA
   }

   \date{Received\ \ Feb.\,9, 2011;\ \ accepted Mar.\,21, 2011}

   \abstract{We present new continuum and line observations, along with
     modelling, of the faint $(6\!-\!8)$\,Myr old T\,Tauri star ET\,Cha
     belonging to the $\eta$ Chamaeleontis cluster. We have
     acquired {\sc Herschel/Pacs} photometric fluxes at 70$\,\mu$m and
     160$\,\mu$m, as well as a detection of the [OI]\,63\,$\mu$m
     fine-structure line in emission, and derived upper limits for
     some other far-IR OI, CII, CO and o-H$_2$O lines.  These
     observations were carried out in the frame of the Open Time
     Key Programme GASPS, where ET\,Cha was selected as one of the
     science demonstration phase targets.  The {\sc Herschel} data is
     complemented by new simultaneous {\sc Andicam} $B\!-\!K$ photometry,
     new {\sc Hst/Cos} and {\sc Hst/Stis} UV-observations, a
     non-detection of CO\,$J\!=\!3\!\to\!2$ with {\sc Apex},
     re-analysis of a {\sc Ucles} high-resolution optical spectrum
     showing forbidden emission lines like [OI]\,6300\,\AA,
     [SII]\,6731\,\AA\ and 6716\,\AA, and [NII]\,6583\,\AA, and a
     compilation of existing broad-band photometric data.  We
     used the thermo-chemical disk code ProDiMo and the Monte-Carlo
     radiative transfer code MCFOST to model the protoplanetary disk
     around ET\,Cha. The paper also introduces a number of physical
     improvements to the ProDiMo disk modelling code concerning the
     treatment of PAH ionisation balance and heating, the heating by
     exothermic chemical reactions, and several non-thermal pumping
     mechanisms for selected gas emission lines. By applying an
     evolutionary strategy to minimise the deviations between model
     predictions and observations, we find a variety of united gas and
     dust models that simultaneously fit all observed line and
     continuum fluxes about equally well. Based on these models we can
     determine the disk dust mass with confidence, $M_{\rm
     dust}\!\approx\!(2-5)\times10^{-8}\,M_\odot$ whereas the
     total disk gas mass is found to be only little constrained, $M_{\rm
     gas}\!\approx\!(5\times10^{-5} - 3\times10^{-3})\,M_{\odot}$.
     Both mass estimates are substantially lower than previously
     reported. In the models, the disk extends from 0.022\,AU (just
     outside of the co-rotation radius) to only about 10\,AU,
     remarkably small for single stars, whereas larger disks are found
     to be inconsistent with the CO\,$J\!=\!3\!\to\!2$
     non-detection. The low velocity component of the [OI]\,6300\,\AA\
     emission line is centred on the stellar
     systematic velocity, and is consistent with being emitted from the
     inner disk. The model is also consistent with the line flux of
     H$_2$\,v$=$1$\to\,$0 S(1) at 2.122\,$\mu$m and with the
     [OI]\,63\,$\mu$m line as seen with {\sc Herschel/Pacs}. An
     additional high-velocity component of the [OI]\,6300\,\AA\
     emission line, however, points to the existence of an additional
     jet/outflow of low velocity $40\!-\!65\,$km/s with mass loss rate
     $\approx\!10^{-9}\rm\,M_\odot/yr$. In relation to our low
     estimations of the disk mass, such a mass loss rate suggests a
     disk lifetime of only $\sim\!0.05-3$\,Myr, substantially shorter
     than the cluster age. If a generic gas/dust ratio of 100 was
     assumed, the disk lifetime would be even shorter, only
     $\sim\!3000$\,yrs. The evolutionary state of this unusual
     protoplanetary disk is discussed.}


   \maketitle

%=====================================================================

\section{Introduction}

Gas-rich dust disks around young stars (hereafter, protoplanetary
disks) provide the raw material to build up new planets. The physical,
thermal, and chemical conditions in the disk, the timescale over which
the gas disperses, and the physical mechanisms contributing to the gas
dispersal are keys to understanding what type of planets can form and
on what timescales.

Significant progress has been made in the past few years in
measuring the dispersal timescale of the dust component of the
disk. Infrared surveys of nearby star-forming regions and associations
have established that the frequency of optically thick dust disks
decreases exponentially with time \citep[e.g.][]{Mamajek2009}. By an
age of 10\,Myr only a few percent of pre-main sequence Sun-like stars
(T\,Tauri stars) still retain an optically thick dust disk \citep[see
e.g.][for reviews]{Hernandez2008,Pascucci2010}. Since the near-mid
infrared excess ($\lambda\!\la\!30\,\mu$m) is sensitive to the
presence of small dust grains, not larger than a few microns in size,
these observations effectively trace the dispersal of small grains
within about 10\,AU from T Tauri stars. Millimetre observations,
tracing colder dust at hundreds of AU from the central star, indicate
a similarly fast clearing for the outer disk, within about
$10\!-\!30$\,Myr for T\,Tauri stars \citep{Carpenter2005}.  There is
growing observational evidence that the dust disk lifetime depends on 
stellar mass. Disks around intermediate-mass ($\ga\!1.5\rm\,M_\odot$)
stars disperse in less than 10\,Myr, whereas disks around low-mass
stars (M dwarfs and brown dwarfs) persist for longer times
\citep{Carpenter2006, Currie2007, Riaz2008}.

Due to observational challenges in detecting gas lines from disks and
difficulties in interpreting them, much less is known about the
evolution of the gas component of the disks. Three observables point to a
dispersal timescale similar to (or possibly shorter than) the dust
dispersal timescale: the exponential decrease with time in the frequency of
accreting stars \citep{Fedele2010}; the non-detections of
infrared gas lines from abundant molecules and atoms in tenuous dust
disks \citep{Hollenbach2005, Pascucci2006}; upper limits on
the H$_2$/dust mass ratio of less than 10 in two $\sim\!12\,$Myr old
edge-on disks \citep{Lecavelier2001, Roberge2005}.

\begin{table*}
\def\z{\hspace*{-1mm}}
\centering
\caption{Observed Line Fluxes $[10^{-18}\rm\,W/m^2]$ with {\sc
         Herschel/Pacs} and {\sc Apex}. Detection are listed as
         $F_L\pm\sigma$ whereas non-detections are listed as $<\!3\sigma$.}
\vspace*{-2mm}\hspace*{-2mm}
\resizebox{18.6cm}{!}{
\begin{tabular}{cccccccccc}
\hline\hline
  [OI]
& [OI]
& [CII]
& o-H$_2$O
& o-H$_2$O
& o-H$_2$O
& \z\z CO\,$J\!=\!36\!\to\!35$\z
& \z CO\,$J\!=\!33\!\to\!32$\z
& \z CO\,$J\!=\!29\!\to\!28$\z
& \z CO\,$J\!=\!3\!\to\!2$\z\\ 
  63.18\,$\mu$m 
& 145.52\,$\mu$m  
& 157.74\,$\mu$m  
& 78.74\,$\mu$m   
& \z 179.53\,$\mu$m\z   
& \z 180.49\,$\mu$m\z   
& 72.84\,$\mu$m   
& 79.36\,$\mu$m   
& 90.16\,$\mu$m 
& 866.96\,$\mu$m \\
\hline
&&&\\[-2ex]
  $30.5\pm3.2$ 
& $<$ 6.0 
& $<$ 9.0 
& $<$ 30
& $<$ 5.0
& $<$ 5.2
& $<$ 8.0
& $<$ 24  
& $<$ 9.6  
& $<$ 0.05 \\
\hline
\end{tabular}}
\label{obsflux}
\end{table*}

The aim of the {\sc Herschel} Open Time Key Program "Gas in Protoplanetary
Systems" (GASPS, Dent\etal in prep.) is to provide new insights into
the chemical and gas temperature structure of protoplanetary disks,
the gas/dust ratio, the gas dispersal timescale, and disk evolution.
%answer to the questions: {\sl Over what timescale does the
%majority of the primordial disk mass disperse?} and {\sl How much gas is
%left during the late epochs of planet formation?} Answering these
%questions will constrain the time available to form giant planets, and
%put new constraints on the availability of volatiles during
%terrestrial planet formation.
GASPS will acquire a large sample of sensitive far-infrared {\sc
Herschel/Pacs} spectra for 240 disks in nearby star-forming regions
and associations that span the critical 1-30\,Myr age range over which
disks are known to disperse. The primary signatures of the gas in the
disk are expected to be the forbidden [OI]\,63.2\,$\mu$m,
[OI]\,145.5\,$\mu$m, and [CII]\,157.7\,$\mu$m lines, as well as some CO
and H$_2$O lines.
%To support the interpretation of the PACS observations, the GASPS team has
%developed sophisticated radiative transfer models
%\citep[MCFOST:][]{Pinte2006}, coupled with gas thermo-chemical models
%\citep[ProDiMo:][]{Woitke2009a, Kamp2009}. 
The first GASPS papers have shown that (1) the [OI]\,63\,$\mu$m line can be
used as primary gas indicator and is often detected toward protoplanetary
disks \citep{Mathews2010}, (2) a combination of far-IR and (sub-)millimetre
gas lines provides a promising tool to estimate the total gas mass of
protoplanetary disks \citep{Pinte2010}, and (3) detailed models of individual 
sources allow to characterise the disk structure and 
shape, and the dust and gas components of protoplanetary disks
\citep{Meeus2010,Thi2010b}.

In this paper, we present an analysis of the circumstellar disk of
ET\,Cha, an approximately 8\,Myr old late-type T\,Tauri star, with the
goals of characterising in detail its dust and gas content. ET\,Cha is
one of the few nearby relatively old stars still possessing an
optically thick dusk disk \citep{Sicilia2009} and still accreting disk
gas \citep{Lawson2004}. TW\,Hya and PDS~66 are two other well-known
old stars with properties similar to ET\,Cha.  Both disks have been
studied in detail and show evidence of evolution with respect to
$1\!-\!2$\,Myr old T Tauri disks in Taurus, for example, depleted
inner disk in TW\,Hya \citep{Calvet2002} and flatter disk structure
for PDS\,66 \citep{Cortes2009}. Both disks are likely to have too low
disk masses to form giant planets at this evolutionary stage. ET\,Cha
would be the third such old disk system where observational data
allows for an in-depth-study of its dust and disk properties.

The paper is structured as follows. Section~2 provides an overview of
the prior knowledge of the source. We then present new
multi-wavelength observations of ET\,Cha in Sect.~3.  Section~4
presents a detailed dust and gas disk model for ET\,Cha. We describe the main
results of our models in Sect.~5. Finally, we discuss some critical
aspects of the modelling, and the implications of both models, in
Sect.~6, before we finish the paper with our conclusions in Sect.~7.

\section{ET\,Cha: an old T\,Tauri star with active accretion}
\label{sec:ETCha}

ET\,Cha (2\,MASS J08431857-7905181, ECHA J0843.3-7905, { 
also referred to as RECX\,15\footnote{The ROSAT survey reported by
\citep{Mamajek1999,Mamajek2000} lists only RECX\,1-12. Lables 13-15
have been used to denote three post-ROSAT stars discovered in or near
the cluster core, including ET\,Cha.}}) is a low-mass T\,Tauri star
that was identified by \citep{Lawson2002} as a member of the nearby,
8\,Myr old $\eta$ Chamaeleontis moving group
\citep{Mamajek2000}\footnote{We note that \citet{Luhman2004} derived
an age of the $\eta$ Cha association of only $6^{+1}_{-2}$\,Myr.}.
The association is located only 97\,pc away from the Sun
\citep{Mamajek1999} and is virtually unaffected by extinction
\citep{Luhman2004}, an ideal set of conditions to study circumstellar
disks in detail. {A slighly smaller distance to ET\,Cha of
94.3\,pc was reported by \citet{vanLeeuwen2007}, but we have used the
earlier and better known value of 97\,pc for the modelling in this
paper.} \citet{Brandeker2006} obtained high-angular resolution images
of ET\,Cha and concluded that it has no companions (brown dwarfs)
outside of 10\,AU (30\,AU).

ET\,Cha is one of the few association members that possess a
circumstellar disk, as indicated by a series of {\sc Spitzer} observations
that revealed the presence of dust including strong mid-infrared
silicate features \citep{Megeath2005, Bouwman2006, Gautier2008,
Sicilia2009}. Despite the age of the system, the infrared colours of
the source are reminiscent of those of much younger (1--2\,Myr)
circumstellar disks. Furthermore, optical spectroscopy of ET\,Cha has
shown that it is undoubtedly accreting, with a very strong and broad
H$_\alpha$ emission line \citep{Lawson2002, Lyo2004,
Luhman2004}. Based on the observed H$_\alpha$ line, the mass accretion
rate of ET\,Cha has been estimated to be $10^{-9}\rm\,M_\odot/yr$
\citep{Lawson2004}, and the disk inclination (by modelling the
H$_\alpha$ line profile) to be about $60\degr$ as measured from
face-on. The spectra also reveal a series of forbidden optical
emission lines ([OI], [SII], [NII]) that unambiguously indicate the
presence of a jet/outflow. In addition, the stellar absorption lines
in these spectra allowed for precise spectral typing of the central
star; all estimates agree with a $\rm M3\!-\!M3.5$ spectral
classification.

Among all members of the $\eta$ Cha association, ET\,Cha shows the
largest variability in the visible of order $0.3\!-\!0.4$\,mag, which
places it at the high end of the variability distribution of WTTS
\citep[see Fig.\,1 in][]{Grankin2008}. The main feature in the
observed lightcurve is a $\sim$\,12\,day period which has been found
in two consecutive years. \citet{Lawson2002} also noted a flare
lasting for about 1.7 days. The more regular 12\,day variations are
most easily interpreted in terms of an accretion hotspot co-rotating
with the star. However, our re-analysis of optical absorption line
profiles (see Sect.~\ref{sec:vsini}) suggests that the stellar
rotational period is much shorter, around 2\,days. Therefore, the
physical origin of the $\sim$\,12\,day period remains uncertain. The
stellar variability casts some doubt on the derivation of stellar
parameters, because the photometry reported by \citet{Lawson2002} was
taken ``near maximum light'', when \eg an accretion hot spot could
contribute significantly to the observed flux. Furthermore, it
prevents using optical and near-infrared colours in case of
non-simultaneous data. For this reason, we have obtained new,
simultaneous photometry, which is presented in
Sect.\,\ref{sec:ANDICAMphotometry}.

\citet{Howat2007} detected a clear gas signature from the disk of
ET\,Cha in form of the H$_2$\,v$=$1$\to\,$0 S(1) line in emission at
2.122\,$\mu$m with {\sc Gemini/Phoenix}, with an integrated line flux
of $\rm(2.5\pm0.1)\times 10^{-18}\,W/m^2$. These high angular
resolution observations showed a narrow line
(FWHM\,$=\!18\pm1.2$\,km/s) centred on the stellar velocity to within
$\sim\!1\,$km/s.  No angular offset between the line and the star was
detected at the level of 4\,AU. Therefore, \citet{Howat2007} argue
that this line is emitted by H$_2$ gas in Keplerian rotation at
$\sim$2\,AU.  \citet{Howat2007} observed 3 other disk-bearing members
of the $\eta$ Cha association, but ET\,Cha was the only one with
detectable H$_2$ emission. \citet{Bary2008} and
\citet{Martin-Zaidi2009, Martin-Zaidi2010} reported on several 
detections of H$_2$-lines toward other sources where the emission is
also likely originating from the disk rather than from an outflow.

\begin{figure}
\centering
\includegraphics[width=90mm,height=70mm]{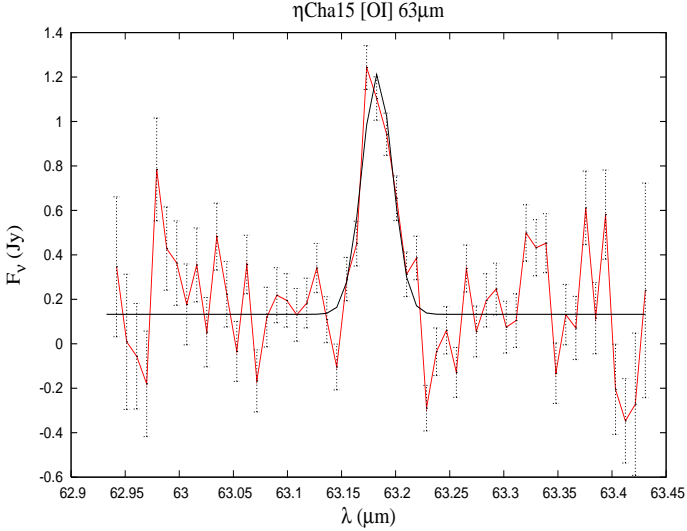}\\
\caption{{\sc Herschel/Pacs} line observations, with overplotted
Gaussian to measure the integrated line flux of
[OI]\,63.18\,$\mu$m. All other line observations are non-detections
as listed in Table~\ref{obsflux}. Errorbars indicate the 1\,$\sigma$ noise
level. The [OI]\,63.18\,$\mu$m line profile is dominated by the
spectral resolution of the PACS spectrometer $R\!\approx\!3000$.}
\label{fig:Herschel}
\vspace*{-1mm}
\end{figure}

%\begin{figure}
%\centering
%\includegraphics[width=40mm]{OI63_spaxel.eps}\\
%\caption{Map of [OI]\,63.2$\,\mu$m line flux toward ET\,Cha. The figure
%         shows the $5\times5$ ``spaxels'' of the PACS spectrometer
%         with the central pixel pointing toward ET\,Cha, covering
%         $45\arcsec\!\times 45\arcsec$ ($4000\,{\rm AU}\times 
%         4000\,{\rm AU}$).
%         {\bf TBD: Bill, where is north? Labels bigger, Color scale?}.}
%\label{fig:spaxel}
%\end{figure}

%\begin{figure*}
%  \centering
%  \begin{tabular}{ccc}      
%    \resizebox{50mm}{!}{\includegraphics[angle=0]{OI63_1_1.eps}} &
%    \resizebox{50mm}{!}{\includegraphics[angle=0]{OI63_1_2.eps}} &
%    \resizebox{50mm}{!}{\includegraphics[angle=0]{OI63_1_3.eps}} \\  
%    \resizebox{50mm}{!}{\includegraphics[angle=0]{OI63_2_1.eps}} &
%    \resizebox{50mm}{!}{\includegraphics[angle=0]{OI63_2_2.eps}} &
%    \resizebox{50mm}{!}{\includegraphics[angle=0]{OI63_2_3.eps}} \\ 
%    \resizebox{50mm}{!}{\includegraphics[angle=0]{OI63_3_1.eps}} &
%    \resizebox{50mm}{!}{\includegraphics[angle=0]{OI63_3_2.eps}} &
%    \resizebox{50mm}{!}{\includegraphics[angle=0]{OI63_3_3.eps}} \\ 
%  \end{tabular}
%  \caption{The 3$\times$3 spectral cube for ET\,Cha. The central
%      spaxel shows a clear identification of the [OI] 63.2\,$\mu$m
%      line. The spaxel [1,2] has a broad feature centered at
%      63.2\,$\mu$m. For the other neighboring spaxels, the mean
%      signal level is at 0.0, with no proper line identification. }
%  \label{fig:spaxel}
%\end{figure*}

\section{Observations and Data Reduction}

\subsection{Herschel/PACS}

{\sc Herschel/Pacs} observations were obtained for ET\,Cha during the
Science Demonstration Phase. The photometric observations were
obtained in the scan map mode in the blue (70\,$\mu$m) and the red
(160\,$\mu$m) filters. Two different scan map angles were used,
45$\degr$ (obsid 1342187338, 133s) and 63$\degr$ (obsid 1342189366,
220s). Both scans were obtained at a medium scan speed of
$20\arcsec$/s, with a cross scan step of $5\arcsec$ and scan leg
length of 3'. The number of scan legs was 8 for the 63$\degr$ scan,
and 4 for the 45$\degr$ scan. For spectroscopic observations, a 1669s
PacsLineSpec (obsid 1342186314) and a 5150s PacsRangeSpec (obsid
1342187019) were obtained. The PacsLineSpec provides two simultaneous
spectra at wavelengths $62.93\!-\!64.43\,\mu$m, and
$180.76\!-\!190.29\,\mu$m. PacsRangeSpec covers six spectral ranges of
$71.81\!-\!73.28\,\mu$m, $78.37\!-\!79.76\,\mu$m,
$89.28\!-\!90.48\,\mu$m, $143.59\!-\!146.53\,\mu$m,
$156.70\!-\!159.47\!\mu$m, and $178.51\!-\!180.96\,\mu$m. Observations
were taken in the chop-nod mode, with a small $2\arcsec$ dither. The
target was centred at the central spaxel of the
$9.4\arcsec\!\times\!9.4\arcsec$ grid of the {\sc Pacs} Integral Field
Unit. The data was reduced using the {\sc Herschel} Interactive
Processing Environment ({\sc Hipe}; Ott 2010) developer build version
3.0.1212, and the data reduction scripts provided at the {\sc
Herschel} data reduction workshop held in January 2010.

For the photometric data, a mosaic was created from the two scan
maps. Aperture photometry was performed using an aperture radius of
16\arcsec in the blue, and 19.2\arcsec in the red. An aperture
correction of 0.922 (blue), and 0.885 (red) was applied to the
photometry. The aperture corrections were obtained from the {\sc Pacs}
PhotChopNod Release Note (Feb.~22, 2010). The flux calibration
uncertainty is estimated to be 5\% in the blue and 10\% in the
red. For the spectroscopic data, we extracted the spectrum from the
central spaxel, and then applied an aperture correction in order to
minimise the flux loss in the neighbouring spaxels. Spectra from the
central spaxel were extracted for both the A and the B nods. We then
applied wavelength binning to each nod spectrum, using a bin size that
is half the width of the instrumental resolution. The final spectrum
is the mean of the wavelength-binned spectra from the two nods. The
absolute flux calibration uncertainty is estimated to be 40\%. We have
detected the [OI]\,63.2$\,\mu$m emission line for ET\,Cha, while all
other lines are undetected (see Table~\ref{obsflux}). We used the IDL
routine MPFITPEAK to fit an error-weighted Gaussian to the observed
[OI]\,63.2$\,\mu$m line, and measured the integrated flux of the
Gaussian line fit. The 1-$\sigma$ error to the line flux was
calculated by setting the height of the Gaussian equal to the
continuum rms value, and the width equal to the instrumental
resolution. The continuum emission at the rest wavelength of
63.18$\,\mu$m was estimated by fitting a first-order polynomial to the
spectral region. 

%The measured FWHM of the [OI]\,63.2$\,\mu$m line is 0.032\,$\mu$m,
%which is a bit larger than 0.021\,$\mu$m, the value expected from the
%nominal spectral resolution of the {\sc Pacs} spectrometer
%($R\!\approx\!3000$ at 60$\,\mu$m).  This indicates that the line
%could be marginally resolved. See futher discussion of this issue in
%Sect.~\ref{sec:PACS}.

%{\bf TBD: Figure~\ref{fig:spaxel} shows the angular distribution of
%the [OI]\,63.2$\,\mu$m line flux around ET\,Cha. The line is detectable
%only in the central spaxel, the line fluxes in the neighboring spaxels
%are below the $3\sigma$ noise level, suggesting that most of the
%[OI]\,63.2$\,\mu$m line is emitted from the innermost $\sim\!800$\,AU.}

\subsection{CTIO/ANDICAM photometry}
\label{sec:ANDICAMphotometry}

\begin{table}
\caption{Photometric data of ET\,Cha.}
\label{tab:photometry}
\vspace*{-2mm}\hspace*{-1mm}\resizebox{91mm}{!}{
\begin{tabular}{c|c|ccc}
  $\lambda\rm\,[\mu m]$ & mag. & $F_\nu\rm\,[Jy]$ 
                        & instrument & ref.\\
\hline
\multicolumn{5}{l}{used data ...}\\
\hline
$0.091-0.111$ & -- & 9.81e-6              & {\sc Fuse} (scaled)  & GH\\
$0.111-0.145$ & -- & 2.85e-4              & {\sc Hst/Cos}        & GH\\
$0.145-0.205$ & -- & 1.69e-4              & {\sc Hst/Cos/Stis}   & GH\\
 0.442 (B) & 15.64 & $0.00195\pm 0.0001$  & {\sc Ctio/Andicam}   & GD\\
  0.55 (V) & 14.68 & $0.0046 \pm 0.00025$ & {\sc Ctio/Andicam}   & GD\\ 
  0.66 (R) & 13.44 & $0.0111 \pm 0.0005$  & {\sc Ctio/Andicam}   & GD\\
  0.82 (I) & 12.23 & $0.033  \pm 0.002$   & {\sc Ctio/Andicam}   & GD\\ 
  1.23 (J) & 10.44 & $0.107  \pm 0.005$   & {\sc Ctio/Andicam}   & GD\\ 
  1.63 (H) & 9.79  & $0.124  \pm 0.006$   & {\sc Ctio/Andicam}   & GD\\ 
  2.19 (K) & 9.32  & $0.125  \pm 0.006$   & {\sc Ctio/Andicam}   & GD\\ 
  3.60     & 8.38  & $0.125  \pm 0.003$   & {\sc Spitzer/Irac}   & M\\
  4.50     & 7.91  & $0.123  \pm 0.001$   & {\sc Spitzer/Irac}   & M\\ 
  5.80     & 7.42  & $0.124  \pm 0.003$   & {\sc Spitzer/Irac}   & M\\ 
  8.00     & 6.51  & $0.162  \pm 0.001$   & {\sc Spitzer/Irac}   & M\\ 
  24.0     & 3.52  & $0.280  \pm 0.003$   & {\sc Spitzer/Mips}   & S\\
$7.6-37$   &       & \multicolumn{2}{l}{{\sc Spitzer/Irs} 
                     low resolution spectrum}                    & B,S\\
  70.0     & -- &    $0.18   \pm 0.02$    
                     & $\!\!\!${\sc Herschel/PacsPhot}$\!\!\!$   & BR\\
 160.0     & -- &    $0.069  \pm 0.007$   
                     & $\!\!\!${\sc Herschel/PacsPhot}$\!\!\!$   & BR\\
 870.0     & -- &        $<0.036$         & {\sc Apex/Laboca}    & NP\\  
\hline
\multicolumn{5}{l}{unused data ...}\\
\hline
  0.45  (B) & 15.07 & $0.00399          $  & MSSSO 2.3m         & L,BR\\
  0.558 (V) & 13.97 & $0.00940          $  & SAAO 1m            & La,BR\\
  0.695 (R) & 12.98 & $0.0199           $  & SAAO 1m            & La,BR\\
  0.90  (I) & 11.77 & $0.0494           $  & SAAO 1m            & La,BR\\
  1.24 (J)  & 10.51 & $0.102   \pm 0.003$  & 2{\sc Mass}        & T,NP\\
  1.65 (H)  &  9.83 & $0.125   \pm 0.004$  & 2{\sc Mass}        & T,NP\\
  2.17 (K)  &  9.43 & $0.114   \pm 0.004$  & 2{\sc Mass}        & T,NP\\
  3.80 (L') &  8.14 & $0.14    \pm 0.006$  & {\sc Vlt/Icsaac}   & H,IP\\
  25.0      & -- &    $0.298   \pm 0.03$   & {\sc Iras}         & I\\
  60.0      & -- &    $0.281   \pm 0.04$   & {\sc Iras}         & I\\
% 24.0      & -- &    $0.2325  \pm 0.0001$ & {\sc Spitzer/Mips} & G\\
% 70.0      & -- &    $0.1733  \pm 0.0003$ & {\sc Spitzer/Mips} & G\\
  70.0      & -- &    $0.137   \pm 0.008$  & {\sc Spitzer/Mips} & S\\
 160.0      & -- &          $<0.12$        & {\sc Spitzer/Mips} & G\\
\end{tabular}}\\[2mm]
{\footnotesize Detections are listed as $F_\nu\pm\sigma$ whereas
  non-detections are listed as $<$\,$3\sigma$. Abbreviations for
  references, data reduction and flux conversion are: GH $\!=\!$ Greg
  Herczeg, this paper; GD $\!=\!$ G.~Duch{\^e}ne, this paper; BR
  $\!=\!$ B.~Riaz, this paper; NP $\!=\!$ N.~Phillips, this paper; IP
  $\!=\!$ I.~Pascucci, this paper; M $\!=\!$ \citet{Megeath2005}; S
  $\!=\!$ \citet{Sicilia2009}; B $\!=\!$ \citet{Bouwman2006}; L
  $\!=\!$ \citet{Lyo2004}; La $\!=\!$ \citet{Lawson2002}; H $\!=\!$
  \citet{Haisch2005}; T $\!=\!$ \citet[2MASS Point Source
    Catalogue][]{2MASS}; I $\!=\!$ \citet[IRAS Faint Source
    Catalogue][]{IRASFSC}; G $\!=\!$ \citet{Gautier2008}.}
\end{table}

We obtained new simultaneous optical/NIR photometry using the
dual-channel ANDICAM instrument on the CTIO 1.3\,m telescope on March
6, 2009. Photometric calibration was performed using the PG\,1047
Landolt field. Both the optical and near-infrared data were reduced
using standard procedures (flat-fielding, cosmetic cleaning, shift-and
add). The photometry on ET\,Cha and the photometric standard was
extracted within a 3\arcsec aperture and airmass and colour
corrections were applying using coefficients from the ANDICAM
website\footnote{http://www.astro.yale.edu/smarts/smarts13m/photometry.html}
in the optical and from \citet{Frogel1998} in the near-infrared. The
resulting photometry is listed in Table\,\ref{tab:photometry} along
with the mid- and far-infrared fluxes adopted in our analysis. The new
photometric fluxes are substantially lower than previously published
fluxes, see Table~\ref{tab:photometry}.

\subsection{APEX/LABOCA photometry}

An upper limit at $870\,\mu$m has been obtained from new continuum
maps of the $\eta$ Chamaeleontis association taken with the LABOCA
bolometer array on {\sc Apex} \citep{Siringo2009}. The data was reduced
using the Bolometer array data Analysis (BoA) software, with a
pipeline optimised for faint compact sources. Fluxes were extracted
within BoA by fitting the amplitude of a beam-sized Gaussian at
specified positions within a map with a pixel scale of
$4.6\arcsec/{\rm px}$. For each target the flux was extracted at
points in a $5\times5$ rectangular grid centred on the target with a
spacing of $37\arcsec$ (twice the beam FWHM). The sample standard
deviation of the 24 off-source measurements is the 1-$\sigma$
uncertainty quoted here. Using aperture photometry instead of beam
fitting, with a variety of aperture sizes, and aperture corrections
computed from the beam, yields very similar results ($9 \pm 11\,{\rm
mJy}/{\rm beam}$ for apertures with $r={\rm HWHM}$).

\subsection{UV observations}

ET\,Cha was observed in the far-ultraviolet with the Cosmic Origins
Spectrograph (COS) and in the near-ultraviolet and blue with the Space
Telescope Imaging Spectrograph (STIS) on board of the Hubble Space
Telescope (HST) as part of the programme ``Disks, Accretion and
Outflows of T\,Tau Stars'' (P.I. G.~Herczeg) on 5 Feb.~2010. The
far-ultraviolet observations were obtained with the G130M and G160M
gratings with 3891s and 4501s integration times, respectively, to
cover the 1140--1790\,\AA\ spectral range with a resolution of
$\sim$18\,000.  The data were reduced with the COS calibration
pipeline CALCOS and individual segments were combined with the IDL
coaddition procedure described by \citet{Danforth2010}.  The
near-ultraviolet and blue observations were obtained with the G230L
and G430L gratings with integrations of 45s and 680s,
respectively, to cover the 1700-5700 \AA\ spectral range with a
resolution of $\sim$2000.  The data were reduced with custom-built
software in IDL.

\begin{figure}[!t]
\centering
\includegraphics[width=88mm]{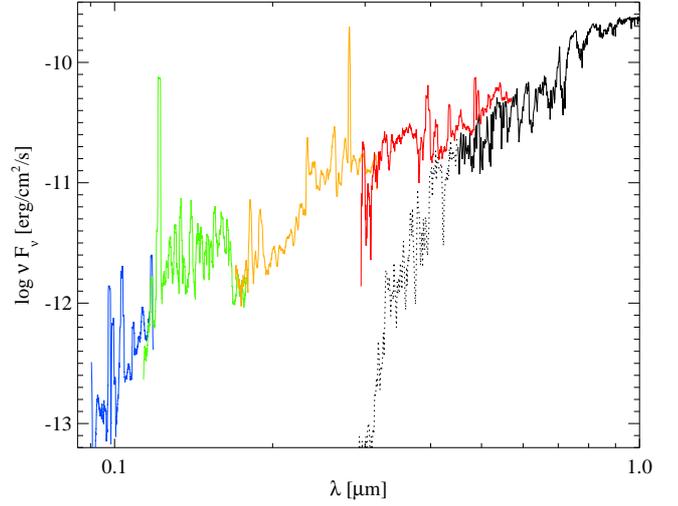}
\vspace*{-1mm}
\caption{Observed UV spectra and the stellar atmosphere
model spectrum for ET\,Cha (see Sect.~\ref{sec:star}). Blue: {\sc
Fuse} spectrum of TW\,Hya divided by 20. Green: {\sc Hst/Cos} spectrum
of ET\,Cha. Orange and red: G230L and G430L {\sc Hst/Stis} spectra of
ET\,Cha. Black: stellar atmosphere model spectrum with $T_{\rm
eff}\!=\!3400\,$K, $\log\,g\!=\!3.89$ and
$R_\star\!=\!0.84\rm\,R_\odot$, at a distance of 97\,pc, plotted as 
dotted line (unused) for $\lambda\!<\!450\,$nm.}
\label{fig:UV}
\end{figure}

A {\sc Fuse} spectrum of ET\,Cha covers the spectral region shortward
of 1140\,\AA\ but has poor signal-to-noise.  Instead, the
912--1140\,\AA\ flux was estimated from a high-quality {\sc Fuse}
spectrum of TW\,Hya \citep{Johns-Krull2007}, scaled to the flux in the
1250--1700\,\AA\ bandpass. All different data have been smoothed to a
common low resolution, see Fig.~\ref{fig:UV}. Integrated fluxes over
three different UV bands as listed in Table~\ref{tab:UV}.

\begin{table}
\vspace*{-1mm}
\centering
\caption{Band integrated UV-fluxes for ET\,Cha.}
\vspace*{-2.5mm}
\begin{tabular}{c|c}
spectral band [\AA] & integrated spectral flux $[\rm\,erg/cm^2/s$]\\
\hline
&\\[-2.1ex]
\ \ $912-1110$ & $8.03\times10^{-14}$\\
$1110-1450$ & $1.39\times10^{-12}$\\
$1450-2050$ & $9.73\times10^{-13}$
\end{tabular}
\label{tab:UV}
\vspace*{-3mm}
\end{table} 

\subsection{Re-analysis of high-resolution optical spectroscopy}
\label{sec:optical}

A high-resolution optical spectrum of ET\,Cha was acquired on June 26,
2002 with the 3.9m Anglo-Australian Telescope {\sc Aat} and University
College London coud{\'e} echelle spectrograph ({\sc Ucles}) as first
published by \citet{Lawson2004}. A 1.5\arcsec slit was used delivering
a resolving power of $\sim$30\,000 ($10$\,km/s at
6300\,\AA) and covering the wavelength range between 4980 and
9220\,\AA. The spectrum was calibrated using dome-flats, bias frames
and ThAr arc frames, making use of standard library routines such as
$\sf ccdproc$ within $\sf IRAF$ \citep[see][for a detailed
description]{Lawson2004}. We have paid particular attention to the
removal of telluric contributions to these lines. We removed the
OI\,6300\,\AA\ telluric contribution using the RECX\,10 spectrum, a
star of similar spectral type but without any evidences of an
accretion disk or outflow \citep{Lyo2003}, meaning that the measured
OI\,6300\,\AA\ emission from RECX\,10 is thoroughly due to the
atmosphere of the Earth.

The average air-masses during observations of ET\,Cha and RECX\,10
are 2.6 and 2.7, respectively. We made a [OI]\,6300\,\AA-map by first
removing the [OI] feature from the RECX\,10 spectrum, and then
subtracting this edited spectrum from the original RECX\,10
spectrum. The ET\,Cha spectrum was divided by the [OI]\,6300\,\AA-map
to remove the OI telluric line. We used the photospheric Li\,I
absorption lines at 6707.76\,\AA\ and 6707.91\,\AA\ to measure a
stellar radial velocity of $\sim$22\,km/s ($\sim$34.6\,km/s after
heliocentric correction).

\begin{figure}
\includegraphics[width=88mm,height=75mm]{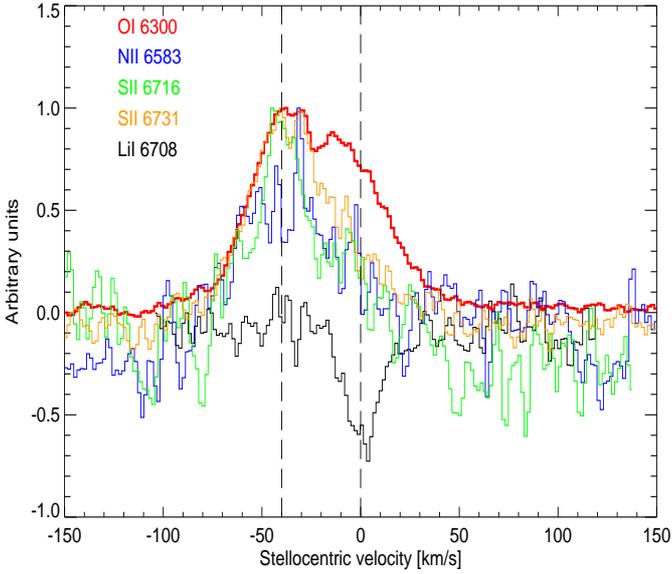}
\caption{Optical line spectra of ET\,Cha. All line profiles are
  continuum subtracted and normalised to line peak
  emission$\,=\,$1. The forbidden oxygen line [OI]\,6300\,\AA\ is the
  strongest emission line by a factor 5 or more, see
  Table~\ref{tab:OpticalLines}, hence the increased noise level in the
  other lines. The dip in the [OI]\,6300\,\AA\ line coincides with the
  position of the telluric component which has been carefully removed,
  see text. The LiI\,6708\,\AA\ absorption line has been used to
  determine the stellar velocity.}
\label{fig:OI6300}
\end{figure}

\begin{table}
\begin{center}
\caption{Measured optical emission line properties.}
\label{tab:OpticalLines}
\vspace*{-2mm}
\def\z{\hspace*{-1mm}}
\begin{tabular}{cc|cccc}
line & \z remarks\z & EW\,[\AA]\z 
        & \z$L_{\rm line}\,[10^{-7}L_\odot]$\z 
        & \z$v_{\rm cen}$\z & \z FWHM\z \\
\hline
\z$\rm[OI]\,6300\,\AA$\z  & HVC & -6.0    & $170-260$   & -42  & $47\pm\!15$\\
\z$\rm[OI]\,6300\,\AA$\z  & LVC & -5.4    & $150-230$   & 0    & $38\pm\!15$\\
\z$\rm[SII]\,6731\,\AA$\z & all & -1.8    &  $50-76$    & -35  & n.a. \\
\z$\rm[SII]\,6716\,\AA$\z & all & -0.5    &  $14-21$    & -32  & n.a. \\
\z$\rm[NII]\,6583\,\AA$\z & all & -0.05   & $1.4-2.1$   & -40  & n.a. \\
\end{tabular}\\[0mm]
\end{center}
{\footnotesize EW$\,=\,$equivalent width, HVC$\,=\,$high-velocity
  component, LVC$\,=\,$low-velocity component.  The luminosity
  interval corresponds to the uncertainty in the red continuum as
  derived from photometry $R\!=\!(13.44\!-\!12.98)\,$mag, compare
  Table~\ref{tab:photometry}. Negative values for the centre velocity
  $v_{\rm cen}$ indicate a blue-shift, n.a. means no values
  derived. $v_{\rm cen}$ and FWHM are in [km/s].}
\end{table}

\begin{figure}
\centering
\includegraphics[angle=270, width=91mm]{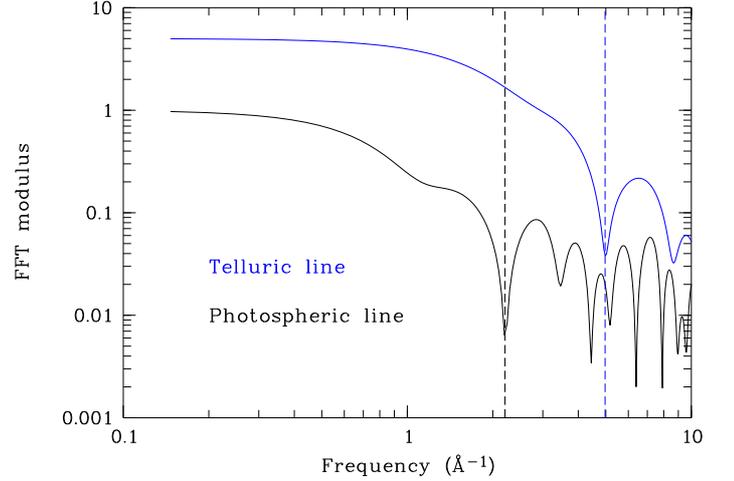}
\caption{ $v_{\rm rot} \sin i$ determination: line profile Fourier
  transforms. The Fourier transform of a photospheric line at 6709
  \AA\ (black line) and a telluric line at 9196 \AA\ (blue line) are
  shown. The location of the first minimum of the photospheric line
  profile provides a measurement of $v_{\rm rot} \sin i$.  The method
  requires that the main line broadening mechanism is rotation velocity,
  that is, the first minimum of the photospheric line must be located at
  a lower frequency than telluric or arc lines, as shown in the
  figure.  Arbitrary normalisation constants have been applied to the
  curves to improve readability.}
\label{Figure:vsiniFft}
\end{figure}

Figure~\ref{fig:OI6300} shows the observed profiles of ET\,Cha in the
oxygen line [OI]\,6300\,\AA, with 3 other optical forbidden emission
lines overplotted that trace outflows, plus a Li\,I absorption line to
determine the systematic stellar velocity. The [OI]\,6300\,\AA\ line
shows a broad component centred around the stellar systematic velocity
(low-velocity component LVC), and a blue component shifted by about
42\,km/s (high-velocity component HVC).  We have fitted the HVC and
LVC by two Gaussian profiles.  Measured equivalent widths are listed
in Table~\ref{tab:OpticalLines}. Equivalent widths for these lines are
converted to line luminosities using the procedure outlined in
\citet{Hartigan1995}, assuming a distance of 97\,pc and zero
visual extinction. As expected for outflows, the [NII]\,6583\,\AA,
[SII]\,6716\,\AA\ and [SII]\,6731\,\AA\ emission lines emphasise the
HVC, and are all blue-shifted by about the same margin.

\begin{figure*}
\centering
\includegraphics[width=13.5cm]{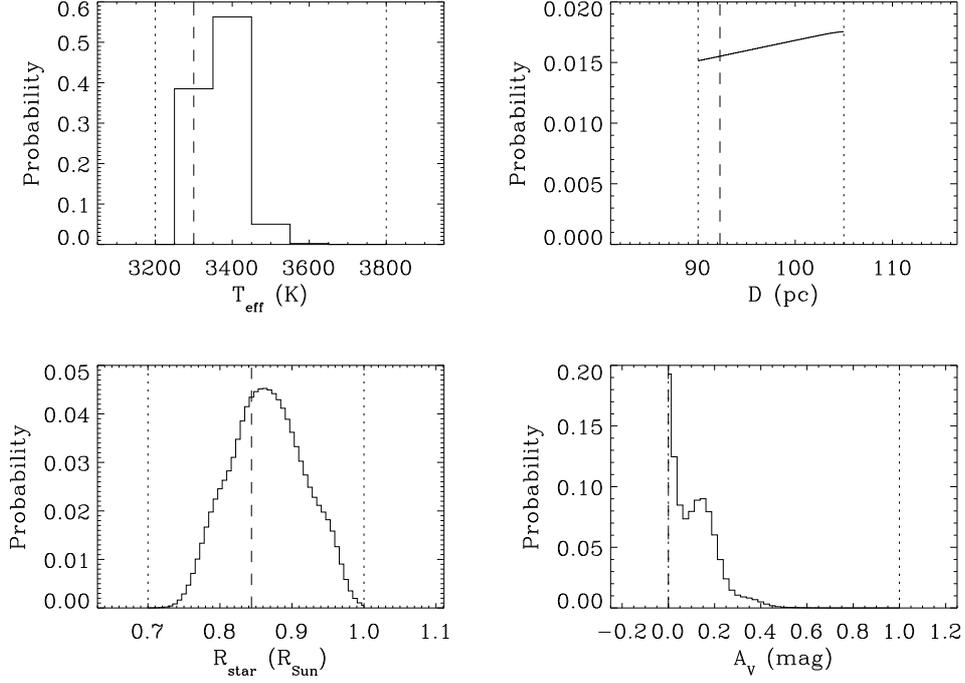}
\caption{Bayesian probability distributions for effective temperature
  $T_{\rm eff}$, visual extinction $A_V$, distance $D$ and stellar
  radius $R_\star$ for ET\,Cha using our new simultaneous $BVRIJ$
  photometry (solid histograms). The vertical dotted lines bracket the
  range of values explored for each parameter while the vertical
  dashed line represent the absolute best fit photospheric model.}
\label{fig:stellar}
\end{figure*}

\subsection{Projected stellar rotational velocity}
\label{sec:vsini}

The high resolution spectrum presented in Sect.~\ref{sec:optical} was
additionally used to determine the projected rotational velocity of the star
$v_{\rm rot} \sin i$. This quantity is typically derived from empirical
relations between the full-width at half-maximum of stellar absorption
lines and $v_{\rm rot} \sin i$ \citep[see e.g.][]{Martinez2010}. However, in
this paper we use the Fourier transform of the line profile
\citep{Gray1992}.  The full power of this method is revealed in case
of very high spectral resolution and signal to noise
\citep{Reiners2003}, but has also been applied successfully
to moderate and low quality data \citep{Mora2001}, using the
following simplifications.

\begin{figure}[!b]
\centering
\includegraphics[width=75mm]{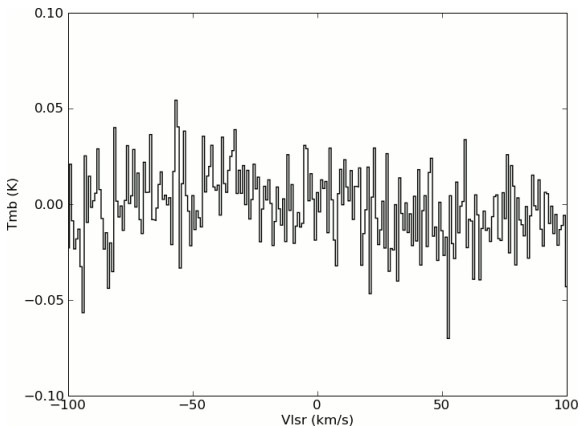}
\caption{Deep APEX spectrum of ET\,Cha at 345.796\,GHz
  (866.963\,$\mu$m), showing the mean beam temperature $T_{\rm mb}$
  versus local standard of rest velocity $v_{\rm lsr}$. The $^{12}$CO
  $J\!=\!3\!\to\!2$ line was not detected.}
\label{fig:CO32}
\end{figure}

The location of the first minimum of the Fourier transform of any
isolated photospheric line profile provides a measurement of $v \sin
i$.  If many lines are available, the average value and standard
deviation can be estimated.  The advantage of this method is that the
measurements are independent of empirical calibrations and synthetic
spectra, provided that enough lines can be identified and the dominant line
broadening mechanism is stellar rotation.  In terms of Fourier
analysis, the latter requirement is equivalent to the following
condition: the first minimum of the Fourier transform of photospheric
line profiles must be at a smaller frequency than the Fourier
transform of lines not affected by rotation velocity (\eg arc or
telluric lines). In addition, the algorithm is robust against the
introduction of noise, and capable of working with low signal to noise
data.

The method has been applied to the visible {\sc Ucles} spectrum of ET\,Cha
as described in Sect.~\ref{sec:optical}. Few lines could be used for
the analysis, due to the low signal to noise ratio ($\sim$10 for
unbinned data). A synthetic Kurucz spectrum with physical parameters
similar to those listed in Sect.~\ref{sec:star} was used to identify
isolated atomic photospheric lines.  After clipping bad lines, only 16
lines remained in the final average used below.
Figure~\ref{Figure:vsiniFft} shows a sample Fourier transform of both
a photospheric and a telluric line to illustrate the method. We obtain
\begin{equation}
  v_{\rm rot} \sin i = (15.9 \pm 2.0)\ {\rm km/s} \ .
\end{equation}

\subsection{APEX $^{12}$CO $J\!=\!3\!\to\!2$}

The source was studied in the $J\!=\!3\!\to\!2$ rotational transition
of $^{12}$CO at 345.796\,GHz using the {\sc Apex} telescope and
heterodyne receiver system twice, in April and July 2010, for a total
on-source time of 100 minutes. A standard on-off switching scheme was
used, with reference position $180\arcsec$ off in Azimuth. With an
aperture efficiency of $\eta_{\rm mb}\!=\!0.6$, the effective area of
the telescope is 67.9\,m$^2$, and the beam size $\Omega_A$ at
345.796\,GHz is 13.7$\arcsec$. The CO\,$J\!=\!3\!\to\!2$ line is
commonly detected from disks of radii 100\,AU or more; however, in the
case of ET\,Cha, the line was not detected (Fig.\ref{fig:CO32}), down
to an rms of 17\,mK (antenna temperature) for 0.425\,km/s wide
velocity bins. Assuming the underlying linewidth of the emitting
region is 7.5\,km/s, as predicted by the best fitting model
(Sect.~\ref{sec:results}) and typical of CO lines from disks, the
upper limit to the line flux is $3\sigma =
5.5\times10^{-20}\rm\,W/m^2$.

%======================================================================
\section{Disk modelling}

The strong near--mid IR continuum excess of ET\,Cha (see
Fig.~\ref{fig:SED}) as well as the narrow H$_2$\,v$=$1$\to\,$0 S(1)
emission line at 2.122\,$\mu$m \citep{Howat2007} indicate the
presence of a gas-rich protoplanetary disk. The optical
high-resolution spectrum with blue-shifted emission lines shows,
however, that in addition to the disk, there also exists an outflow
that might contribute to the gas emission lines. In the following, we
present a combined gas+dust disk model to explore in how far a disk
model alone can explain all photometric and spectroscopic observations
of ET\,Cha. The possible contribution of an outflow is discussed in
Appendix~\ref{sec:outflow}.

The disk modelling procedure in this paper consists of three
phases. In phase~1, we make a Bayesian analysis, based on the new
$BVRIJ$ photometric data, to obtain best fitting values for the
stellar parameters. In phase~2, we fix a few more parameters like the
inner disk radius based on physical assumptions, e.g., according to
stellar luminosity and dust sublimation temperature, and in
phase~3, we use the thermo-chemical code ProDiMo \citep{Woitke2009a,
Kamp2009} to calculate the gas- and dust temperature structure in the
disk, the chemical and ice composition, and to fit all remaining disk,
dust, and gas parameters of the model to the observations as good as
possible.

\subsection{Stellar parameters}
\label{sec:star}

While our new $JHK$ photometry is consistent with the 2\,{\sc Mass}
photometry, the $VRI$ photometry is about $0.5\!-\!0.7$\,mag fainter
than the data published in \citep{Lawson2002}. Indeed, our
simultaneous photometry appears to represent the 'faint' state for
ET\,Cha which, in the accretion hotspot scenario, is purely
photospheric.

To estimate the stellar properties, we therefore adopt our new
simultaneous photometry. We only fit the $BVRIJ$ fluxes to ensure that
our estimates are not biased by the near-IR excess from the disk. We
computed a grid of photospheric models, varying $T_{\rm eff}$ (using
the $\log g\!=\!4.5$ NextGen models from \citep{Allard1997}, $A_V$
\citep[using the $R_V\!=\!3.1$-law from][]{Cardelli1989}, $R_\star$
and the distance $d$ to ET\,Cha within generous ranges. The optical
extinction is defined by $A_V\!=\!2.5\log_{10}({\rm e})\,\tau_V$ where
$\tau_V$ is the optical depth in the visual due to interstellar dust
extinction.  We note that the last two parameters are degenerate.

The results are shown in Fig.~\ref{fig:stellar}. A very good fit is
obtained with $T_{\rm eff}\!=\!3300-3400$\,K, $d\!=\!97-100$\,pc,
$R_\star\!=\!0.84\!-\!0.89\,R_\odot$ (corresponding to
$L_\star\!=\!0.085\,L_\odot$) and $A_V\!\leq\!0.2$\,mag. This is in
excellent agreement with the findings of \citet{Luhman2004}, including
the effective temperature which they derived exclusively from their
spectrum and not from photometry. For the disk modelling, we adopt
$T_{\rm eff}\!=\!3400$\,K and $R_\star\!=\!0.84\,R_\odot$, which is
both the most probable combination of parameters and the closest to
the spectroscopically derived stellar properties.

Finally, we note that conducting the same analysis on the previous
photometric dataset (obtained in the ``bright'' state) yields both a
hotter central star ($T_{\rm eff}\!\approx\!\!4000$\,K) and a
substantial foreground extinction ($A_V\!=\!0.75$\,mag), both of which
are inconsistent with our prior knowledge of the source (see
Sect.~\ref{sec:ETCha}).

In the following, we use the nearest NextGen stellar input spectrum
for $T_{\rm eff}\!=\!3400\,$K, $L_\star\!=\!0.085\,\rm L_\odot$ and
solar metallicities (resulting in $\log\,g\!=\!3.89$ and
$R_\star\!=\!0.84\,\rm R_\odot$ for $M_\star\!=\!0.2\,\rm M_\odot$).
According to the stellar evolutionary models of \citet{Siess2000}, our
choice of $T_{\rm eff}$ and $L_\star$ is consistent with a
$6\!-\!8$\,Myr old star of mass $(0.2\!-\!0.3)\,\rm M_\odot$ and solar
metallicities.

\subsection{Stellar UV-excess}

\begin{figure}
\vspace*{-7mm}
\hspace*{-3mm}\includegraphics[width=9.6cm]{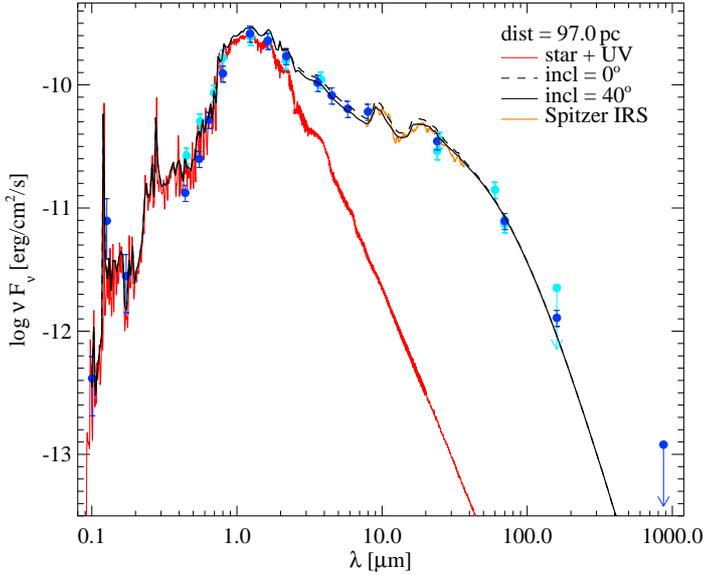}
\vspace*{-10mm}
\caption{Comparison of model SED to observations. Blue bullets
  are photometric data points used for the fitting (see
  Table~\ref{tab:photometry}). Light blue symbols refer to other,
  unused photometry points. Symbols with downward arrows indicate
  $3\sigma$ upper limits. The black line is the SED from our
  best-fitting model. The red line shows the assumed
  stellar input spectrum, completed by piecewise constant fluxes
  $F_\nu$ in three UV bands as obtained by integrating over our 
  UV observations (Table~\ref{tab:UV}).}
\vspace*{-2mm}
\label{fig:SED}
\end{figure}

In the UV, stellar activity and mass accretion create an excess
emission with respect to classical stellar atmosphere
models. Therefore we use our composite observed UV spectrum of ET\,Cha
(Fig.~\ref{fig:UV}) at $\lambda\!<\!450\,$nm as input
instead. Although strong Ly$\alpha$ emission from ET\,Cha was detected
with {\sc Hst/Cos}, no emission was detected near line centre because
of neutral hydrogen in our line of sight to the star.  Based on an
analysis of a similar Ly$\alpha$ emission profile from TW\,Hya
\citep{Herczeg2004}, we roughly estimate that the total flux in this
line is twice the detected value. Therefore, we have doubled the
detected flux between 1210\,\AA\ and 1220\,\AA\ for the construction
of our UV input spectrum ``as seen by the disk''. This yields an
integrated flux of $2.22\times10^{-12}\rm\,erg/cm^2/s$ in the
1110--1450\,\AA\ band, in contrast to Table~\ref{tab:UV}.  We have
converted these integrated UV fluxes into additional photometry
points, see Table~\ref{tab:photometry}, and plotted these points with
large errorbars in Fig.~\ref{fig:SED}.  These data result in a
fractional UV luminosity $f_{\rm UV}\!=\!L_{\rm UV}/L_\star\!=\!0.017$
with the UV luminosity $L_{\rm UV}$ being integrated from 912\,\AA\ to
2500\,\AA\ as introduced by \citet{Woitke2010}. The ``extra UV'' is
not important for the spectral energy distribution (SED) modelling, but
is essential for the gas modelling, because the UV irradiation is a
decisive factor for the gas heating and chemistry in the disk.

\subsection{Disk inclination and inner radius}
\label{sec:diskpara}

For the disk modelling in Sect.~\ref{sec:diskfit} we have fixed two
parameters to reduce the dimension of parameter space for the fitting
problem, namely the disk inclination and inner radius.

First, the disk inclination is assumed to be $i\!=\!40\degr$ as
measured from face-on orientation. As long as the disk is not
intersecting the line of sight to the star, this parameter does not
have a large impact on the model results concerning both SED-analysis
and line flux predictions, and we do not have clear observations, e.g.,
images, that would allow us to determine this quantity
unambiguously. Large values for the disk inclination are supported by
the H$_\alpha$ line analysis \citep[$60\degr$,][]{Lawson2004}, the low
outflow velocities observed (Sect.~\ref{sec:optical}), and by the
strongly rotation broadened stellar absorption lines
(Sect.~\ref{sec:vsini}), favouring an edge-on rather than a face-on
disk orientation.

However, our disk modelling suggests large disk scale heights in the
inner disk regions, needed to reproduce the strong near-IR excess (see
Fig.~\ref{fig:struc}). For inclinations of 60\degr and higher, the
observer's line of sight to the central star would therefore intercept
the disk. This would cause a dramatic reduction of the observed fluxes
at optical and UV wavelengths, as well as a strong reddening of the
optical colours. This is inconsistent with our initial assumption that
the star is seen directly, used to estimate the stellar properties. In
principle, one could increase both the intrinsic stellar luminosity
and effective temperature of the star to compensate exactly for this
reddening, but to match the observed SED, the required $T_{\rm eff}$
is well beyond values consistent with the M3--M3.5 spectroscopic
classification of ET\,Cha. For inclinations $\ga\!60\degr$ we
furthermore observe that the 10\,$\mu$m and $20\,\mu$m silicate
features are no longer in emission, which clearly contradicts of
observed SED of ET\,Cha.

Second, concerning the disk inner radius, the strong and continuous
near-IR fluxes of ET\,Cha suggest a disk which extends inward close to
the star, with no dust holes or gaps. T\,Tauri disks are generally
assumed to be truncated by the stellar magnetosphere near the
co-rotation radius, with material accreting along magnetic field lines
onto high-latitude regions of the star
\citep{Koenigl1991,Shu1994}. From the analysis of optical absorption
lines of ET\,Cha (Sect.~\ref{sec:vsini}) we have inferred a projected
stellar rotation velocity of $v_{\rm rot} \sin i\!=\!(15.9 \pm
2.0)\,$km/s.  At $i\!=\!40\degr$, this translates into $v_{\rm
rot}\!=\!(25 \pm 3)\,$km/s (about 12\% of the break-up velocity) and,
by assuming $M_\star\!=\!0.2\rm\,M_\odot$ and
$R_\star\!=\!0.84\,R_\odot$ (Sect.~\ref{sec:star}), into a stellar
rotation period of $P\!=\!(1.7 \pm 0.2)$\,days, and a co-rotation
radius of $(0.015 \pm 0.002)$\,AU.

However, the dust temperature at the co-rotation radius ($0.015$\,AU)
is larger than 1500\,K, associated with the sublimation temperature of
silicates. Therefore, we have assumed a slightly larger inner disk
radius, $R_{\rm in}\!=\!0.022\,$AU, where the dust temperature is
about $1500\,$K.

\subsection{A single disk model}
\label{sec:diskmodel}

For a single disk model, we use the radiation thermo-chemical disk
code ProDiMo \citep{Woitke2009a, Kamp2009} to calculate the dust
continuum radiative transfer, and the gas thermal balance and
chemistry throughout the disk. We use 10 elements, 76 gas phase and
solid ice species, and 992 reactions including a detailed treatment of
UV-photorates \citep[see][]{Kamp2009}, H$_2$ formation of grain
surfaces, vibrationally excited H$_2^\star$ chemistry, and ice
formation (adsorption, thermal desorption, photo-desorption, and
cosmic-ray desorption) for a limited number of ice species
\citep[see][for details]{Woitke2009a}.  We also use ProDiMo in this
paper to compute all observables including the SED, images, and gas
emission line fluxes and profiles. Latest improvements to the ProDiMo
model include X-rays chemistry and heating, a parametric
prescription for dust settling, UV fluorescent pumping, PAH ionisation
and heating/cooling, [OI]\,6300\,\AA\ pumping by OH-photodissociation,
H$_2$-pumping by its formation on grain surfaces, formal solutions of
the line transfer problem, and chemical heating. These improvements
are explained in the Appendices \ref{sec:settling} to
\ref{sec:chemheat}.
%\ref{sec:fluor}, \ref{sec:PAHion}, \ref{sec:OHpump}, \ref{sec:linetrans} 

Our modelling of ET\,Cha is based on a prescribed disk density
structure, using power-laws for the surface density distribution and
disk flaring, see Appendix~\ref{sec:fixstruc}.  This approach allows
for a rapid model computation \citep[avoiding the structure iteration
loop, see Fig.~1 in][]{Woitke2009a}, and is hence more suitable for a
deep search for fitting values in parameter space. In this mode, the
disk code uses altogether 25 physical parameters, most of which are
considered as fixed for the modelling of ET\,Cha (for instance the
stellar properties, see Table~\ref{tab:Parameter}).

\subsection{Parameter fitting procedure}
\label{sec:diskfit}

The determination of the remaining free disk, gas and dust parameters,
like the total disk gas mass for instance, is a key objective of the
modelling of individual targets.
%\footnote{Other tasks are, for example,
%to provide detailed insight into the physics and chemistry of
%protoplanetary disks in general for data interpretation, or the
%prediction of trends, \ie the expected behaviour and correlations of
%observables and model parameters for analysis.}. 
These parameters are ``determined'' in this paper by considering the
values (better the ranges of values) that lead to a satisfying
match between predictions and observations in the frame of the model,
henceforth called ``solutions''. Practically all determinations of
properties of astrophysical objects work that way, even if it's just a simple
one-dimensional dependence between property and observable, because
the desired quantities are rarely accessible via direct
observations. This is the well-known problem of model inversion in
astronomy \citep{Lucy1994}, with all the usual shortcomings and
concerns which can be subdivided into four families:
\begin{itemize}
\item[1.] Concerns about the quality of the model itself (missing
      physics, poorly determined input quantities like cross-sections
      and rate coefficients, numerical issues like grid resolution,
      etc.).\\*[-1.8ex]
\item[2.] Concerns about the completeness of solutions found in parameter
      space, for instance, how can we be sure of having found the best
      solution?\\*[-1.8ex]
\item[3.] Under-determination, \ie a weak dependence of observables on
      model parameters in combination with considerable uncertainties
      in the observables.\\*[-1.8ex]
\item[4.] Non-uniqueness of inversion procedure (multiple solutions)
\end{itemize} 
Given the complexity of the disk model at hand, with $8\!-\!14$ free
parameters for the modelling of the disk of ET\,Cha, we have to take
into account the possibility that different solutions exist which
practically achieve the same degree of agreement between model results
and observations. Whereas in a low-dimension parameter space this
seems to a be a weird, seldom special case, it occurs frequently in
high dimensions, with numerous local minima. The manifold of
solutions in a multi-dimensional parameter space can be (and usually
is) astonishingly rich and complicated in structure.

Furthermore, it is important to note that, in a high-dimen\-sional
parameter space, an exhaustive search is practically impossible. This
is even more so for our disk modelling, since one complete model takes
about 1~CPU~hour on a single 3\,GHz processor machine. For 10
parameters, with 20 well-selected values around a main solution each,
one would need to run $20^{10}\!\approx\!10^{13}$ models which would
take about $1.2\times10^9$\,CPU\,yrs.

Therefore, we are not able to fully devitalise the concerns (1) to
(4), but have to face the fact the any parameter determination by
model inversion in high-dimensional parameter space is an
intrinsically uncertain business. Our strategy in this paper is as
follows.  We use an evolutionary strategy to find a handful of
well-fitting disk models, corresponding to different local minima in
parameter space. Among these, we select a ``best-fitting'' model, the
properties of which are described in detail in
Appendix~\ref{sec:bestmodel}. The best-fitting model is then re-run
with different choices of input physics in Appendix~\ref{sec:inputphysics},
to explore the principal effects of \eg X-rays and the treatment of
H$_2$-formation.

We then conclude about the confidence intervals of derived parameter
values in Sect.~\ref{sec:results} by considering (i) small deviations
of single parameters around the best fitting solution, (ii) overall
experience from fitting by hand and variance of different solutions
found by the evolutionary strategy, and (iii) direct constraints from
physical arguments like optical depth and column densities. We admit
that this modelling procedure is not entirely satisfactory. A more
complete discussion of our modelling strategy will be the topic of a
forthcoming paper.

\begin{figure*}
\vspace*{-2mm}
\hspace*{-2mm}\begin{tabular}{p{56mm}p{56mm}p{56mm}}
\includegraphics[width=65mm,height=57mm]{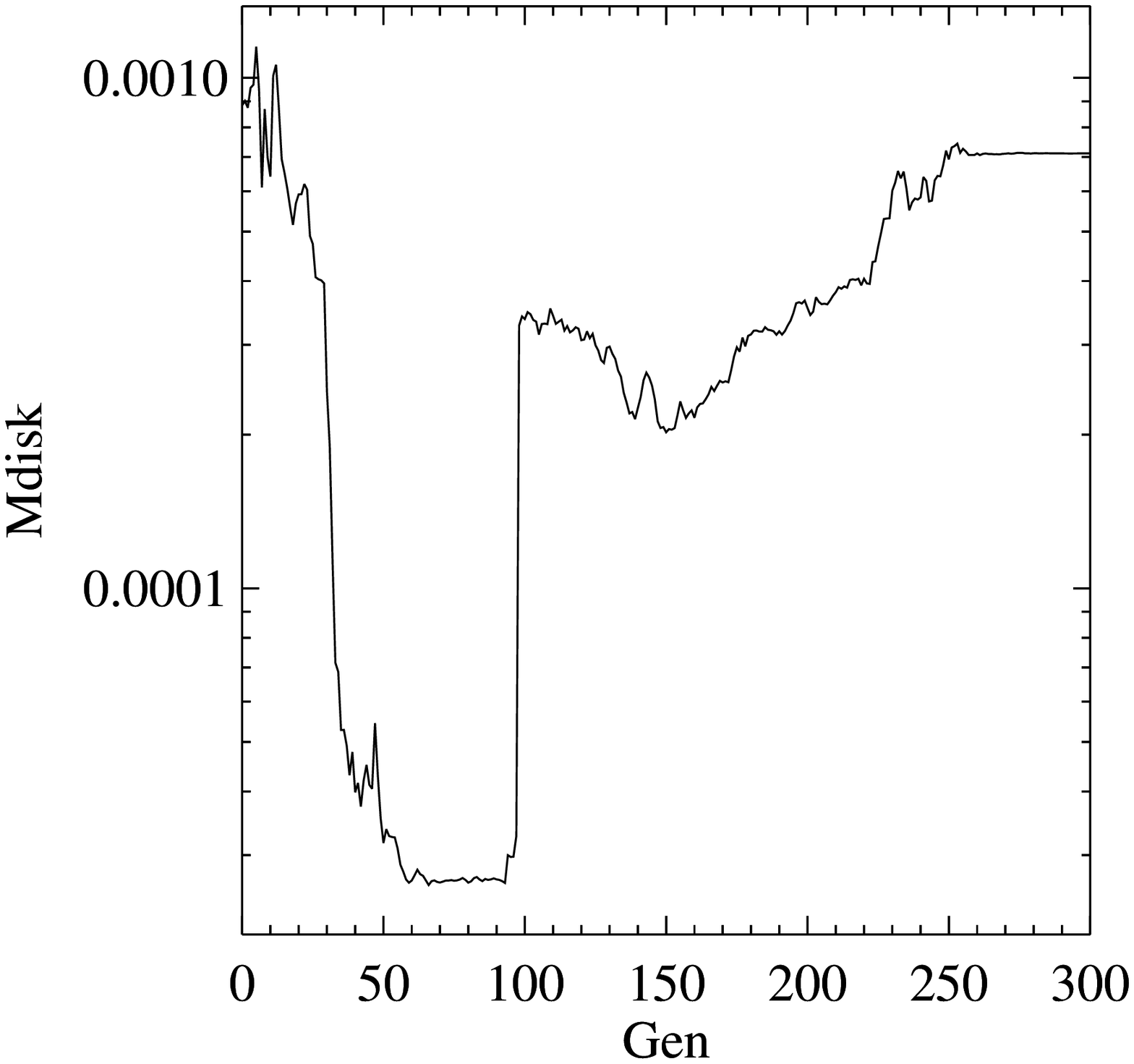} &
\includegraphics[width=65mm,height=57mm]{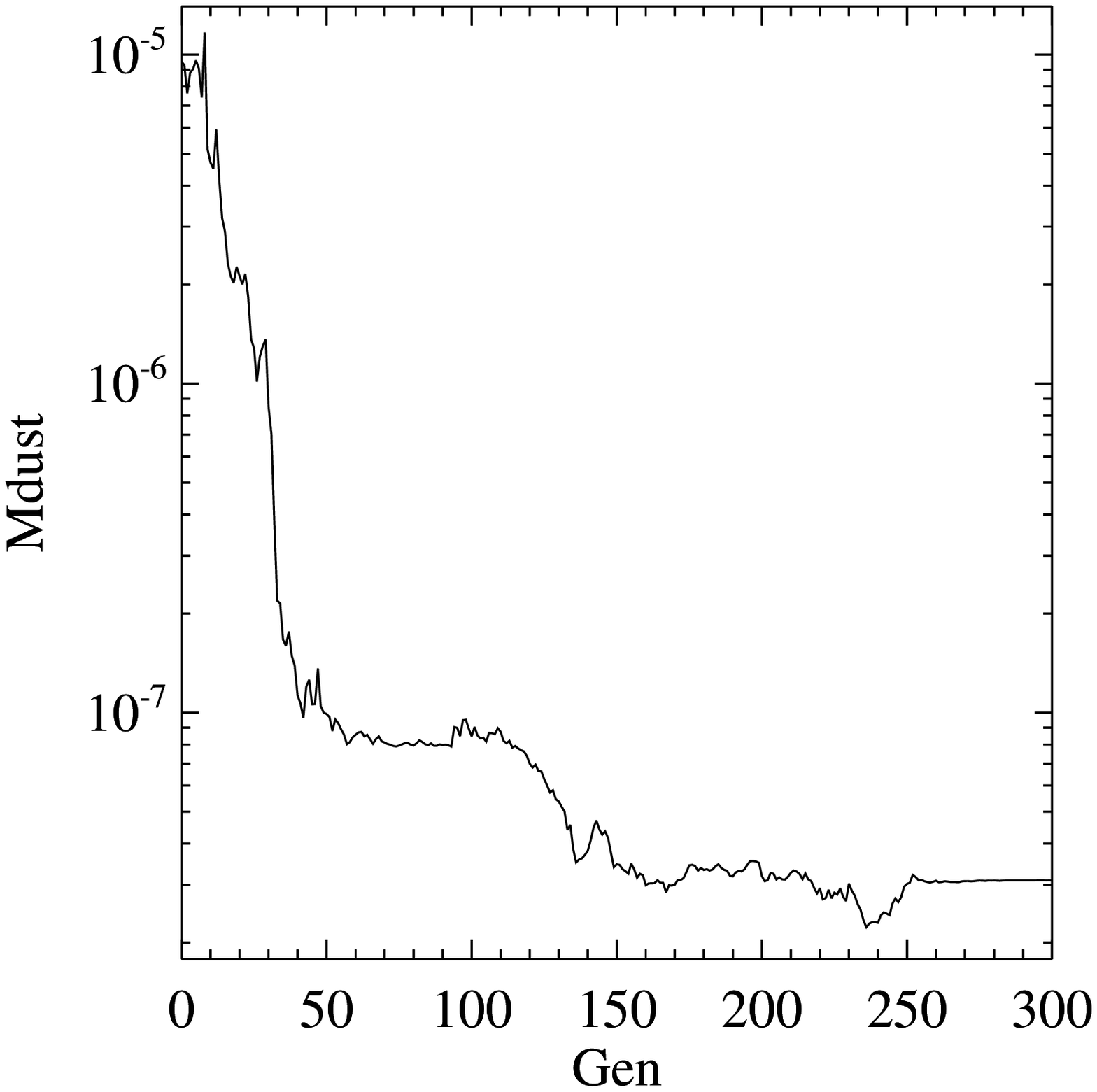} &
\includegraphics[width=65mm,height=57mm]{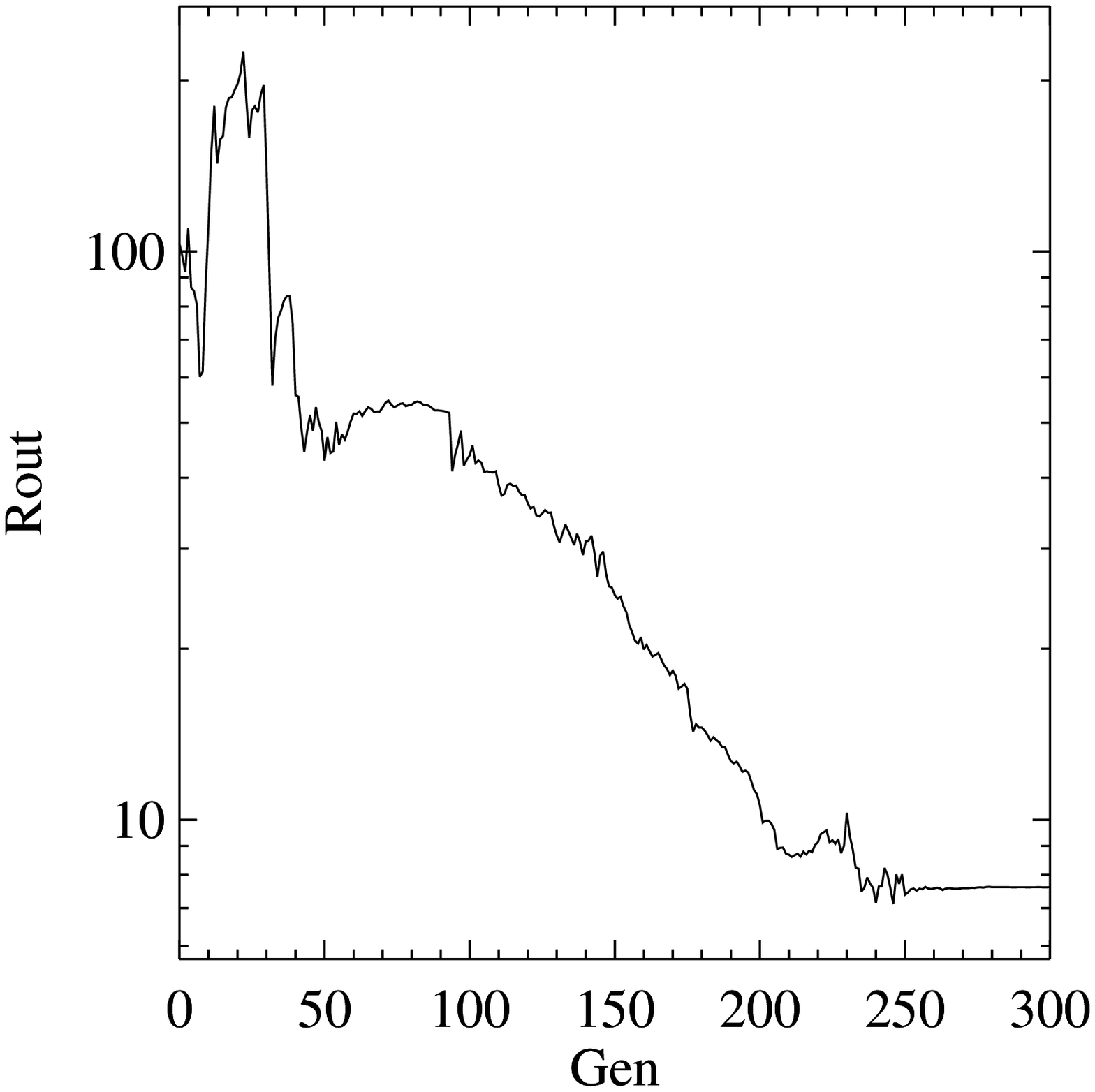} \\[-3mm]
\includegraphics[width=65mm,height=57mm]{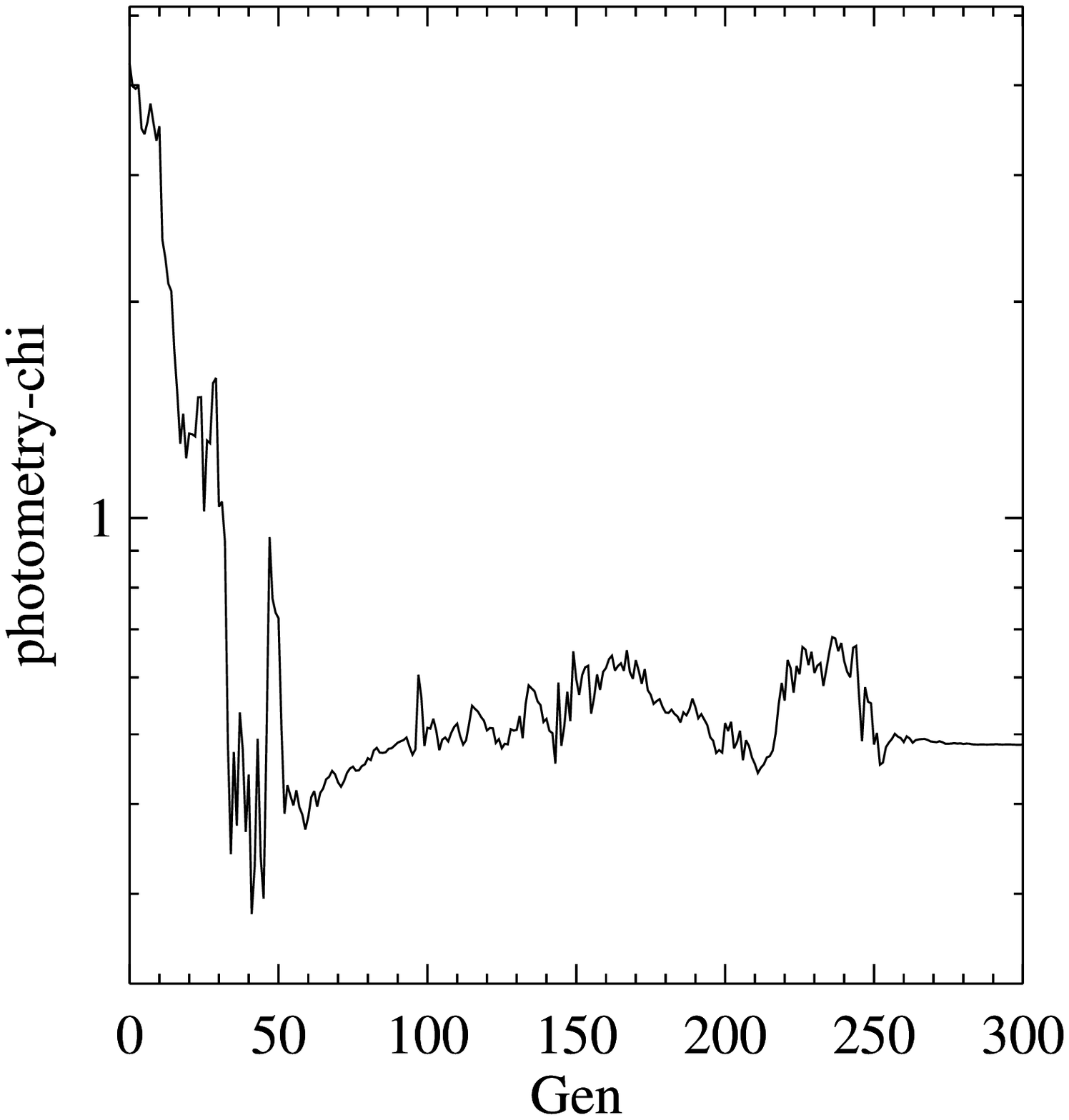} &
\includegraphics[width=65mm,height=57mm]{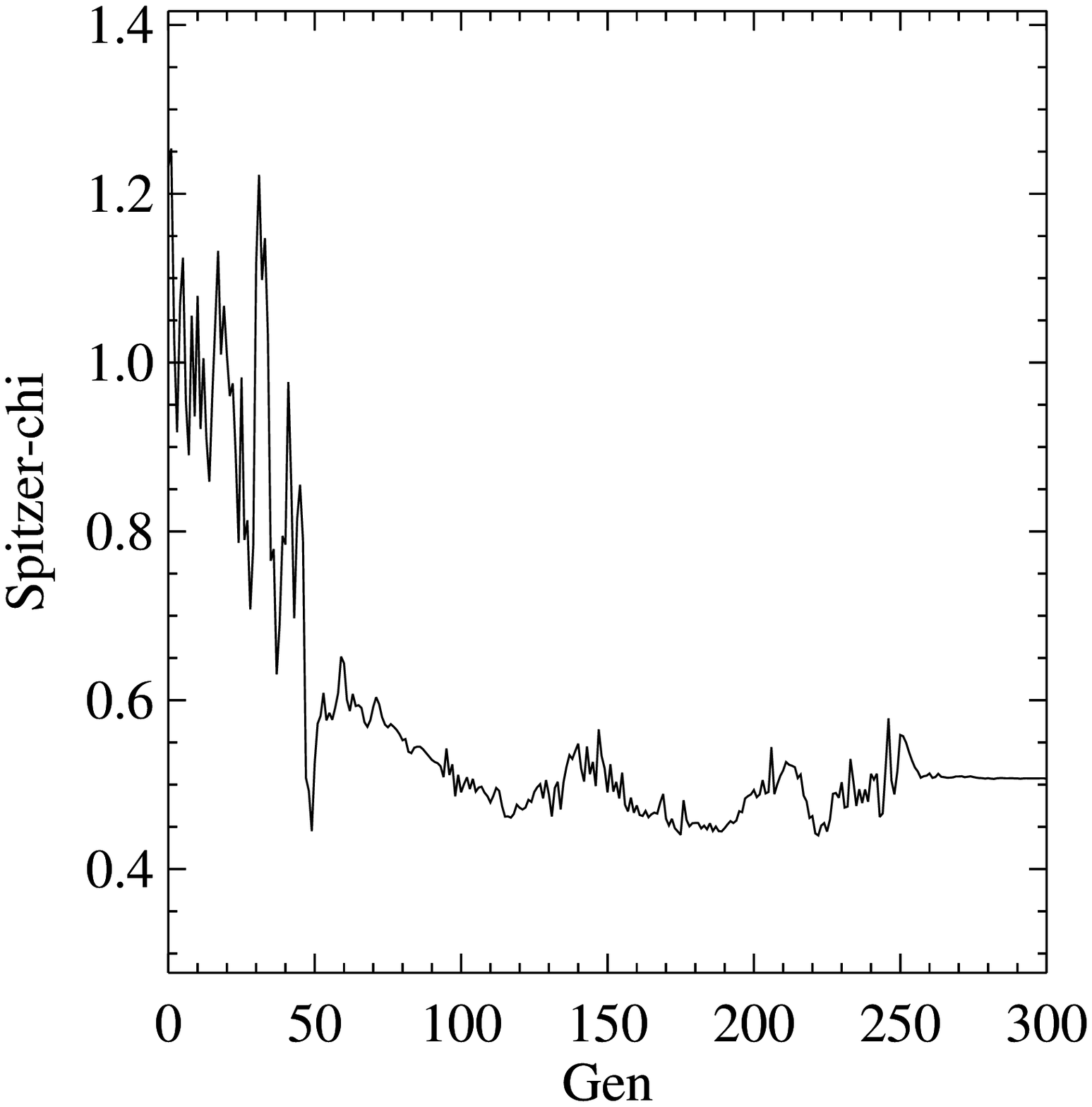} &
\includegraphics[width=65mm,height=57mm]{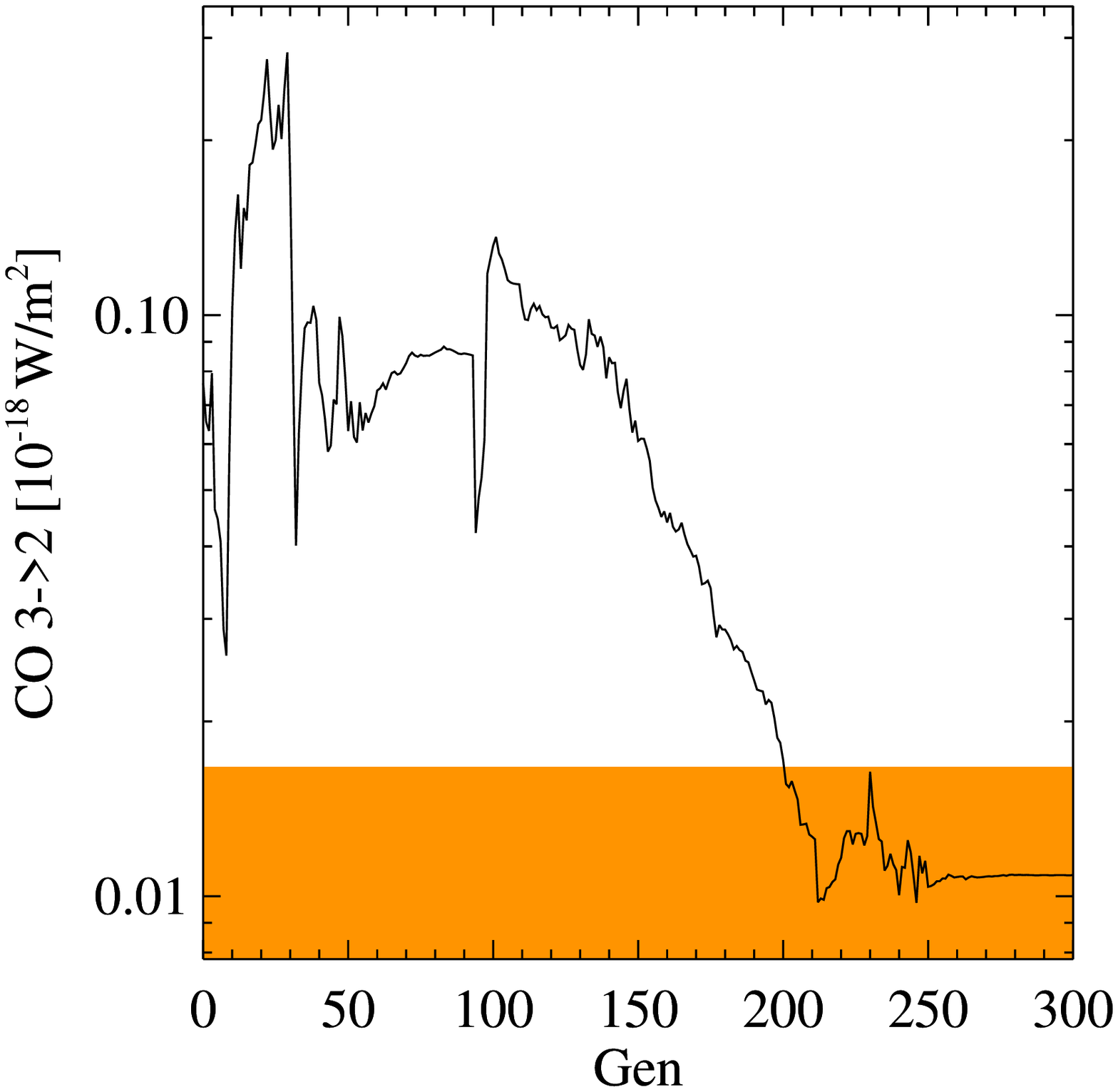} \\[-3mm]
\includegraphics[width=65mm,height=57mm]{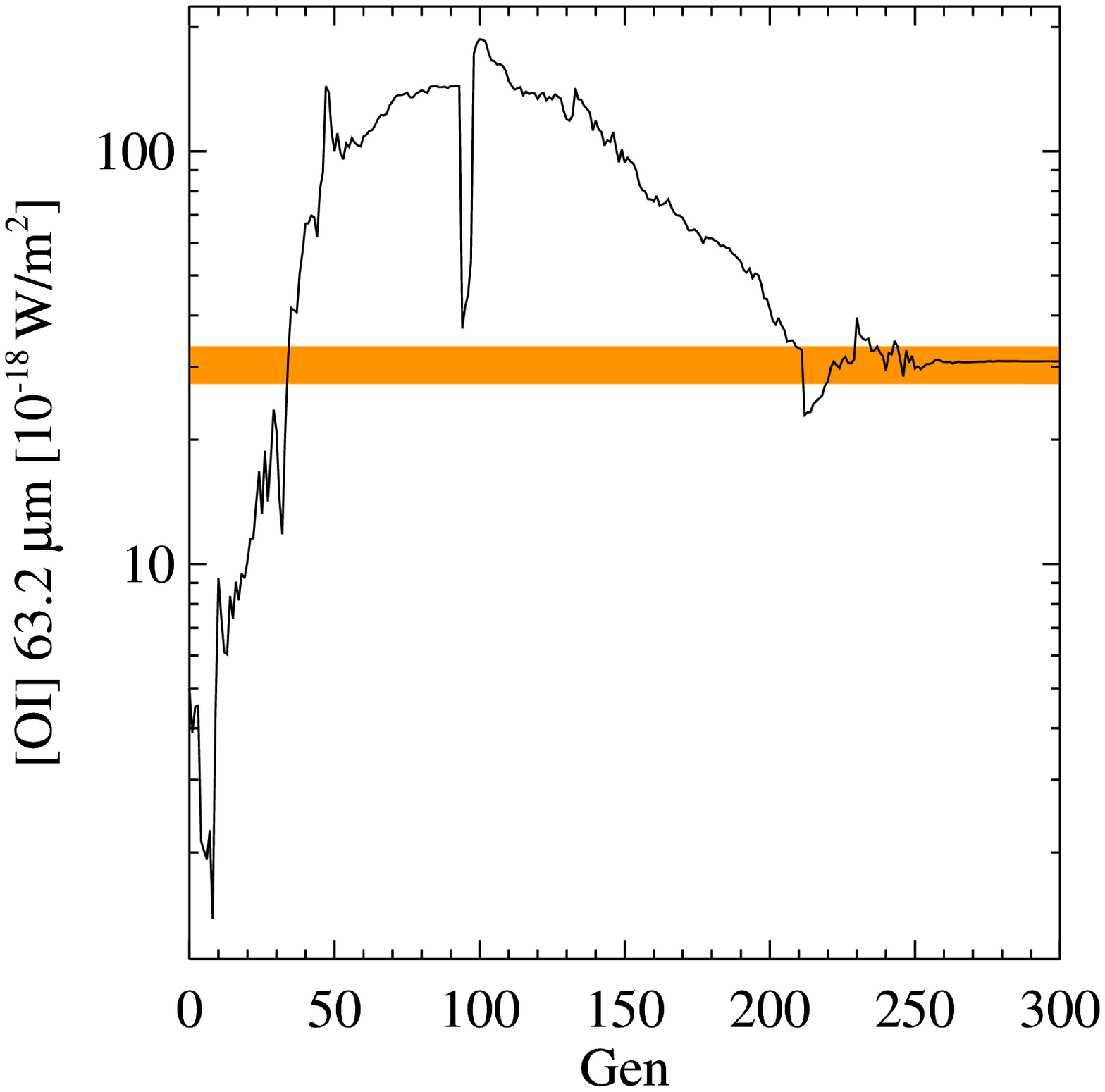} &
\includegraphics[width=65mm,height=57mm]{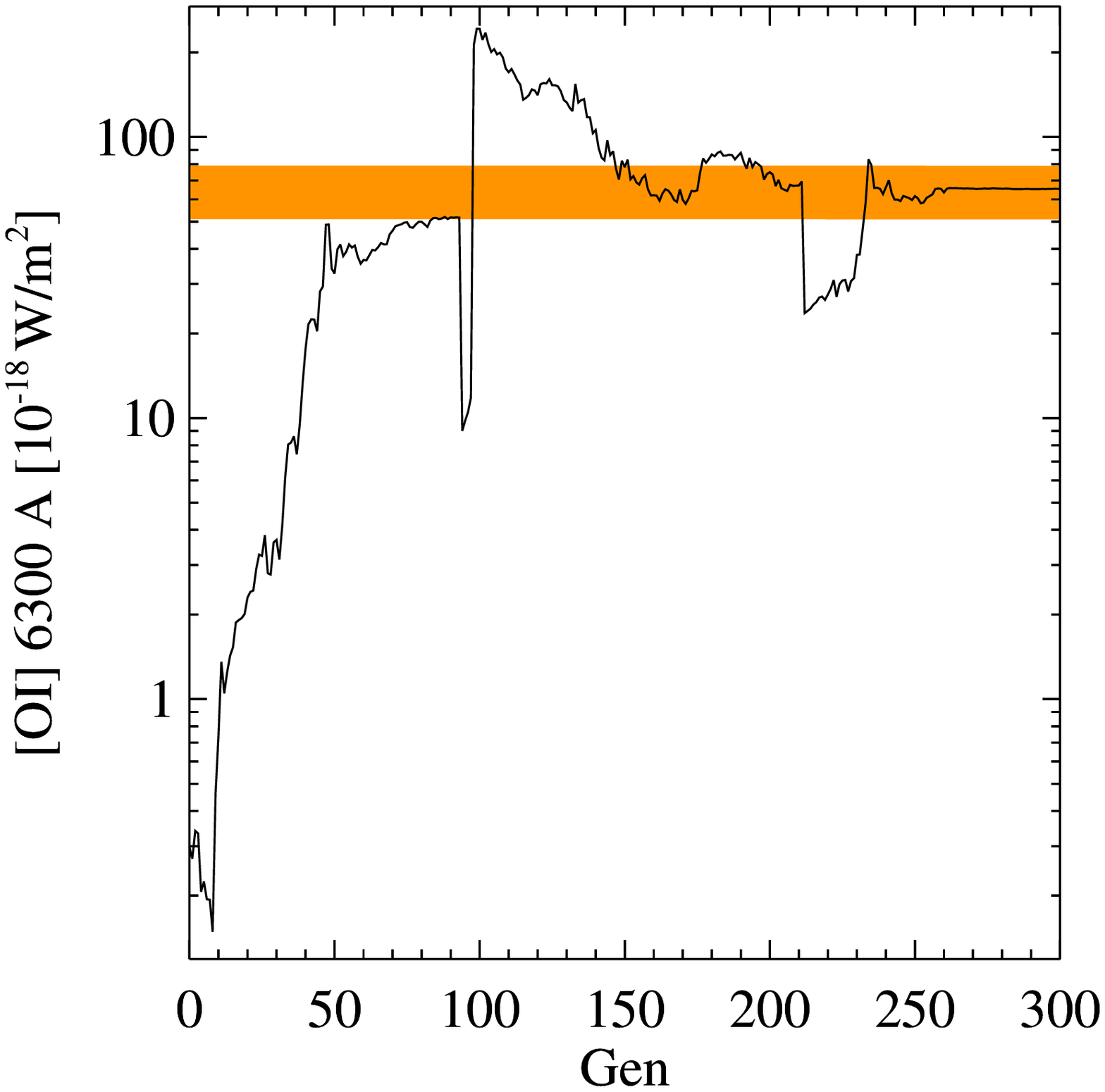} &
\includegraphics[width=65mm,height=57mm]{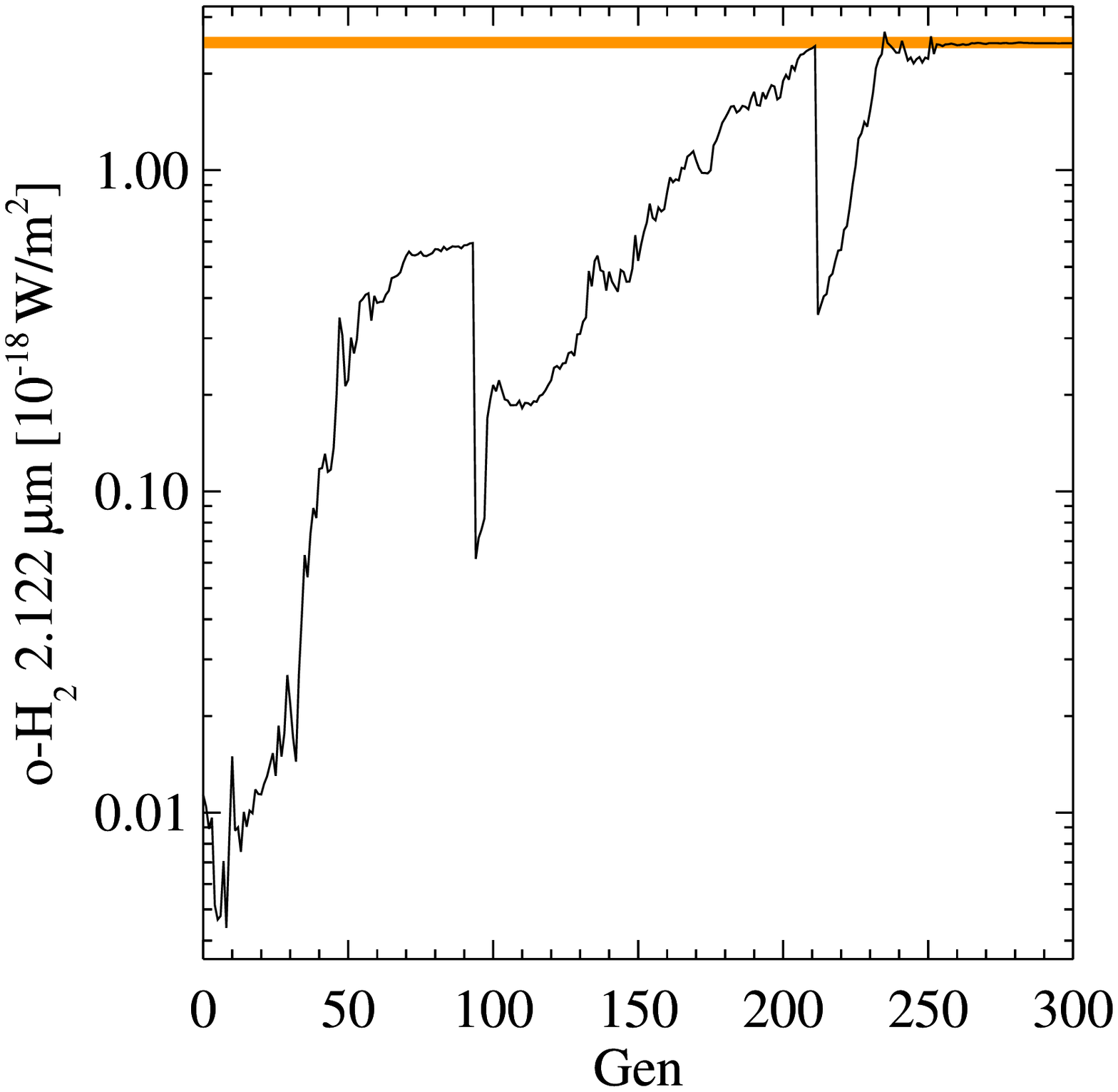}
\end{tabular}
\caption{Auto-optimisation of model parameters to fit all continuum
  and line observations of ET\,Cha. The first row shows the evolution
  of some model parameters as function of generation (total gas mass
  $M_{\rm gas}\,[M_\odot]$, total dust mass $M_{\rm dust}\,[M_\odot]$,
  outer disk radius $R_{\rm out}$\,[AU]). The two l.h.s. plots in the
  second row show the resulting $\chi$ for photometry and {\sc
  Spitzer} spectrum. The other four plots shows the resulting line
  fluxes as labelled, where the orange bars indicate the observed line
  flux ranges with $1\sigma$ errors. At generation 94, the
  evolutionary strategy got stuck in an unsatisfactory solution (local
  minimum) with well-fitting SED, but very bad line fits $\chi_{\rm
  Line}\!\approx\!5$, \ie a mean deviation between model and observed
  line fluxes of (e$^5\!\!\approx$150)$\,\sigma$. We have manually
  changed parameters at generation 95, increasing the disk mass by a
  factor 10. The evolutionary strategy then found a well-fitting model
  which can reproduce all available line and continuum observations
  within about $1\,\sigma$ on average.}
\label{fig:evolution}
\end{figure*}

\subsection{Fit quality and evolutionary strategy}

Mathematically, all model inversion techniques can be formulated in terms
of a certain strategy to minimise $\chi^2$, \ie to find a minimum of
the deviations between model predictions and observations in parameter
space. In this paper, we use the following logarithmic measure of these
deviations as
\begin{eqnarray}
  &&\displaystyle\chi^2 = \frac{1}{N}\sum\limits_{i=1}^N \Delta_i^2 
       \quad\mbox{with}\nonumber\\
  &&\Delta_i = \left\{\begin{array}{ccc}\displaystyle
          \frac{\log\big(F^i_{\rm mod}/F^i_{\rm obs}\big)}
               {\sigma_F^i/F^i_{\rm obs}}
          & , & F^i_{\rm obs} \ge 3\,\sigma_F^i \\[5mm] \displaystyle
         \frac{F^i_{\rm mod}}{3\,\sigma_F^i}
          & , & F^i_{\rm obs} <   3\,\sigma_F^i \ .
         \end{array}\right. 
  \label{eq:chi1}
\end{eqnarray}
$F^i_{\rm obs}$ and $\sigma_F^i$ are an observed flux and its
uncertainty, and $F^i_{\rm mod}$ is the predicted flux by the model. The
logarithmic nature of our $\chi^2$ is motivated by the need to assign
an equally large (bad) number to $\chi$ if a model flux is a factor 10
too large as if it is a factor 10 too small, analog to considering
deviations in magnitudes\footnote{Defining a magnitude as $m_{\rm
obs}^i\propto\log F_{\rm obs}^i$, the error thereof is
$\sigma_m^i=\partial\log(F_{\rm obs}^i)/\partial F_{\rm obs}^i\times\sigma_F^i 
= \sigma_F^i/F^i_{\rm obs}$ and hence the deviation
$\Delta_i=(m_{\rm obs}^i-m_{\rm mod}^i)/\sigma_m^i$ is equivalent to
Eq.~(\ref{eq:chi1}).}.

Equation~(\ref{eq:chi1}) is applied separately to all photometric
data, all $\lambda$-points in the {\sc Spitzer} spectrum, and to the
line fluxes of CO\,$J\!=\!3\!\to\!2$, [OI]\,63\,$\mu$m,
[OI]\,6300\,\AA\ (LVC) and o-H$_2$\,2.122\,$\mu$m, resulting in
$\chi^2_{\rm Phot}$, $\chi^2_{\rm Spit}$, and $\chi^2_{\rm Line}$,
respectively. The total fit quality of a model is then calculated as
\begin{equation}
  \chi^2 = w^{\rm Phot}\chi^2_{\rm Phot}
         + w^{\rm Spit}\chi^2_{\rm Spit}
         + w^{\rm Line}\chi^2_{\rm Line}
  \label{eq:chi2}
\end{equation}
with adjustable weights $w^{\rm Phot}$, $w^{\rm Spit}$ and $w^{\rm Line}$,
normalised as $w^{\rm Phot}+w^{\rm Spit}+w^{\rm Line}\!=\!1$.

We have applied the $(1,11)$\,-\,evolutionary strategy with adaptive
step-size control of \citet{Rechenberg2000} to minimise $\chi$, \ie to
find best-fitting values of our remaining free gas, dust and disk
parameters. The strategy uses 1 parent producing 11 offsprings with
slightly modified parameters, the best of which will become the parent
of the next generation (the parent always dies). The step-size
$\delta$ is transmitted to the children and treated as additional
parameter to be optimised. After some experiments, we have chosen
$w^{\rm Phot}\!=\!0.35$, $w^{\rm Spit}\!=\!0.45$ and $w^{\rm
Line}\!=\!0.2$ for optimum performance of the evolutionary
strategy.

Figure~\ref{fig:evolution} visualises one exemplary run of the
evolutionary strategy, showing the changing model parameters and
results over 300 generations. We started from a generic disk setup
with $M_{\rm gas}\!=\!10^{-3}M_\odot$, gas/dust ratio of 100, and an
outer disk radius of $R_{\rm out}\!=\!100\,$AU. This model fails badly
to explain the SED and, in particular, the line fluxes.
CO\,$J\!=\!3\!\to\!2$ is about 10 times too strong and both
[OI]\,6300\,\AA\ and o-H$_2$\,2.122\,$\mu$m are more than a factor of
100 too weak. However, after about 300 generations, the model has
achieved a good fit of all available continuum and line
observations. The final parameter values are far from these initial
guesses, yielding a disk that is less than 10\,AU in radius and much
less massive. All models from about generation 50 onwards fit the dust
observations about equally well, the SEDs from these models are almost
indistinguishable by eye when plotted as in Fig.~\ref{fig:SED}. A good
fit to the line observations was achieved only from generation 200
onward, after the disk radius has shrunk to about $R_{\rm
out}\!\la\!10\,$AU while continuing to fit the dust observations about
equally well. Thus, the line observations can help to break the
degeneracy of SED-fitting.

\begin{table}
\begin{center}
\caption{Parameters of the best-fitting disk model.}
\label{tab:Parameter}
\begin{tabular}{l|c|c}
\\[-3.8ex]
\hline
 quantity & symbol & value\\
\hline 
\hline 
stellar mass                      & $M_{\star}$   & $0.2\,M_\odot$\\
effective temperature             & $T_{\rm eff}$ & $3400\,$K\\
stellar luminosity                & $L_{\star}$   & $0.085\,L_\odot$\\
%UV excess                         & $f_{\rm UV}$  & 0.02\\
%UV power-law index                & $p_{\rm UV}$  & $-\,0.5$\\
\hline
&&\\[-2.2ex]
disk gas mass$^\star$      & $M_{\rm gas}$& $6.1\times 10^{-4}\,M_\odot$\\
inner disk radius                 & $\Rin$       & 0.022\,AU\\
outer disk radius$^\star$         & $\Rout$      & 8.2\,AU\\
column density power index$^\star$& $\epsilon$   & -0.020\\
reference scale height$^\star$    & $H_0$        & 0.011\,AU\\
reference radius                  & $r_0$        & 0.1\,AU\\
flaring power index$^\star$       & $\beta$      & 1.09\\ 
\hline
&&\\[-2.2ex]
disk dust mass$^\star$     & $M_{\rm dust}$ & $2.6\times10^{-8}\,M_\odot$\\
minimum dust particle radius      & $\amin$        & $0.05\,\mu$m\\
maximum dust particle radius      & $\amax$        & $1\,$mm\\
dust size dist.\ power index      & $p$            & 4.1\\
minimum settling particle size    & $a_{\rm s}$    & 0\\
dust settling power index         & $s$            & 0\\
dust material mass density        & $\rho_{\rm gr}$& 3\,g\,cm$^{-3}$\\
dust composition                  & $\rm Mg_2SiO_4$ & 32.9\%\\
(volume fractions)                & amorph.\,carbon & 24.4\%\\
                                  & $\rm MgFeSiO_4$ & 23.0\%\\
                                  & $\rm SiO_2$     &  8.8\%\\
                                  & $\rm MgSiO_3$   &  7.6\%\\
                                  & cryst.\,silicate&  3.3\%\\
\hline 
&&\\[-2.2ex]
strength of incident ISM UV       & $\chi^{\rm ISM}$ & 1\\
cosmic ray H$_2$ ionisation rate  & $\zeta_{\rm CR}$   
                                           & $5\times 10^{-17}$~s$^{-1}$\\
PAH abundance rel. to ISM$^\star$ & $f_{\rm PAH}$    & 0.081\\
chemical heating efficiency$^\star$ & $\gamma^{\rm chem}$  & 0.55\\
$\alpha$ viscosity parameter      & $\alpha$         & 0\\
\hline 
&&\\[-2.2ex]
disk inclination                  & $i$ & $40\degr$\\
distance                          & $d$ & 97\,pc\\
\hline
\end{tabular}
\end{center}
\vspace*{-2mm} {\footnotesize Parameters marked with $^\star$ have
been varied in the evolutionary optimisation run depicted in
Fig.~\ref{fig:evolution}. The values of the other parameters have been
assumed, fitted by hand, or have been obtained from additional
evolutionary optimisation runs not discussed here.}
\end{table}

\begin{table}
\centering
\caption{Computed line fluxes $\rm[10^{-18}\,W/m^2]$,
at $d\!=\!97$\,pc and $i\!=\!40\degr$ from the best-fitting
disk model, in comparison to the observations. Non-detections are listed
as $<\!3\sigma$ upper limits.}
\label{tab:ModelLines}
\vspace*{-1mm}
\begin{tabular}{lr|c|cc}
line        & $\rm\lambda\,[\mu m]$           & observed & model\\
\hline
$\rm[OI]\ ^3P_1\to\, ^3P_2$        & 63.18    & $30.5\pm3.2$  
   & 34.5   \\
$\rm[OI]\ ^3P_0\to\, ^3P_1$        & 145.52   & $<6.0$        
   & 2.6    \\
$\rm[OI]\ ^1D_2\to\, ^3P_2$ (HVC)\hspace*{-2mm}  
                                   & 0.6300   & $73\pm25$    
   & --     \\
$\rm[OI]\ ^1D_2\to\, ^3P_2$ (LVC)\hspace*{-2mm}  
                                   & 0.6300   & $65\pm25$    
   & 69.6   \\
$\rm[CII]\ ^2P_{3/2}\to\, ^2P_{1/2}$ & 157.74 & $<9.0$        
   & 0.11  \\
%CO\,$J\!=\!1\!\to\!0$              & 2600.76  & --            
%   & 0.0017 & 0.0015\\
%CO\,$J\!=\!2\!\to\!1$              & 1300.40  & --            
%   & 0.017  & 0.015\\
CO\,$J\!=\!3\!\to\!2$              & 866.96   & $<0.05$
   & 0.014  \\
%CO\,$J\!=\!18\!\to\!17$            & 144.78   & --            
%   & 0.069  \\
CO\,$J\!=\!29\!\to\!28$            & 90.16    & $<9.6$       
   & 4.9  \\
CO\,$J\!=\!33\!\to\!32$            & 79.36    & $<24$       
   & 3.3  \\
CO\,$J\!=\!36\!\to\!35$            & 72.84    & $<8.0$            
   & 2.6  \\
o-H$_2\rm\ v\!=\!1\!\to\!0\ S(1)$  & 2.122    & $2.5\pm0.1$   
   & 2.4   \\ 
%o-H$_2\rm O\ 1_{10}\!\to\!1_{01}$  & 538.29   & --       
%   & 0.011  & 0.004\\
o-H$_2\rm O\ 2_{21}\!\to\!2_{12}$  & 180.49   & $<5.2$ 
   & 1.1  \\
o-H$_2\rm O\ 2_{12}\!\to\!1_{01}$  & 179.53   & $<5.0$
   & 1.4   \\
%o-H$_2\rm O\ 3_{03}\!\to\!2_{12}$  & 174.63   & --       
%   & 0.17   & 0.068\\
%o-H$_2\rm O\ 2_{21}\!\to\!1_{10}$  & 108.07   & --       
%   & 0.20   & 0.092\\
o-H$_2\rm O\ 4_{32}\!\to\!3_{12}$  & 78.74    & $<30$
   & 11.1   \\
%p-H$_2\rm O\ 2_{02}\!\to\!1_{11}$  & 303.46   & --       
%   & 0.007  & 0.002\\
%p-H$_2\rm O\ 1_{11}\!\to\!0_{00}$  & 269.27   & --       
%   & 0.017  & 0.006\\
%p-H$_2\rm O\ 4_{13}\!\to\!3_{22}$  & 144.52   & --       
%   & 0.005  & 0.003\\
%p-H$_2\rm O\ 3_{13}\!\to\!2_{02}$  & 138.53   & --       
%   & 0.042  & 0.019\\
%p-H$_2\rm O\ 2_{20}\!\to\!1_{11}$  & 100.98   & --       
%   & 0.041  & 0.020\\
p-H$_2\rm O\ 3_{22}\!\to\!2_{11}$  & 89.99    & $<9.6$
   & 6.4  \\
\end{tabular}
\end{table}

\section{Results}
\label{sec:results}

We identify the result of the evolutionary run depicted in
Fig.~\ref{fig:evolution} as our main ``best-fitting'' model. The
resulting parameters are listed in Table~\ref{tab:Parameter}, the
computed line fluxes are summarised in Table~\ref{tab:ModelLines} and
the resulting SED is plotted in Fig.~\ref{fig:SED}. More details about
the internal physical and chemical structure of this disk model are
shown in Appendix~\ref{sec:bestmodel}.

However, the identification of a best-fitting model was not at all
straightforward. Altogether 17 runs of the evolutionary strategy have
been executed with about 50 to 300 generations each, choosing
different parameters to be varied, different initial guesses of the
model parameters, or using different setups of the evolutionary
strategy.  Not all of these runs succeeded to find well-fitting
solutions. Some almost equally well-fitting solutions are listed in
Table~\ref{tab:Parameter2}. Based on these diverse model inversion
results, we have to be very careful with conclusions about the
disk properties of ET\,Cha. The following section summarises our
confidence intervals for the various disk shape, dust and gas
parameters and discusses which observations are key for their
determination.

\begin{table}
\begin{center}
\caption{Parameters of different, about equally well-fitting models
from different runs of the evolutionary strategy.}
\label{tab:Parameter2}
{\ }\\*[-3ex]
\begin{tabular}{c|cccc}
\hline
\multicolumn{1}{c}{parameter} 
  & model 1 & model 2 & model 3 & model 4 \\
\hline 
\\*[-2.2ex]
$M_{\rm gas}\rm\,[10^{-4}\,M_\odot]$  & 0.088 & 0.65 & 6.1    & 25\\
$M_{\rm dust}\rm\,[10^{-8}\,M_\odot]$ & 3.3   & 2.6  & 2.6    & 3.7\\
$\Rout$\,[AU]                    & 5.9    & 7.5      & 8.2    & 8.3\\
$H_0$\,[AU]                      & 0.0103 & 0.0096   & 0.0110 & 0.0108\\
$\epsilon$                       & 1.16   & 0.046    & $-0.020$ & 0.008\\
$\beta$                          & 1.33   & 1.15     & 1.09   & 1.07\\ 
$p$                              & 3.2    & 3.9  & 4.1$^\star$ & 4.2\\
$a_{\rm s}\rm\,[\mu m]$ & 0.05$^\star$ & 0.01$^\star$ & 0$^\star$ & 0$^\star$\\
$s$                              & 0.01   & 0.13 & 0$^\star$ & 0$^\star$ \\
$f_{\rm PAH}$                    & 0.081  & 0.098    & 0.081  & 0.081\\
$\gamma^{\rm chem}$              & 0.14   & 0.20     & 0.55   & 0.09\\
\hline 
\multicolumn{1}{c}{results} & & & &\\
\hline
&&\\[-2.2ex]
$\chi_{\rm Phot}$                   & 0.48   & 0.78     & 0.50  & 0.45\\
$\chi_{\rm Spit}$                   & 0.91   & 0.51     & 0.51  & 0.50\\
$\chi_{\rm Line}$                   & 1.03   & 1.04     & 1.25  & 1.02\\
\hline
\end{tabular}
\end{center}
{\footnotesize  Each final model is
located in a solid local $\chi$-minimum, the essence after
trying a few thousand models. Parameter symbols are explained in
Table~\ref{tab:Parameter}. Other fixed model parameters are as listed in 
Table~\ref{tab:Parameter}. Model 3 was selected as best-fitting
model. Parameters marked with $^\star$ have been fixed during the
evolution.}
\end{table}

\subsection{Dust mass} 

ET\,Cha's spectral energy distribution (SED) is characterised by a
strong near and mid IR excess relative to the star, similar to other
much more massive T\,Tauri disks, see Fig.~\ref{fig:SED}. However, in
the far-IR, the fluxes are very faint. 69\,mJy at 160,$\mu$m was
too weak to be detected by {\sc Spitzer}, and we have not detected the
object at $870\,\mu$m with {\sc Apex}. Thus, the new {\sc Herschel}
photometry points at $70\,\mu$m and $160\,\mu$m allow for the most
complete SED analysis of the source to date.

In all computed disk models for ET\,Cha, the disk is optically thin at
$160\,\mu$m.  The observed flux at distance $d$ is hence given by
\begin{equation}
  F_\nu = \frac{\kappa^{\rm abs}_\nu
                 B_\nu\big(\langle T_{\rm dust}\rangle\big)\,
                 M_{\rm dust}}{d^2}
  \label{eq:F160}
\end{equation}
where $\langle T_{\rm dust}\rangle$ is the dust mass averaged dust
temperature, $\kappa^{\rm abs}_\nu$ the dust absorption coefficient in
$\rm [cm^2/g(dust)]$ and $B_\nu$ the Planck function. In the
best-fitting model we measure $\langle T_{\rm dust}\rangle\approx
56$\,K, and $\kappa^{\rm abs}_\nu=34\rm\,cm^2/g(dust)$ at 160\,$\mu$m,
which, according to Eq.~(\ref{eq:F160}), results in a 160\,$\mu$m flux
of 48\,mJy which is in good agreement with both, the computed flux
from the full disk model (51\,mJy) and the observed value of 68\,mJy.

Therefore, the observed 160\,$\mu$m-flux leaves no doubt that the mass of the
dust in the disk of ET\,Cha must be low, $M_{\rm
dust}\!\approx\!2.6\times10^{-8}\,M_\odot$ according to our best fitting
model. More massive disks produce too strong 160\,$\mu$m continuum
fluxes. Although there are certainly some ambiguities in the dust mass
determination, \eg if other dust size parameters are used, leading to
a different $\kappa^{\rm abs}_\nu$, but among all parameters, $M_{\rm
dust}$ certainly belongs to those which are only little influenced by
other parameters and can be determined with some confidence. Across all
SED-fitting disk models, with varying dust composition and size
parameters, the value for $M_{\rm dust}$ ranges in
$(2-5)\times10^{-8}\,M_\odot$.

\subsection{Dust disk characteristics and particle sizes}

According to the low dust mass we derive, the disk of ET\,Cha has only
small optical depths across the disk. Our best-fitting model
has vertical dust absorption optical depths $<$1 for
$\lambda\!\ga\!4\,\mu$m, except for the optically thick $10\,\mu$m and
$20\,\mu$m silicate features. This is valid for all radii since the
column density exponent $\epsilon\!\approx\!0$.  However, the radial
optical depths are much larger. The midplane radial dust extinction
optical depth at $1\,\mu$m is about 150. Therefore, the disk of
ET\,Cha is on the borderline between optically thin and thick. It has
a complex dust temperature structure due to radial shielding effects
(see Fig.~\ref{fig:diskmodel1}), but the vertical dust emission could
be treated in the optically thin limit at most wavelengths.

Some conclusions about the dust particle sizes can be drawn from the shape
of the 10\,$\mu$m and 20\,$\mu$m silicate features, which are clearly
seen in emission for ET\,Cha (Fig.\ref{fig:SED}). We need a large
amplitude of opacity variation across these features (see
Fig.\ref{fig:dustopac}) to model the SED of ET\,Cha, which favours
small, (sub-)\,micron sized dust particles. Since the peaks are
optically thick, warm and small grains must be located in front of
cooler dust along the line of sight. We notice that it is easier to
fit the observed silicate features with small values of the column
density power-law index $\epsilon$, \ie a roughly flat surface density
distribution, and small disk inclinations. For larger $\epsilon$ or
larger disk inclinations (closer to edge-on), the silicate features
weaken and eventually vanish.

If dust and gas are well-mixed (no dust settling, \eg
models 3 and 4 in Table~\ref{tab:Parameter2}), the models clearly
favour very small particles. We can fit the entire SED with a uniform
dust population that is either truncated at about 1\,$\mu$m, or,
alternatively, with a continuous size distribution ranging from
0.05\,$\mu$m to 1\,mm, but with an unusually large power-law index of
$p\!=\!4.1$. The second approach allows for slightly better SED-fits.
These findings with ProDiMo have been carefully checked against the
Monte Carlo radiative transfer code MCFOST
\citep{Pinte2006,Pinte2009}, showing a very good agreement in
calculated dust temperatures and continuum fluxes.

However, if dust settling is taken into account (in the approximate
way explained in Appendix~\ref{sec:settling}),
Table~\ref{tab:Parameter2} demonstrates that smaller $p\!=\!3.9$ or even
$p\!=\!3.2$ are also possible, in which case the volume-integrated dust
size distribution in ET\,Cha would not be unusual at all, close to the
default value of $p\!=\!3.5$. Our honest conclusion about ET\,Cha is
hence that it's dust must predominantly be made of (sub-)\,micron
particles {\it at the disk surface}, where the silicate emission
features form. We have therefore decided to refrain from quoting any
errorbars to our results concerning the dust size parameters.

\begin{figure*}
\centering
\begin{tabular}{ccc}
  \hspace*{-2mm}\includegraphics[height=59.5mm,
    trim=0mm 0mm 8mm 0mm, clip]{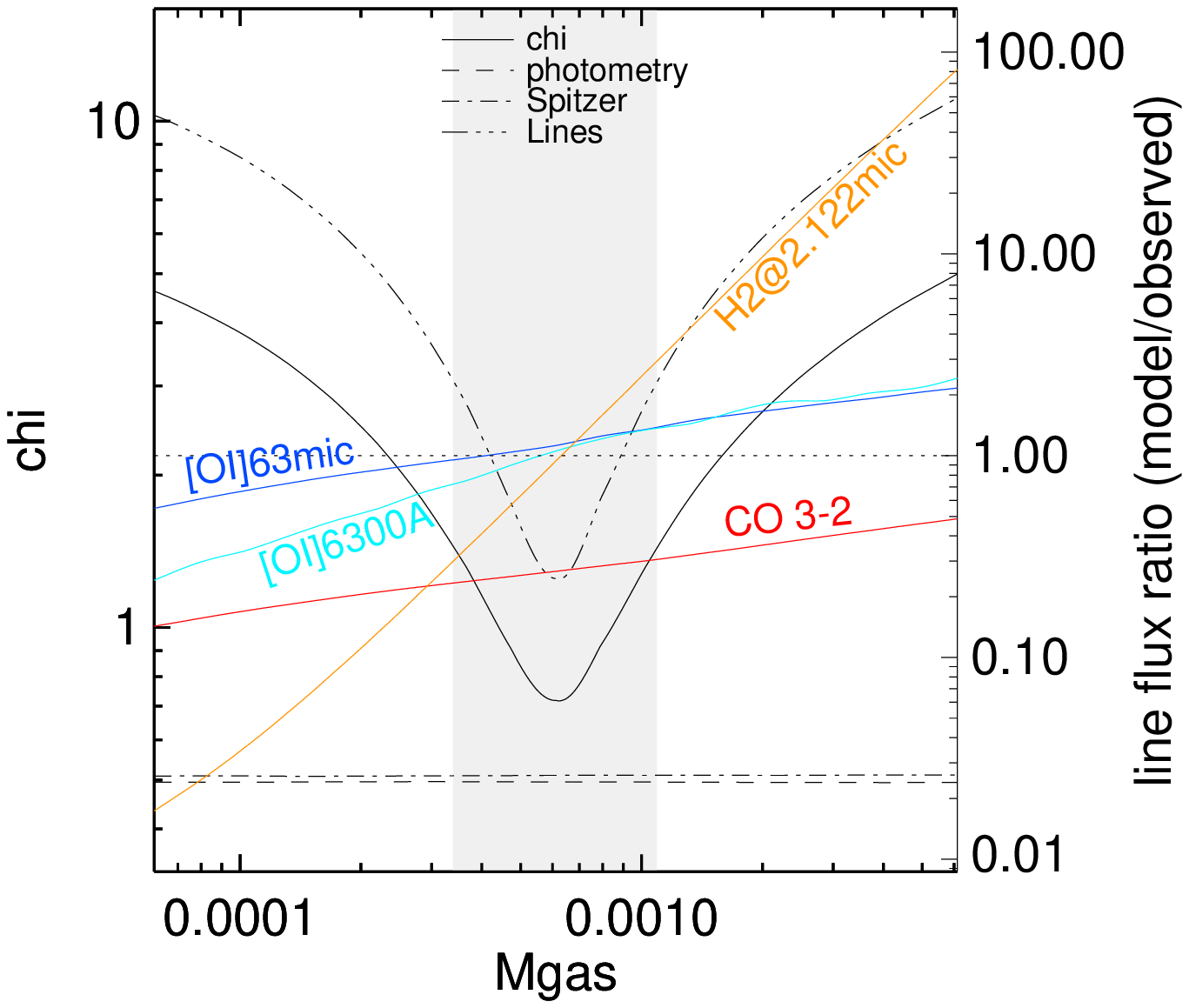} &
  \hspace*{-4mm}\includegraphics[height=59.5mm,
    trim=10mm 0mm 15mm 0mm, clip]{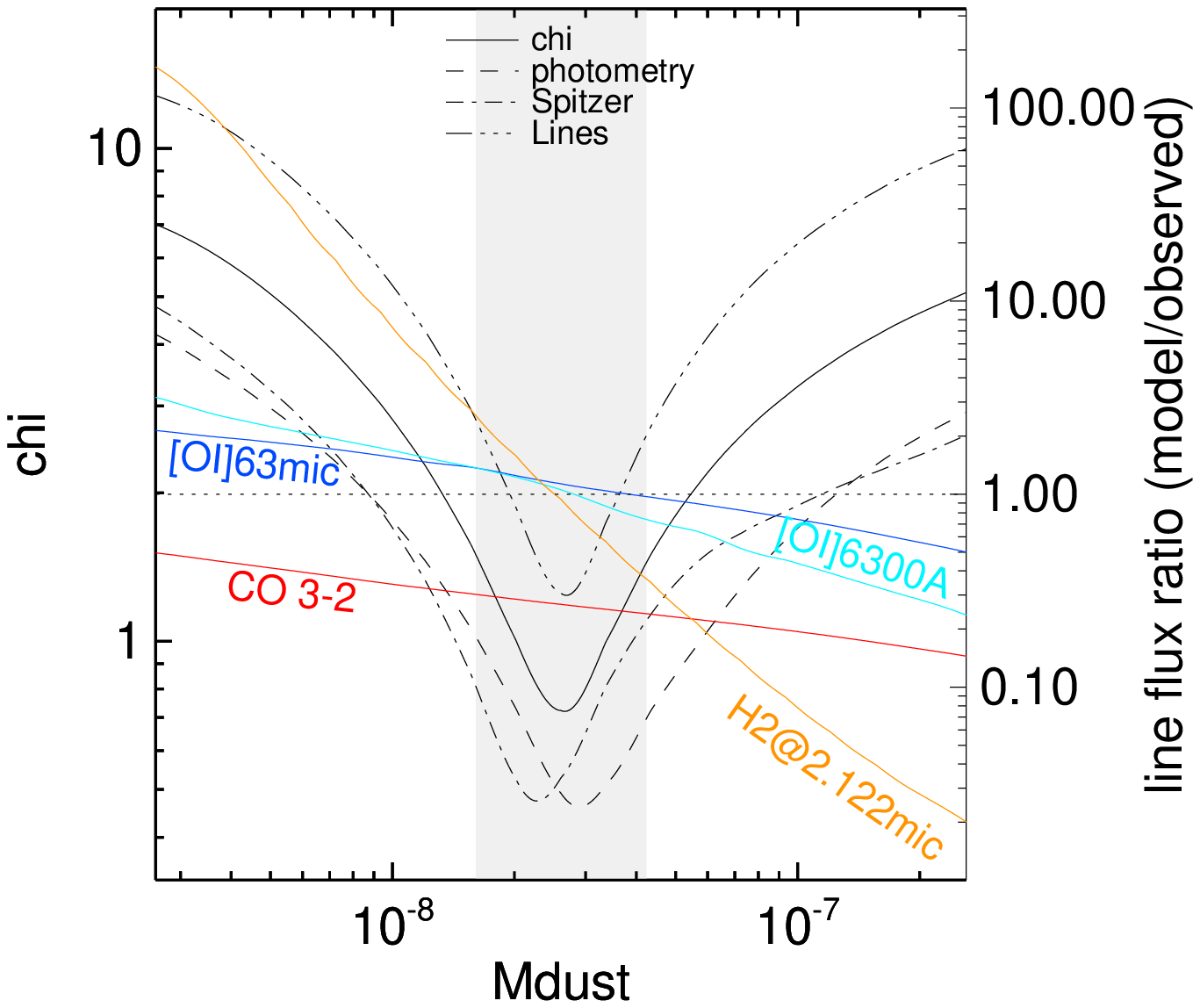} &
  \hspace*{-3mm}\includegraphics[height=59.5mm,
    trim=10mm 0mm 0mm 0mm, clip]{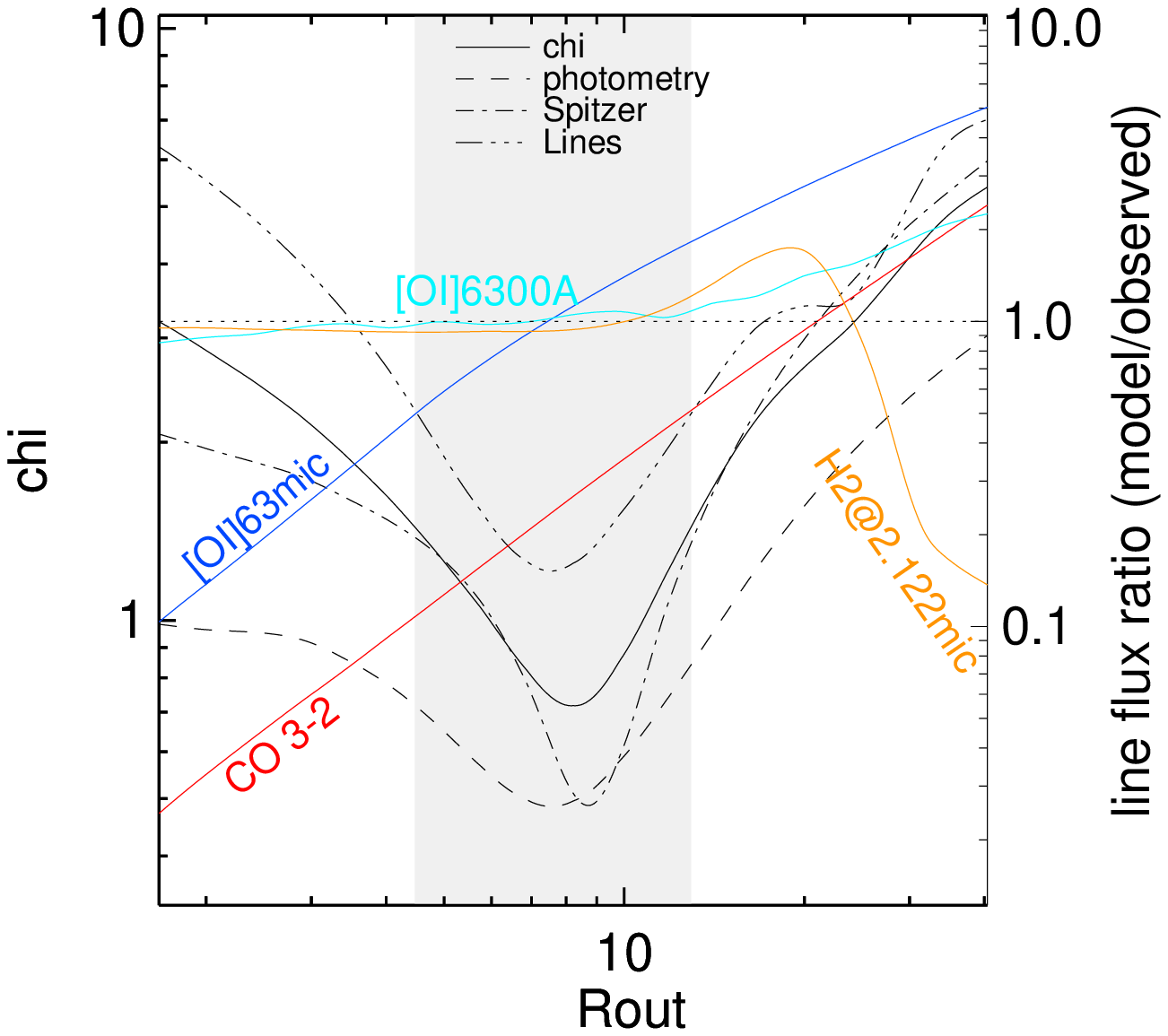} \\[-2mm]
\end{tabular}
\caption{Systematic variation of single parameters around the values
  of the best-fitting disk model, showing the dependencies on total
  disk gas mass $M_{\rm gas}\rm\,[M_{\odot}]$ {\bf (l.h.s.)}, on total
  dust mass $M_{\rm dust}\rm\,[M_{\odot}]$ {\bf (middle)} and on outer
  disk radius $R_{\rm out}\,$[AU] {\bf (r.h.s.)}.  The full black line
  shows the total $\chi$, and the dotted, dashed and dashed-dotted
  lines show its constituents, $\chi_{\rm Phot}$, $\chi_{\rm Spit}$
  and $\chi_{\rm Line}$, see Eq.~(\ref{eq:chi2}).  The computed line
  flux ratios with respect to the observations $F_{\rm line}/F_{\rm
  obs}$ ($F_{\rm line}/(3\sigma)$ for CO\,$3\!\to\!2$) are shown by
  the coloured lines as labelled, with tickmarks on the right axis).
  The grey shaded box marks the parameter-interval where $\chi$ is not
  larger than twice its minimum.}
\vspace*{-1mm}
\label{fig:errors}
\end{figure*}

\subsection{Dust composition}

\begin{figure}
\vspace*{-6mm}
\hspace*{-2mm}\includegraphics[width=9.2cm]{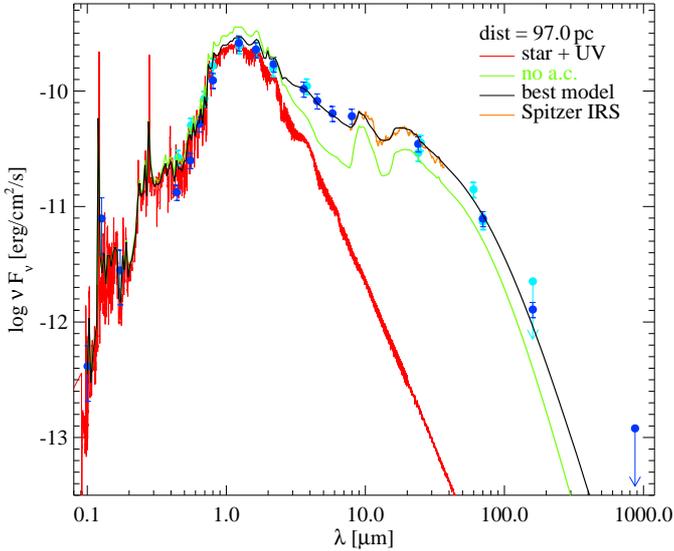}
\vspace*{-9mm}
\caption{Comparison between predicted SEDs of two models with
  different dust material composition. The black model shows once more
  the SED of our best fitting model, with dust composition as listed
  in Table~\ref{tab:Parameter}. In the green model, we have omitted the
  amorphous carbon. All model parameters are identical otherwise.}
\label{fig:SED_nocarbon}
\end{figure}

In our best fitting model, we assume an effective mix of about 33\%
amorphous fosterite $\rm Mg_2SiO_4$ \citep{Jaeger2003}, 24\% amorphous
carbon \citep{Zubko2004}, 23\% amorphous olivine $\rm MgFeSiO_4$
\citep{Dorschner1995}, 9\% amorphous silica $\rm SiO_2$
\citep{Posch2003}, 8\% amorphous enstatite $\rm MgSiO_3$
\citep{Dorschner1995}, and 3\% crystalline fosterite $\rm Mg_2SiO_4$
\citep{Servoin1973}. The citations indicate our sources for the
complex refractory indexes of the various pure materials. The dust
absorption and scattering opacities are calculated by applying
effective mixing theory \citep{Bruggeman1935} and Mie theory, based on
these optical data and volume fractions, and assuming that the
chemical dust composition and unsettled dust size distribution is
unique throughout the disk. The inclusion of crystalline fosterite is
motivated by the fine-structure of the observed second silicate
feature at 20\,$\mu$m \citep{Bouwman2006}, see \citep{Sicilia2009},
which shows several narrow peaks close to the fosterite peak positions
in the data of \citet{Servoin1973}. The resulting volume fractions are
a by-product of our automated fitting procedure from additional runs
of the evolutionary strategy not shown in
Table~\ref{tab:Parameter2}. We do not claim, however, to have
determined the dust composition of ET\,Cha, as our method is focused
on fitting the overall shape of SED rather than individual dust
features.

The inclusion of 24\% amorphous carbon, however, was an important step
to understand the SED of ET\,Cha. A comparison model without amorphous
carbon (green line in Fig.~\ref{fig:SED_nocarbon}) demonstrates
its impact on the SED. First, amorphous carbon reduces the dust albedo
at UV to near-IR wavelengths (see Fig.~\ref{fig:dustopac}). Pure
laboratory silicates have an albedo of about 80\%\,-\,99\% around
$1\,\mu$m. This leads to a substantial starlight amplification via
scattering by the disk at UV to near-IR wavelengths, which is
inconsistent with the photometric data. Second, amorphous carbon
enhances the absorption and thermal emission in the near-mid IR.  As a
consequence, the disk is more effectively heated by star light, and
produces more thermal emission shortward of the 10\,$\mu$m silicate
feature, just where ET\,Cha is very bright, resulting in a much better
fit of the SED if we include amorphous carbon.  However, any other
kind of impurities or inclusions, metallic iron for instance, would
cause a similar increase of the dust absorption opacities at optical
to near-IR wavelengths, amorphous carbon is just one of the
options. The formation of small alien inclusions in the dust material
has been demonstrated by \citet{Helling2006a} and \citet{Helling2006b}
to be a natural consequence of the refractory dust formation
process in somewhat different oxygen-rich environments, namely in
brown dwarf atmospheres. Concerning ET\,Cha we conclude that strong
near-IR dust absorption opacities are needed to fit the SED,
substantially stronger than those of pure amorphous or crystalline
silicates which have a ``glassy'' character.

\begin{figure*}
\centering
\begin{tabular}{cccc}
 %\hspace*{ -8mm}\includegraphics[width=51mm]{LineCO32.eps} &
  \hspace*{ -9mm}\includegraphics[width=65mm]{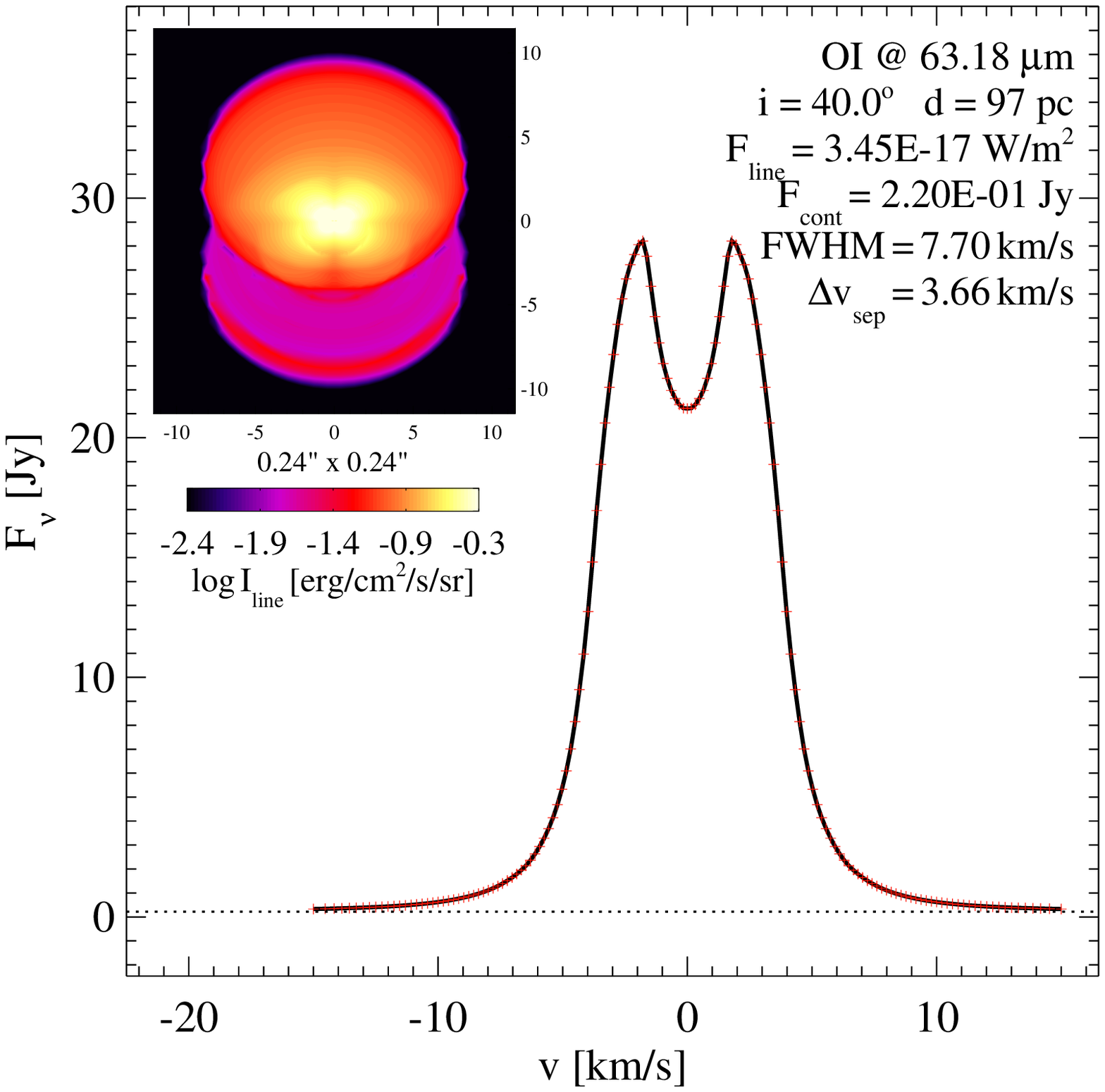} &
  \hspace*{-12mm}\includegraphics[width=65mm]{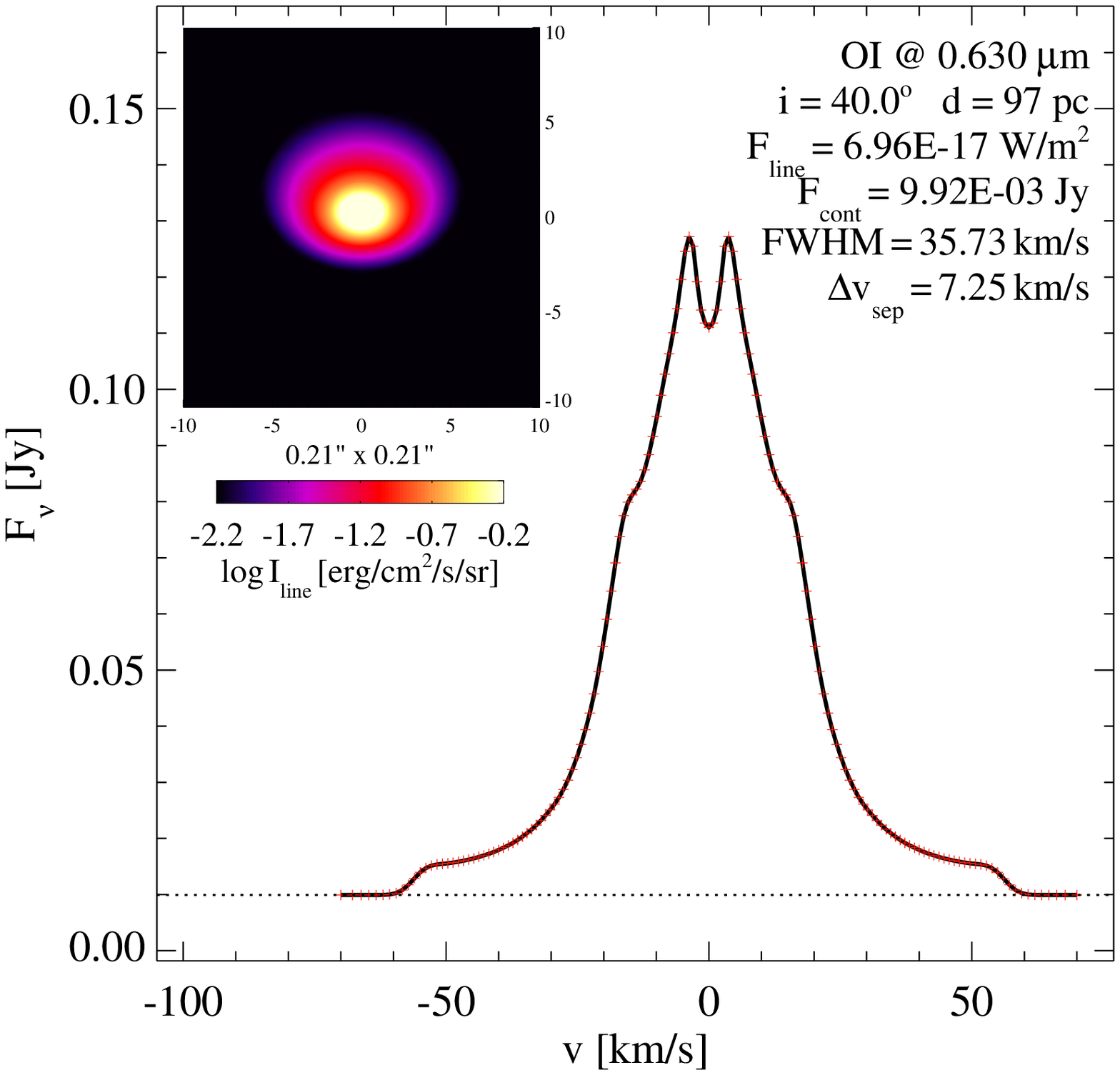} &
  \hspace*{-11mm}\includegraphics[width=65mm]{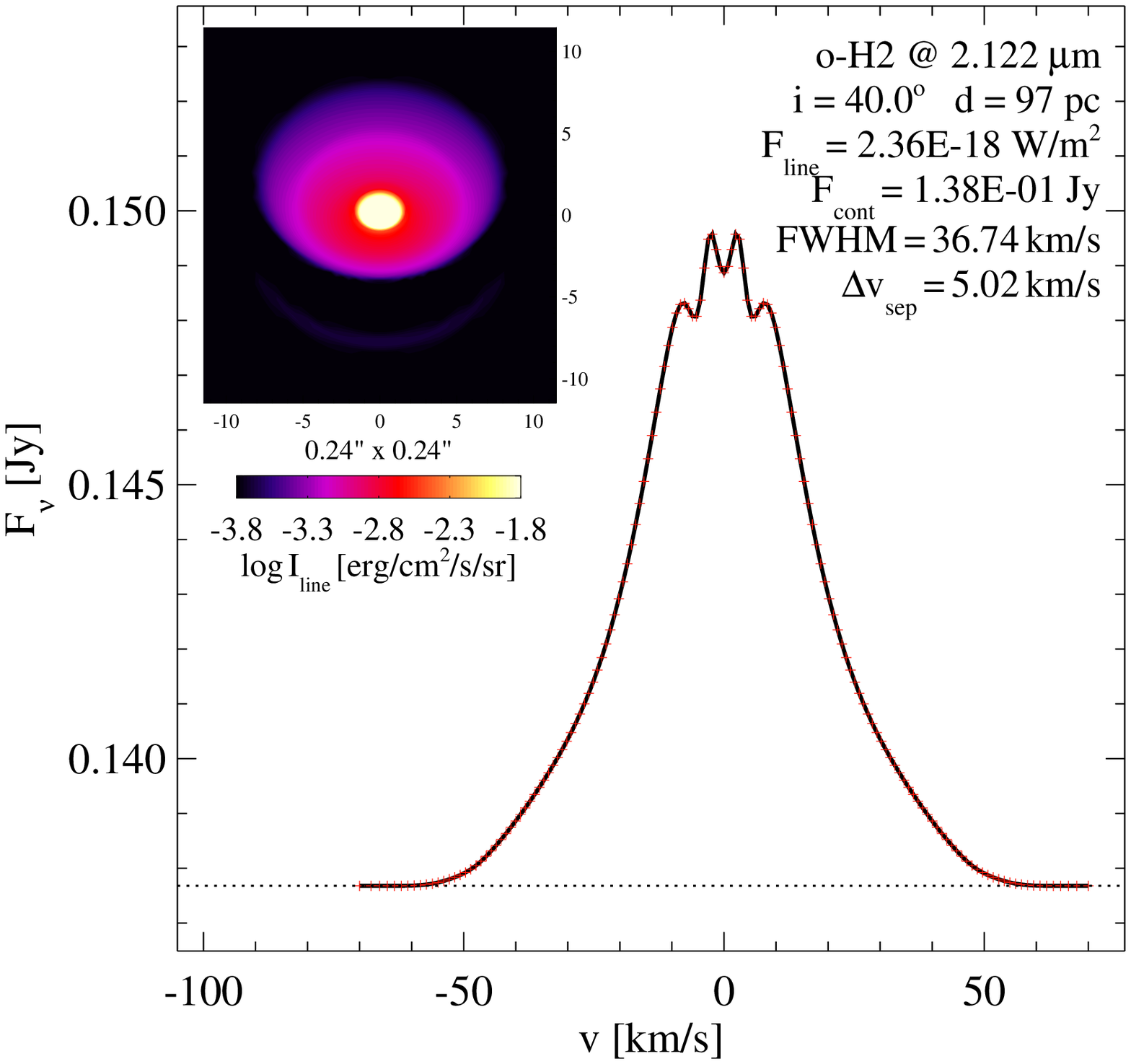} \\[-1mm]
 %\hspace*{ -9mm}\includegraphics[width=52mm]{LineAnalysisCO32.eps} &
  \hspace*{ -6mm}\includegraphics[width=67mm]{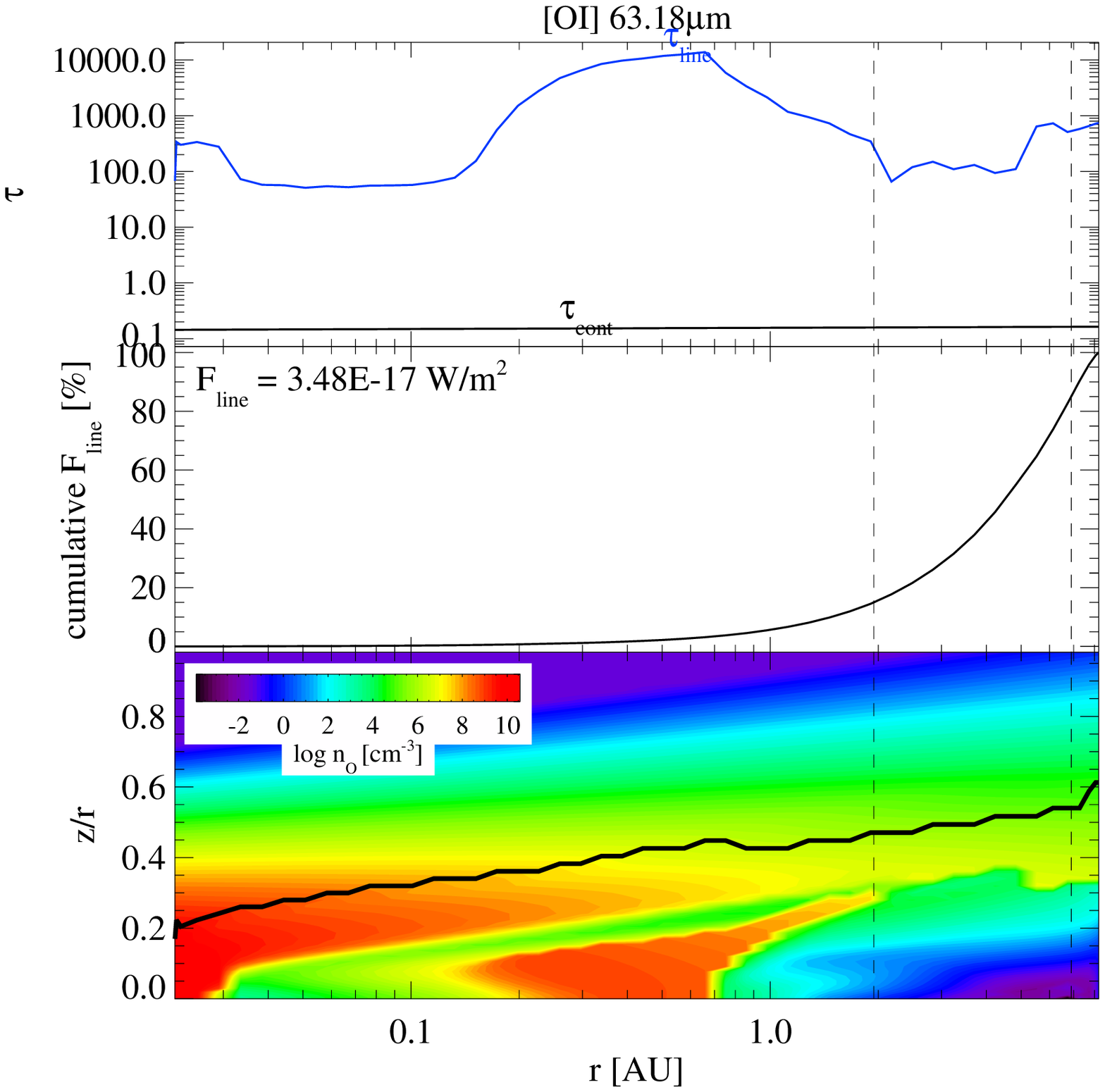} &
  \hspace*{ -9mm}\includegraphics[width=67mm]{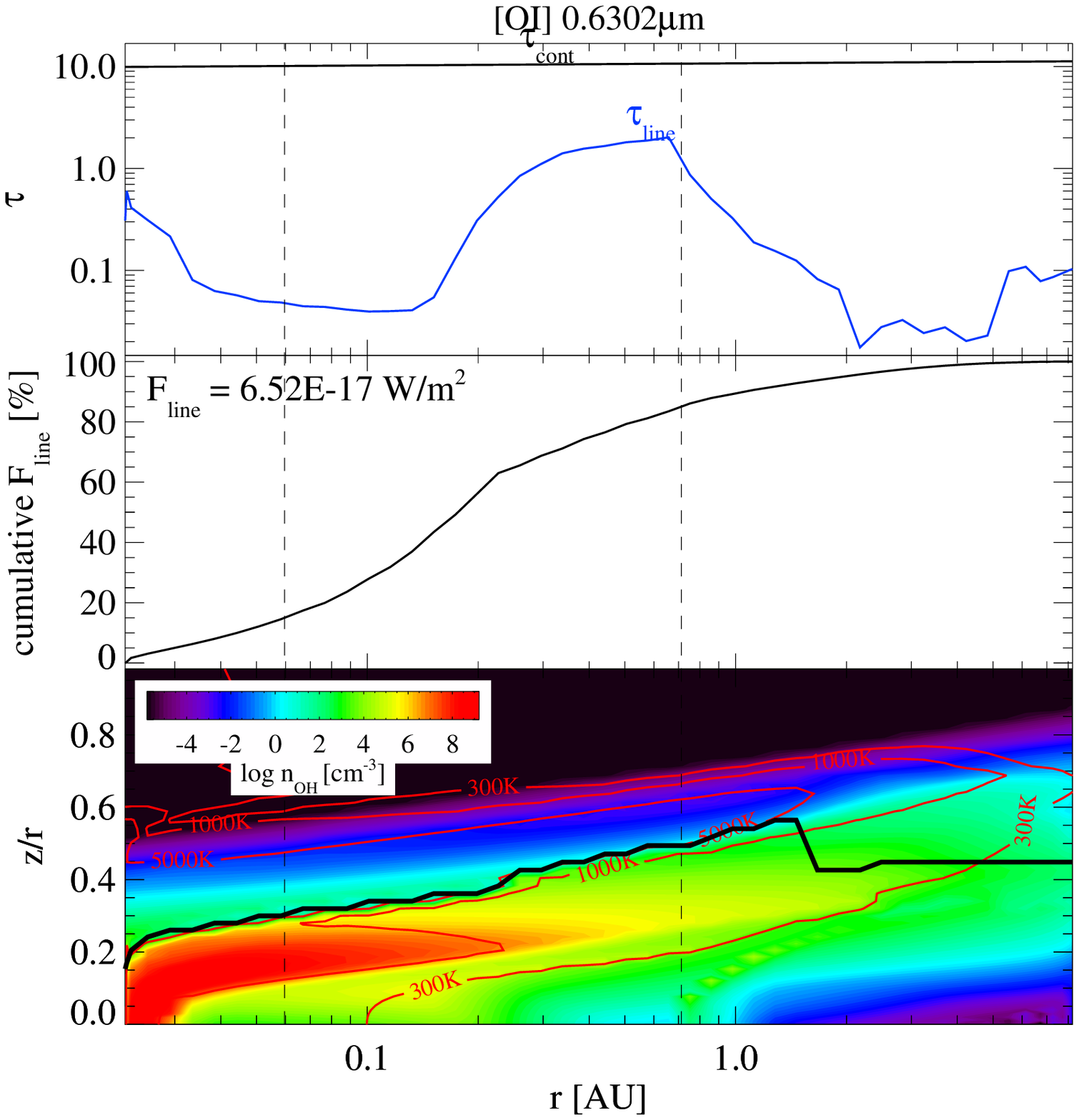} &
  \hspace*{ -9mm}\includegraphics[width=67mm]{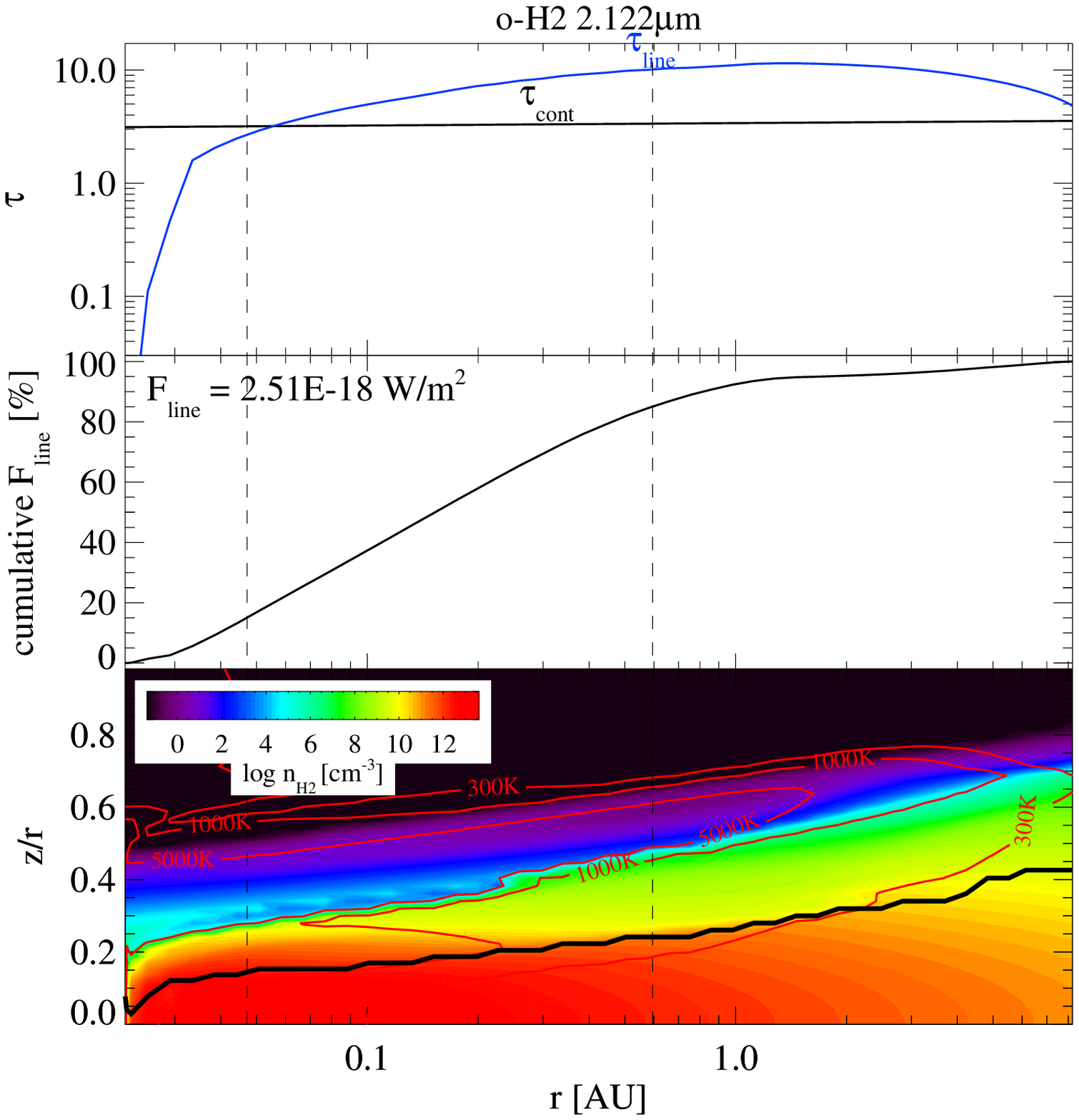} 
\end{tabular}
\caption{Line profiles, intensity maps, and spatial origin of some
  emission lines according to the best-fitting disk model. The {\bf
  upper row of figures} shows the calculated continuum level, line
  profile, and continuum-subtracted frequency-integrated line
  intensity-map for [OI]\,63.2$\,\mu$m (left), [OI]\,6300\,\AA\
  (middle) and o-H$_2$v=1$\to$0\,S(1)\,2.122\,$\mu$m (right), based
  on formal line transfer solutions at $\rm
  inclination\!=\!40\degr$. Note the different scaling of velocity
  axis for the different lines. The {\bf lower row of figures}
  visualises the vertical line and continuum optical depths as
  function of radius, the cumulative line flux, and the species
  density $\rm [cm^{-3}]$.  The vertical dashed lines indicate where
  the line flux is coming from, bracketing 70\% of the cummulative
  line flux in radius.  The thick black lines in the lowest boxes mark
  those grid cells that contribute most to the vertical line flux; 
  they mark the geometrical depth of the line formation region. For the
  analysis figures in the lower row, we have directly applied the
  upwards escape probability formalism (without formal solutions of
  line transfer).  These analyses are hence only approximate in nature
  and valid for strictly upward line propagation only ($\rm
  inclination\!=\!0\degr$). The line fluxes calculated in this
  way deviate from the proper results under $40\degr$ by less than 8\%.}
\label{fig:lines}
\end{figure*}

\subsection{Outer disk radius}
\label{sec:Rout}

Any outer disk radius, at least up to 200\,AU, works fine to fit the
SED alone.  However, the non-detection of CO$\,J\!=\!3\!\to\!2$ and
the modest [OI]\,63\,$\mu$m line flux put severe constraints on
$R_{\rm out}$. As Figs.~\ref{fig:evolution} and \ref{fig:errors}
demonstrate, the CO line flux depends very strongly and robustly on
$R_{\rm out}$. All calculated models with $R_{\rm out}\!>\!25$\,AU
would violate the $3\sigma$ upper limit of CO$\,J\!=\!3\!\to\!2$.
Only models with $R_{\rm out}\!<\!10$\,AU are consistent with the
$1\sigma$ CO line flux upper limit. The CO lines are extremely
optically thick, $\tau_{\rm line}\!\approx\!10^5$ (see
Sect.~\ref{sec:lines}).  Therefore, even if the CO abundance was
reduced by a factor of $10^3$ throughout the disk, for example by very
efficient CO ice formation, the conclusion that
the disk of ET\,Cha must be small would still be valid. The relatively
robust determination of $R_{\rm out}\!\approx\!(6-9)$\,AU is
demonstrated by Fig.~\ref{fig:errors} and Table~\ref{tab:Parameter2}.
Smaller values of $R_{\rm out}$ are actually inconsistent with the
measured [OI]\,63\,$\mu$m line flux (unless interpreted as originating
form an outflow, see Sect.~\ref{sec:outflow}), because the outer disk
is responsible for the [OI]\,63\,$\mu$m line emission (see
Sect.~\ref{sec:lines}).

\subsection{Spatial origin and characteristics of gas emission lines}
\label{sec:lines}

Before we continue to conclude about the determination of the disk gas
mass of ET\,Cha, we first have to clarify where the observed
spectral lines come from and what they tell us.
Figure~\ref{fig:lines} shows the calculated line fluxes and profiles
of the three detected lines [OI]\,63\,$\mu$m, [OI]\,6300\,\AA\ (LVC)
and o-H$_2$\,2.122\,$\mu$m. We also plot here the vertical line and
continuum optical depths and the cumulative line flux as function of
radius, to facilitate the discussion of the lines' spatial
origin. These plots are based on the best-fitting disk model, but the
drawn conclusions about the spatial origin of the spectral lines are
quite general and valid for all calculated disk models that fit the
observations of ET\,Cha.\\[-2.2ex]

{\ }\\*[-2.2ex]\indent
{\sffamily\itshape The far-IR [OI]\,63\,$\mu$m line} probes the outer
disk layers, about $2\!-\!7$\,AU in this tiny model, and is optically
thick with vertical optical depths $\tau_{\rm
line}\!\approx\!100\!-\!1000$ in this radial region\footnote{For more
typically extended disks, $R_{\rm out}\!=\!100\!-\!500$\,AU, the
[OI]\,63\,$\mu$m line mostly originates in $10\!-\!100$\,AU, slightly
larger in case of Herbig\,Ae disks \citep{Kamp2010}.}. The continuum is
optically thin. Since the line is collisionally excited, with an
excitation energy of about 228\,K, the line flux probes first and
foremost the existence of warm gas ($T_{\rm gas}\!\ga\!50\,$K) in the
disk surface, here at relative heights
$z/r\!=\!0.4\!-\!0.5$. In this region, PAH heating is usually the most
important heating process. Therefore, this line shows a strong
correlation with the assumed PAH abundance, $f_{\rm PAH}$, UV
irradiation, and disk flaring.

{\ }\\*[-2.2ex]\indent
{\sffamily\itshape The optical [OI]\,6300\,\AA\ line} originates from
hot gas in the inner disk, about $0.06\!-\!0.7$\,AU in this model. As
a forbidden line, with very small Einstein coefficient
$A_{41}\!\approx\!6.5\times10^{-3}\rm\,s^{-1}$, it is mostly optically
thin $\tau_{\rm line}\!\approx\!0.1\!-\!1$. With an excitation energy
of $\sim$15900\,K, it probes the existence of hot and dense gas
$T_{\rm gas}\!\ga\!3000\,$K in front of an optically thick
continuum. To our surprise, this line is {\it not} significantly
excited by OH photo-dissociation in the model. Neglecting this pumping
effect (see Appendix~\ref{sec:OHpump}) weakens the [OI]\,6300\,\AA\
line flux by only 14\%. Thus, the [OI]\,6300\,\AA\ line comes from the
{\it bottom of the hot atomic layer}, from about $z/r\!=\!0.3\!-\!0.4$
in this model. As soon as molecules form, for example OH and CO, the
temperature drops significantly by molecular line cooling, and the
[OI]\,6300\,\AA\ line cannot be excited any longer. Since the line is
optically thin, its flux reacts quite sensitively on disk mass. We
note that the predicted line $\rm FWHM\!=\!35.7\,km/s$ is in excellent
agreement with the observations, $\rm FWHM\!=\!(38\pm15)\,km/s$,
meaning that the spatial origin of the [OI]\,6300\,\AA\ line is about
correct in the model.

{\ }\\*[-2.2ex]\indent {\sffamily\itshape The near-IR
o-H$_2$\,v=1$\to$0\,S(1)\,2.122\,$\mu$m line} also probes the inner
disk regions, $0.05\!-\!0.6$\,AU in this model. The line is quite
optically thick, $\tau_{\rm line}\!\approx\!5\!-\!10$ in this region,
where the dust is also optically thick. Since its Einstein coefficient
is extremely small, $A_{61}\!\approx\!7\times10^{-7}\rm\,s^{-1}$, the
line needs large H$_2$ column densities of order $10^{23}\rm\,cm^{-2}$
to become visible over the continuum level. Such column densities are
only reached in quite deep layers, $z/r\!\approx\!0.15$, so this line
forms in much deeper layers than the [OI]\,6300\,\AA\ line.  The
temperature contrast between gas and dust is already quite small in
these layers (Fig.~\ref{fig:diskmodel1}), which limits the
o-H$_2$\,2.122\,$\mu$m line flux. The excitation energy of the line is
about 7000\,K, \ie $T_{\rm gas}\!\ga\!1000\,$K is required to
collisionally excite it.  In the best-fitting model such conditions,
$T_{\rm gas}\!\ga\!1000\,{\rm K}\!\gg\!T_{\rm dust}$, are provided by
the effect of exothermic chemical reactions (see
Appendix~\ref{sec:chemheat}) which is active in the warm and dense gas
close to the inner rim, deep but not too deep, just where this line
forms. Therefore, we see a clear correlation between the
o-H$_2$\,2.122\,$\mu$m line flux and the assumed heating efficiency of
exothermic reactions $\gamma^{\rm chem}$. The line is also
substantially pumped by H$_2$-formation on grains (see
Appendix~\ref{sec:H2pump}). When neglecting this effect, the line
attains a flux of only 35\% of the value from the full model.  We
conclude that the H$_2$ line is sensitive to a temperature contrast
between gas and dust, and H$_2$-formation, in quite deep disk layers
close to the inner rim.  However, the predicted line is too broad
($\rm FWHM\!=\!37\,$km/s) as compared to the observations ($\rm
FWHM\!=\!18\pm1.2$\,km/s). We were unable to find any disk model,
among the $\sim\!20000$ models computed, that shows such a narrow
o-H$_2$\,2.122\,$\mu$m line. Thus, the line forms too close to the
star in the model, and we must be very careful when drawing our
conclusions about the nature of ET\,Cha from the observed
o-H$_2$\,2.122\,$\mu$m line.

{\ }\\*[-2.2ex]\indent
{\sffamily\itshape The sub-mm CO$\,J\!=\!3\!\to\!2$ line} (not
depicted in Fig.~\ref{fig:lines}) probes the size of the disk and the
gas temperature in the outermost disk layers. It is massively
optically thick, $\tau_{\rm line}\!\ga\!10^5$ in the line forming
region 2\,-\,7\,AU, where the dust is optically thin. Its flux is
roughly proportional to the projected disk area times the gas
temperature at relative heights $z/r\!\approx\!0.4\!-\!0.5$ in these
outermost parts of the disk.

\subsection{Disk gas mass and disk shape}

The determination of the total disk gas mass of ET\,Cha in this paper
is based on three detected gas emission lines, namely
[OI]\,63\,$\mu$m, [OI]\,6300\,\AA\ (LVC), and o-H$_2$\,v=1$\to$0
S(1)\,2.122\,$\mu$m, which probe complementary radial and vertical
disk regions. However, the first line, [OI]\,63\,$\mu$m, is massively
optically thick, and the latter two lines mainly probe the hot gas in
the inner disk regions.  CO$\,J\!=\!3\!\to\!2$ was not detected.
These circumstances immediately suggest that a robust gas mass
determination of ET\,Cha is difficult.

Table~\ref{tab:Parameter2} shows that various solutions in form of
well-fitting disk models can be found where the total gas mass differs
by up to two orders of magnitudes (factor 300). A careless reading of
Table~\ref{tab:Parameter2} would suggest that the particular values
for $M_{\rm gas}$, $\epsilon$ and $\beta$ do not matter much, but this
is not true. Careful inspection of the solutions in
Table~\ref{tab:Parameter2} by systematic variation of selected
parameters (see \eg Fig.~\ref{fig:errors} for the best-fitting model)
shows that these solutions all represent well-defined local minima,
where small changes of {\it any} parameter leads to a considerable
deterioration of the combined line + continuum fit quality $\chi$. If
one would measure the uncertainty of gas mass determination from these
dependencies alone (for instance the deviation $\Delta M_{\rm gas}$
where $\chi$ doubles) one would arrive at relatively small errors,
about 40\%. But such an error estimate would be misleading as it would
not take into account the complicated manifold of local minima in
parameter space.

All solutions in Table~\ref{tab:Parameter2} fit our entire set of line
and continuum observations about equally well. The different values
for the gas mass, however, come in certain combinations with other
disk shape parameters, like the column density power-law exponent
$\epsilon$ and the flaring power $\beta$. Certain fine-tuned
combinations of $M_{\rm gas}$, $\epsilon$ and $\beta$ do apparently
all provide the proper density and temperature conditions in the disk
that result in almost exactly the same observables, for example small
$M_{\rm gas}$ in combination with large $\epsilon$ and large $\beta$,
or vice versa.

These relations can be understood by minimum column densities of warm
H$_2$ and atomic oxygen in the inner disk parts that are inevitably
required to make the [OI]\,6300\,\AA\ (LVC) and o-H$_2$\,2.122\,$\mu$m
lines visible over the strong continuum at optical and near IR
wavelengths. For example, in the least massive model~1 in
Table~\ref{tab:Parameter2}, there is so little gas in the disk that a
relatively large value of $\epsilon\!=\!1.16$ is required to
concentrate the mass in the inner disk parts and so to provide the
necessary column densities there. All line fluxes are in good
agreement with the observations then, but the model has problems to
find a good fit of the 10\,$\mu$m and 20\,$\mu$m silicate features.
Also the flaring index of $\beta\!=\!1.33$ is quite extreme, leading to
a relative disk height of about $z/r\!=\!1.2$ in the outer disk parts
-- such a ``disk'' would be taller than wide.  From these arguments,
we will discard model~1 from our selection of valid solutions in the
following, and claim a minimum gas mass of about
$5\times10^{-5}\rm\,M_{\odot}$ to achieve the necessary minimum gas
column densities in the inner disk regions and to fit the silicate
dust emission features simultaneously.

Concerning the other direction, we find it hard to provide any solid
argument why the disk gas mass must not exceed a certain maximum
value.  In combination with our quite robust dust mass determination
of $M_{\rm dust}\!\approx\!3\times10^{-8}\,M_\odot$, however, even our
minimum gas mass of $5\times10^{-5}\rm\,M_{\odot}$ already implies a
very high gas/dust ratio of 2000. An even higher gas mass would translate
into an accordingly higher gas/dust ratio. The largest value we found
with our evolutionary strategy is $2.5\times10^{-3}\rm\,M_{\odot}$.

Summarising our modelling efforts, after having tried about 20000 disk
models and various choices of input physics, we conclude that the disk
gas mass of ET\,Cha is only little constrained, our confidence
interval is about $M_{\rm gas}\!\approx\!(5\times10^{-5} -
3\times10^{-3})\,M_{\odot}$.  The surface density exponent $\epsilon$
and the flaring parameter $\beta$ are likewise poorly constrained,
although the models favour small $\epsilon\!\approx\!0$ to fit the
silicate dust emission features.

In view of the outflow discussion (Sect.~\ref{sec:outflow}), we note
that the unclear physical origin of [OI]\,63\,$\mu$m obviously
increases our uncertainty in gas mass determination. We have set up a
similar evolutionary optimisation run as depicted in
Fig.~\ref{fig:evolution}, but now treating the [OI]\,63\,$\mu$m line
flux, as emitted by the disk, as an upper limit with
$<\!2\times10^{-18}\rm\,W/m^2$. This run resulted in a disk of similar
mass as compared to the best-fitting model, $M_{\rm gas}\!=\!5.5\times
10^{-4}\,M_\odot$, but of even smaller size $R_{\rm out}\!=\!2.1\,$AU,
with an [OI]\,63\,$\mu$m line flux of
$\sim\!4\times10^{-18}\rm\,W/m^2$. We conclude that the
[OI]\,63\,$\mu$m emission line flux is not crucial for the disk gas
mass determination of ET\,Cha, because this disk is tiny and the hot
gas is responsible for our detected [OI]\,6300\,\AA\ (LVC) and
o-H$_2$\,2.122\,$\mu$m lines is located in the inner regions. For
other, more extended protoplanetary disks, the [OI]\,63\,$\mu$m line
will be less optically thick and hence more useful for the purpose 
of gas mass determination \citep{Kamp2010,Pinte2010,Woitke2010}.

\section{Discussion}

\subsection{Scale heights and unidentified heating}

From the SED fitting, we get surprisingly robust results concerning
the assumed vertical disk scale height, $H_0\!=\!(0.008-0.011)\,$AU at
reference radius $r_0\!=\!0.1\,$AU, throughout all SED-fitting models.
In the model, the reference scale height in combination with the
flaring power $\beta$ determines how much star light is captured by
the disk.  This light (unless scattered away) is absorbed by the dust
in the disk surface and then thermally re-emitted, heating also the
inner disk parts. Thus, $H_0$ regulates the dust temperature in the
disk. Since ET\,Cha is very bright in the $(3\!-\!8)\,\mu$m region, we
require large relative scale heights of the order of 10\% close to
the star to create enough dust emission from warm and hot grains to
fit the near and mid IR.
  
Appendix~\ref{sec:struc} (see lower left plot in Fig.~\ref{fig:struc})
demonstrates, however, that our prescribed scale heights are
significantly larger, by a factor of 2-3, than those derived from
self-consistent models where vertical hydrostatic equilibrium is
assumed. This mismatch can be interpreted in three ways: (1) the close
midplane regions of T\,Tauri disks are not in hydrostatic equilibrium,
(2) the midplane temperatures are actually 4-9 times higher than
assumed in the model (assuming $H\!\propto\!c_T\!\propto\!\sqrt T$),
or (3) there is an additional dust heating process active in the close
disk midplane regions that leads to an additional energy flux through
the dust component resulting in more observable near-mid IR photons
without changing the temperatures much. Possibility~1 cannot be
excluded per se. Possibility~2 seems unrealistic because it would
cause the dust to evaporate. We favour possibility~3.
Appendix~\ref{sec:dustnonRE} shows that the inclusion of non-radiative
dust heating via inelastic gas-dust collisions (thermal accommodation,
driven by gas-dust temperature differences created through exothermic
chemical reactions) leads to similar effects than increasing the scale
height. Because of the $\rho^2$-scaling of chemical reactions, this
additional heating affects the dust preferentially in high-density
regions, increasing the production of near IR photons needed to fit
the SED of ET\,Cha with smaller scale heights.  By means of an extra
run of the evolutionary strategy, we found out that we can reduce the
scale height by about 35\% to fit the SED, if we include this effect.

More unidentified dust heating processes may be active in the close
midplanes of T\,Tauri disks. However, viscous heating \citep[according
to the formulation by][]{Frank1992} is {\it not} doing a particularly
good job in explaining the scale height inconsistencies of ET\,Cha, see
Appendix~\ref{sec:viscous}. Viscous heating dumps additional energy
$\propto\!\rho$, i.e., after volume-integration, preferentially into
the outer cold regions, leading to the production of more far IR
photons. Moreover, viscous heating simply has little effect on the
resulting disk temperatures and observable continuum flux in case of
low-mass disks as for ET\,Cha. In fact, the viscous heating
according to the formulation by \citep{Frank1992} produces artifacts
in the uppermost tenuous disk layers, where the viscous heating
$\propto\!\rho$ cannot be balanced by any cooling $\propto\!\rho^2$.

\subsection{Effects of other physical processes}

Appendix~\ref{sec:inputphysics} discusses a number of further physical
processes and their influence on the model results that have not been
mentioned so far and have not been included in the models discussed so
far. To summarise, we find that

(i) X-rays have very little effect (see Appendix~\ref{sec:Xray}) and
the neglecting of X-rays in our main model for ET\,Cha is fully
justified. An X-ray luminosity of $L_X\!=\!6\times10^{28}\,$erg/s, as
observed for ET\,Cha \citep{Lopez2010}, turns out to be much less
important for the disk heating and ionisation as compared to the
strong FUV of ET\,Cha, $L_{\rm UV}\!=\!6.5\times10^{31}\,$erg/s as
measured between 912 and 2500\AA, based on our {\sc Hst/Cos} and {\sc
Hst/Stis} observations.

(ii) The treatment of H$_2$-formation on grain surfaces is one of the
most important yet quite uncertain processes for astrochemical
modelling. Appendix~\ref{sec:H2form} shows that different approaches
to calculate the H$_2$-formation rate lead to a systematic uncertainty
in the model for the computation of the o-H$_2$\,2.122\,$\mu$m about a
factor of 5. Other spectral lines are less effected.

\subsection{Disk inclination and outflow velocity}

The low blue-shifts ($-42\,$km/s) seen in the optical emission lines
(Fig.\,\ref{fig:OI6300}) are quite unusual for the jets/outflows of
T\,Tauri stars.  Typical outflow velocities are $\ga\!100\,$km/s
\citep{Hartigan1995}. A possible explanation could be the projection
effect in case of a strongly inclined disk (i.e., close to
edge-on). Adopting 100\,km/s as a lower limit for the outflow velocity
of ET\,Cha would suggest a disk inclination of $>\!65\degr$.

However, as argued in Sect.~\ref{sec:diskpara}, we find that
inclinations in the range $0\degr\!-\!50\degr$ are in agreement with
the observations, but larger inclinations would result in a partial
obscuration of the star by the disk, with dramatic effects on the
SED. The SED analysis therefore leads to the conclusion that the
inclination of the system, as measured from face-on, is $50\degr$ or
less which, in turn, translates to an outflow velocity of 65\,km/s at
most. If this interpretation is correct, this would make the outflow
from ET\,Cha one of the slowest known outflows from a T\,Tauri star.

\subsection{Outflow and disk lifetime}
\label{sec:outflow}

Our analysis of the optical [OI]\,6300\,\AA\ and [SII]\,6731\,\AA\
emission line profiles and fluxes (see Appendix~\ref{app:outflow})
results in an estimate of the outflow mass-loss rate of ET\,Cha of
$\dot{M}_{\rm outflow}\!\approx\!10^{-9}\rm\,M_\odot/yr$. Such an
outflow can contribute to the [OI]\,63\,$\mu$m emission, which would
render the [OI]\,63\,$\mu$m line flux, as emitted from the disk,
smaller as assumed in Sect.~\ref{sec:diskfit}.  In summary,
Appendix~\ref{app:outflow} shows that both approaches, the simple
energetic outflow analysis by \citet{Hollenbach1985}, as well as the
shock models computed by \citep{Hartigan2004}, suggest that the
outflow from ET\,Cha does contribute a substantial, if not dominant,
fraction of the observed [OI]\,63.2$\mu$m emission line flux. However,
without spatially or velocity resolved observations of the outflow,
and its subsequent detailed modelling (rather than using some
``template'' shock models), it is presently impossible to assess the
exact contribution of the outflow to the line emission.

Our estimate of the outflow rate of ET\,Cha is large compared to the
total disk mass we derive, $M_{\rm
disk}\!\approx\!6\times10^{-4}\rm\,M_{\odot}$, as it suggests a disk
lifetime of only $M_{\rm gas}/\dot{M}_{\rm
outflow}\!\approx\!0.7$\,Myr, inconsistent with the age of the
Chamaeleontis cluster of $\sim\!8\,$Myr. If we would assume a generic
gas/dust ratio of 100 and take our dust mass determination for
granted, $M_{\rm dust}\!\approx\!3\times10^{-8}\rm\,M_{\odot}$, it is
even worse. The disk lifetime would be even shorter in this case, only
3000\,yrs, way too short to be feasible, unless we are just observing
a temporary but short-lived 100$\times$ peak in outflow rate.

We also notice that the mass accretion rate of ET\,Cha was estimated
by \citet{Lawson2004} to be equally large, $\dot{M}_{\rm
acc}\!\approx\!10^{-9}\rm\,M_\odot/yr$ as the outflow mass loss rate,
leading to similar lifetime inconsistencies. Furthermore, a branching
ratio of $\dot{M}_{\rm outflow}\,/\,\dot{M}_{\rm acc}\!\sim\!1$ is
highly unusual, a few percent seems to be a well-established value for
T\,Tauri stars \citep[see e.g.][]{Hartigan1995}. 

\citet{Murphy2011} reported on highly variable $H_{\alpha}$ equivalent
widths and, accordingly, mass accretion rate, for the old T\,Tauri stars in
the $\eta$\,Cha cluster. A factor of 100 variation in the accretion
rate is observed in one newly-identified halo member of $\eta$\,Cha.
But only a simultaneous reduction of both $\dot{M}_{\rm
acc}$ and $\dot{M}_{\rm outflow}$ would help to resolve the
aforementioned lifetime inconsistency, which doesn't seem very likely.

One possible explanation would be a massive but short-lived outflow
due to a flare from a former epoch, when the mass accretion rate was
at least 10 times higher. An outflow of 100\,km/s would need
$\sim\!50$\,yrs to reach a distance of 1000\,AU. This would be in
agreement with the somewhat slow outflow velocity we derive,
$42\!-\!60$\,km/s, because the outflow might have slowed down ever
since.  However, the optical emission lines do not show any evidence
for a red-shifted HVC, as one would expect for a symmetric bi-polar
outflow. The only plausible explanation of the missing red-shifted
HVCs is that these components from the far side are attenuated by the
dust in the disk. According to our disk model, the disk is optically
thick at 6300\,\AA\ up to the outer radius $\sim\!10$\,AU (see
Fig.~\ref{fig:lines}). Therefore, the line emitting region of the
blue-shifted optical emission lines must be quite small, less than
5\,AU if seen under 60$\degr$ disk inclination, which corresponds to
an age of only 0.25\,yr.

\section{Summary and conclusions}

This paper has reported on new observations of ET\,Cha with several
instruments: {\sc Herschel/Pasc}, {\sc Ctio/Andicam}, {\sc
Hst/Cos/Stis}, and {\sc Apex}.  In combination with published data
from {\sc Spitzer}, {\sc Gemini/Phoenix} and {\sc Aat/Ucles}, we have
collected an unprecedented observational data set about this object,
including photometry, UV spectra, high-resolution optical spectrum,
near and mid IR spectra, and far IR and sub-mm line fluxes.

We have calculated united gas and dust models for the disk of ET\,Cha
that can simultaneously fit all line and continuum observations,
except for a too broad o-H$_2$\,2.122\,$\mu$m emission line profile.
The observations also show some blue-shifted components of optical
emission lines that point to an outflow and are not included in the
models.

This paper has explored the parameter space of the disk models by
using an evolutionary strategy to minimise the discrepancies between
model predictions and observations. The paper has also introduced a
number of basic improvements to the ProDiMo disk modelling code
concerning the treatment of PAH ionisation balance and heating,
heating by exothermic chemical reactions, several non-thermal pumping
mechanisms for selected gas emission lines, and formal solutions of
the line transfer problem at given inclination
(Appendix~\ref{sec:newProDiMo}).

From the disk modelling we find a rich variety of fitting disk models
that can explain our observations about equally well.  Some of the
model parameters (like the dust mass and the outer radius) can be
determined with some confidence whereas other parameters (like the
disk gas mass) are poorly constrained:\\*[-2.5ex]{\ }
\begin{itemize}
\item The new {\sc Herschel/Pacs} photometric fluxes at 70$\,\mu$m and
  160$\,\mu$m constrain the disk dust mass of ET\,Cha to be about
  $M_{\rm dust} = (2-5)\times10^{-8}\,M_\odot$, putting the object at
  the borderline between optically thin and optically
  thick.\\*[-1.3ex]

\item Then strong near IR excess of ET\,Cha can be fitted with a disk
  that is truncated at $R_{\rm in}\!\approx\!0.022\,$AU (where $T_{\rm
  dust}\!\approx\!1500\,$K) which is located slightly outside of the
  co-rotation radius of 0.015\,AU. The latter is calculated according
  to the assumption that the star rotates with a period
  $P=1.7\,$days as suggested by our $v_{\rm rot}\sin(i)$ analysis
  of rotationally broadened stellar absorption lines.\\*[-1.3ex]

\item From the {\sc Apex} CO\,$J\!=\!3\!\to\!2$ non-detection, we can
  infer, with confidence, that the disk of ET\,Cha must be tiny in
  radius. The models favour an outer disk radius as small as $R_{\rm
  out}\!\approx\!(6-9)\,$AU. All disk models with $R_{\rm
  out}\!\ga\!25\,$AU would violate the $3\sigma$ CO\,$J\!=\!3\!\to\!2$
  non-detection limit, independent of chemical details.\\*[-1.3ex]

\item The SED-fitting suggests that the dust grains in the surface of
  the protoplanetary disk of ET\,Cha (where the near-mid IR continuum
  forms) must be small in radius (sub-micron sized) and opaque in the
  optical and near-mid IR, \ie absorption must dominate over
  scattering opacities. In the models, these spectral properties are
  provided by the inclusion of about 25\% amorphous carbon, but other
  options like metallic iron are also possible.\\*[-1.3ex]

\item The disk gas mass of ET\,Cha is poorly constrained by our line
  observations CO\,$J\!=\!3\!\to\!2$ (non-detection),
  [OI]\,63\,$\mu$m, [OI]\,6300\,\AA\ (LVC), and
  o-H$_2$v=1$\to$0\,S(1)\,2.122\,$\mu$m. We find a variety of about
  equally well fitting disk models with total gas masses $M_{\rm
  gas}\!=\!(5\times10^{-5}-3\times10^{-3})\,M_\odot$.  The forbidden
  lines of [OI]\,6300\,\AA\ (LVC), and o-H$_2$\,2.122\,$\mu$m are
  close to optically thin and hence quite useful for gas mass
  determination, but originate from hot gas in the inner disk regions
  only.  [OI]\,63\,$\mu$m is massively optically thick and does hence
  not discriminate much between the various disk models for the case
  of ET\,Cha. We would need to observe additional far-IR or sub-mm
  spectral lines that originate from the outer disk parts to determine
  the gas mass with more confidence. However, in order to explain the
  [OI]\,6300\,\AA\ (LVC) and o-H$_2$\,2.122\,$\mu$m lines with a disk
  model, we require line optical depths $\tau_{\rm line}\!\ga\!1$
  which translates into total vertical column densities in the inner
  disk regions of about $N_{\rm O}\!\ga\!10^{21}\rm\,cm^{-2}$ and
  $N_{\rm H2}\!\ga\!10^{24}\rm\,cm^{-2}$ for atomic oxygen and
  molecular hydrogen, respectively. Such column densities are
  incompatible with $M_{\rm gas}\!<\!5\times10^{-5}\,M_\odot$.\\*[-1.3ex]

\item The wide range of fitting gas masses is related to particular
  values of the disk shape parameters, namely the column density
  exponent $\epsilon$ and the disk flaring power $\beta$. Large gas masses
  need to be combined with small values for $\epsilon$ and $\beta$, and vice
  versa.\\*[-1.3ex]

\item From our SED modelling of ET\,Cha we derive disk scale heights 
  of the order of 10\% relative to radius close to the star, which is 
  about a factor of 2-3 larger than the scale heights inferred 
  form self-consistent models that assume hydrostatic equilibrium.
  This discrepancy can partly be explained by an additional
  non-radiative heating of the dust close to the star, for example
  via exothermic chemical reactions.

\end{itemize}
These results suggest a surprisingly high value for the overall
gas/dust mass ratio of ET\,Cha of at least 2000, or even 20000.
Whether or not the gas responsible for the [OI]\,6300\,\AA\ and
o-H$_2$\,2.122\,$\mu$m emissions is still physically connected to the
disk is not entirely clear, but the observations show that this gas is
at least not moving much with respect to the disk.  Possibly, the
overwhelming majority of dust particles responsible for the near to
far-IR emission of the star has already been transformed into larger pebbles
or solid bodies, which have negligible opacities.

The fluxes in blue-shifted emission line components like
[OI]\,6300\,\AA\ (HVC) suggest an outflow with mass-loss rate
$\dot{M}_{\rm outflow}\!\approx\!10^{-9}\rm\,M_\odot/yr$, on a similar
level as the reported mass accretion rate of ET\,Cha. The low
velocities ($\approx\!-42\,$km/s) suggests a high inclination angle of
at least 60$\degr$ (rather more). However, inclinations in excess of
50$\degr$ are inconsistent with our SED-modelling, which favours
$i\!\la\!40\degr$, from which we determine an outflow velocity of
65\,km/s at most \citep[typical values are in excess of 100\,km/s for
T\,Tauri outflows][]{Hartigan1995}. If this interpretation is correct,
ET\,Cha would possess one of the slowest known outflows from a
T\,Tauri star.

An outflow with mass loss rate $\dot{M}_{\rm
outflow}\!\approx\!10^{-9}\rm\,M_\odot/yr$ is likely to contribute
significantly to the [OI]\,63$\mu$m line flux as observed with {\sc
Herschel/Pacs}. Given the observational data we have collected in this
paper, we cannot discriminate between outflow or disk origin of
[OI]\,63$\mu$m.

To conclude, ET\,Cha seems to be an extraordinary and puzzling object
concerning the evolution of protoplanetary disks. Despite its age of
about $(6-8)$\,Myr, there is evidence of active accretion onto the
central star, on a similar level as the derived outflow mass-loss
rate. According to the low disk masses we derive, the disk lifetime
$M_{\rm gas}/\dot{M}_{\rm acc}\approx(0.05\!-\!3)$\,Myr is
inconsistent with cluster age.  If a generic gas/dust ratio of 100 was
assumed, the disk lifetime, based on our relatively robust
determination of the disk dust mass, would be even shorter, only
$\sim\!3000$\,yrs. Either the object is actually much younger than the
age of the Chamaeleontis cluster or the object is going through a
phase of unusually high mass accretion rate and outflow.

\begin{acknowledgements}
  We thank Catherine Dougados for fruitful discussions about the
  properties of disk outflows and the interpretation of optical
  emission lines.  W.-F.~Thi acknowledges a SUPA astrobiology
  fellowship. I.~Pascucci and B.~Riaz acknowledge NASA/JPL for funding
  support. The LAOG group acknowledges PNPS, CNES and ANR (contract
  ANR-07-BLAN-0221) for financial support.  
  I.~de~Gregorio-Monsalvo is partially supported by Ministerio de
  Ciencia e Innovaci{\'o}n (Spain), grant AYA 2008-06189-C03
  (including FEDER funds), and by Consejer{\'i}a de Innovaci{\'o}n y
  Ciencia y Empresa of Junta de Andaluc{\'i}a, (Spain).
%
%  G.~Meeus, C.~Eiroa, I.~Mendigut\'ia and B.~Montesinos are partly
%  supported by Spanish grant AYA 2008-01727.
%
%  C.~Pinte acknowledges the funding from the EC 7$^{th}$ Framework 
%  Program as a Marie Curie Intra-European Fellow (PIEF-GA-2008-220891). 
%
%  D.R.~Ardila, S.D.~Brittain, C.A.~Grady, I.~Pascucci, B.~Riaz,
%  G.~Sandell and C.D.~Howards, J.-P.~Williams, G.~Matthews,
%  A.~Roberge, W.~Danchi acknowledge NASA/JPL for funding support.
%
%  E.~Solano and J.M.~Alacid acknowledge the funding from the Spanish
%  MICINN through grant AYA2008-02156.
%
\end{acknowledgements}

\bibliography{reference}

%\begin{thebibliography}{}
%\end{thebibliography}

%=====================================================================
\Online

\begin{appendix}
\section{New features in ProDiMo}
\label{sec:newProDiMo}

\subsection{Fixed disk structure}
\label{sec:fixstruc}

In contrast to earlier publications, we use a parametrised
description for the shape and distribution of gas and dust in
the disk in this paper
\begin{equation}
  \rho(r,z) = \rho_0\,
              \Big(\frac{r}{r_0}\Big)^{-\textstyle\epsilon}
              \frac{H_0}{H(r)}\,
              \exp\Big(-\frac{z^2}{2\,H(r)^{\,2}}\Big)
\end{equation}
between an inner and outer disk radius, $R_{\rm in}$ and $R_{\rm
out}$, respectively, with sharp edges. $\rho(r,z)$ is the local gas
mass density. $H(r)$ is the vertical scale height of the disk,
assuming to vary with radius as
\begin{equation}
  H(r) = H_0\,\Big(\frac{r}{r_0}\Big)^{\,\beta} \ .
\end{equation}
$H_0$ is the reference scale height at reference radius $r_0$.
$\epsilon$ is the column density power-law index and $\beta$ the
flaring power. The constant $\rho_0$ is adjusted such that the
integrated disk mass $2\pi \iint \rho(x,z)\,dz\;r\,dr$ equals $M_{\rm
disk}$.

\subsection{Dust size distribution and dust settling}
\label{sec:settling}

The dust grains are assumed to have a power-law size distribution in
the unsettled case as
\begin{equation}
  f(a) \propto a^{\,-p}
  \label{eq:dustsizedist}
\end{equation}
between minimum and maximum grain radius, $\amin$ and $\amax$,
respectively.  The free constant in Eq.\,(\ref{eq:dustsizedist}) is
adjusted to result in the prescribed unsettled dust/gas mass ratio
$\rho_{\rm d}/\rho$.

A very simple recipe has been implemented to account for the major
effects of vertical dust settling. We assume that the dust grains
are distributed vertically with a smaller scale height 
\begin{equation}
  H'(a,r) = H(r)\cdot\max\big\{1,a/a_{\rm s}\big\}^{-s/2}
  \label{eq:settleH}
\end{equation} 
where $H(r)$ is the gas scale height, and $s$ and $a_{\rm s}$ are two
free parameters. Since ProDiMo can self-consistently calculate the
vertical stratification of gas from the resulting gas temperatures and
mean molecular weights, in which case $H(r)$ does not exist,
Eq.~(\ref{eq:settleH}) can not be used directly in the general
case. Instead, we make use of the equation that defines $H(r)$, namely
$c_T\!=\!H(r)\,\Omega(r)$ and write
\begin{equation}
  c_T'(a) = c_T \cdot\max\big\{1,a/a_{\rm s}\big\}^{-s/2} \ ,
  \label{eq:settle_cT}
\end{equation}
where $c_T$ is the local gas isothermal sound speed, $c_T'(a)$ is the 
reduced variant for dust size $a$, $\Omega\!=\!v_{\rm
Kepler}/r$ is the angular velocity and $v_{\rm Kepler}$ the
Keplerian velocity. Equation~(\ref{eq:settle_cT}) is then used to 
calculate the vertical distribution of dust particles of different sizes
with respect to the already determined gas stratification by treating
them like an independent fluids with smaller $c_T$ as compared to the gas
\begin{equation}
  \frac{1}{\rho'(a)} \frac{dp'(a)}{dz} = -g_z = \frac{1}{\rho} \frac{dp}{dz}
  \label{eq:settle_rho}
\end{equation}
where $p\!=\!c_T^2\,\rho$ belongs to the gas, and
$p'(a)\!=\!c^{'2}_T(a)\,\rho'(a)$ belongs to dust particles in
$[a,a+da]$. $g_z$ is the local z-component of gravity. The solution of
the differential Eq.~(\ref{eq:settle_rho}) is
\begin{equation}
  f'(a,z) = \frac{{\rm const}(a)}{c^{'2}_T(a,z)} \exp\left(
            \frac{c_T^2(z)}{c^{'2}_T(a,z)}
            \ln\big(c_T^2(z) f(a,z)\big)\right) \ ,
\end{equation}
where $f'(a,z)\,da=\rho'(a)/m_{\rm d}(a)$ is the settled dust size
distribution function, and $c_T(z)$ and $f(a,z)\!\propto\!\rho(z)$ are
considered as known. The constant ${\rm const}(a)$ is then determined
to make sure that the vertical column density of dust particles for
every size is conserved
\begin{equation}
  \int f'(a,z)\,dz = \int f(a,z)\,dz \ .
\end{equation}
According to these assumptions, all dust quantities that used to be
constant, such as the local dust/gas mass ratio $\rho_d/\rho$, the dust
moments $\langle a\rangle$, $\langle a^2\rangle$, $\langle
a^3\rangle$, and the dust opacities per mass, become spatially
dependent quantities. The dust absorption and scattering opacities are
calculated by applying effective mixing \citep{Bruggeman1935} and Mie
theory, based on $f'(a,z)$.

%=====================================================================
\subsection{PAH ionisation equilibrium and PAH-heating:}
\label{sec:PAHion}

We consider a typical size of PAH molecules with $N_{\rm C}\!=\!54$
carbon atoms and $N_{\rm H}\!=\!18$ hydrogen atoms (circumcoronene),
motivated by studies that PAHs with much less carbon atoms would not
be stable around young stars on timescales of a few Myr. Much larger
PAHs are not consistent with the spatial extent of observed PAH
emission in various bands \citep[see e.g.][]{Visser2007}. We include
PAH$^-$, PAH, PAH$^+$, PAH$^{2+}$ and PAH$^{3+}$ as additional
specimen in the chemical reaction network. Circumcoronene is probably
among the smallest PAHs that can survive in a disk around Herbig~Ae
stars \citep{Visser2007}. The following processes are considered in
detail: PAH-photoionisation \citep{Tielens2005}, electron
recombination and some charge exchange reactions
\citep{Wolfire2008,Flower2003}. These reactions do not change the
basic PAH lattice, but only affect the charging of the PAH
molecules. Hence, the total amount of PAHs is conserved and treated
like an element with given element abundance
\begin{equation}
  \epsilon\,({\rm PAH}) = f_{\rm PAH}\;X^{\rm ISM}_{\rm PAH}\;
                        \frac{50}{N_{\rm C}} \ ,
\end{equation}
where $X^{\rm ISM}_{\rm PAH}=3\times 10^{-7}$ being the standard ISM
particle abundance with respect to hydrogen nuclei
\citep{Tielens2008}, and $f_{\rm PAH}$ is the fraction thereof assumed
to be present in the disk.

Concerning the photo-ionisation reactions, we consider the PAH
absorption cross sections $\sigma_{\rm PAH}^k(\nu)\rm\,[cm^2]$ from
\citet[][see their Eqs.\,6--12 and A2--A3]{Li2001} with recent updates
for the resonance parameters from \citet{Draine2007}. We use their
neutral PAH cross section for charge $k\!=\!0$, and the charged cross
section otherwise. The photo-ionisation rates $\rm[s^{-1}]$ for PAH
molecules with charge $k$ are calculated as
\begin{equation}
  R^k_{\rm ph}(r,z) = \frac{4\pi}{hc}\int_{912\,\AA}^{\lambda\rm
          thr} \!\!\sigma_{\rm PAH}^k(\nu) \;\nu J_\nu(r,z)\,
          Y^k_\nu\, s_\nu(r,z) \;d\lambda
\end{equation}
where $\nu\,$[Hz] is the frequency,
$J_\nu(r,z)\rm\,[erg/cm^2/s/Hz/sr]$ the mean intensity as computed by
the dust continuum radiative transfer, and $s_\nu(r,z)$ the following
self-shielding factor
\begin{eqnarray}
  \tau_\nu^{\rm PAH}(r,z) &\approx& 
     \epsilon({\rm PAH})\;\langle\sigma\rangle_{\rm PAH}(\nu)\; 
     {\rm Min}\left\{N^{\rm ver}_{\HH}(r,z), 
                     N^{\rm rad}_{\HH}(r,z)\right\}\\[-1mm]
  s_\nu(r,z) &=& \exp\left({-\tau_\nu^{\rm PAH}(r,z)}\right) \ .
\end{eqnarray}
$N^{\rm ver}_{\HH}$ and $N^{\rm rad}_{\HH}(r,z)$ are the radial inward
and vertical upward hydrogen nuclei column densities [cm$^{-2}$] in
the disk, as measured from point $(r,z)$. For simplicity, we put
$\langle\sigma\rangle_{\rm PAH}(\nu)=0.5\left(
\sigma_{\rm PAH}^{\,0}(\nu)+\sigma_{\rm PAH}^{+1}(\nu)\right)$.
The photo-electron yield
is taken from \citep{Jochims1996}
\begin{equation}
  Y^k_\nu = \left\{\begin{array}{cl}
    1                       & ,\;h\nu>IP^{\,k}+9.2\,{\rm eV}\\
    \displaystyle
    \frac{h\nu-IP^{\,k}}{9.2\rm\,eV} 
                            & ,\;IP^{\,k}+9.2\,{\rm eV}>h\nu>IP^{\,k}\\*[2mm]
    0                       & ,\;h\nu<IP^{\,k} 
  \end{array}\right. 
\end{equation}
According to \citet{Weingartner2001}, the ionisation potentials
$IP^{\,k}$ depend on PAH-size and charge $k$. Using their Eqs.\,(1)
and (2) for $N_{\rm C}\!=\!54$, the results are
$IP^{\,-1}\!=\!3.10\,$eV, $IP^{\;0}\!=\!6.24\,$eV,
$IP^{\,+1}\!=\!9.38\,$eV and $IP^{\,+2}\!=\!12.5\,$eV. The formula is
an approximation that reproduces the ionisation potential of benzene
as the smallest PAH and graphite as infinitely large PAH. The threshold
wavelengths are given by $\lambda_{\rm thr}^k=hc/IP^k$.

With these photo-ionisation and recombination rates, the local
particle densities of PAH$^-$, PAH, PAH$^+$, PAH$^{2+}$ and PAH$^{3+}$
are calculated consistently with the gas-phase and ice chemistry, with
a considerable influence on the resultant local electron density.
Once these particle densities $n_{\rm PAH}^k$ have been determined,
the total PAH heating rate $\rm[erg/cm^3/s]$ can be calculated as
\begin{equation}
  \Gamma_{\rm PAH} = \frac{4\pi}{hc} \sum_k n_{\rm PAH}^k
 \hspace*{-1mm}\int\limits_{912\,\AA}^{\lambda_{\rm thr}^k}\hspace*{-1mm} 
           \sigma_{\rm PAH}^k(\nu) \;\nu J_\nu\, Y^k_\nu\, s_\nu\, 
           \big(h\nu-IP^{\,k}\big)\;d\lambda
\end{equation}
The PAH recombination cooling rate is calculated as
\begin{equation}
  \Lambda_{\rm PAH} = \sum_k n_{\rm PAH}^k\,n_e\, 
            k_{\rm PAH}^k(\Tg)\,\big(1.5\,k\Tg\big)
\end{equation}
where $k_{\rm PAH}^k(\Tg)\rm\,[cm^3\,s^{-1}]$ is the PAH recombination
rate coefficient, $n_e$ is the electron density and $\Tg$ the gas
temperature.  As a general result, we obtain a top-down layered PAH
charge structure, with PAH$^{2+}$ in the uppermost layers with reduced
PAH-heating, but PAH$^-$ and intensified PAH-heating close to the
midplane. For massive disks (unlike ET\,Cha), where the electron
density virtually vanishes in the deepest layers, the PAH charge becomes
neutral again in the midplane.

\begin{table}
\caption{Larger non-LTE model atoms. References are as follows:
  $\Lambda\!=\,$\citet{Lambda2005}, 
  NIST$\,=\,$\citet{NIST2009}
  K$\,=\,$\citet{Krems2006},
  S$\,=\,$\citet{Storzer2000}}
\label{tab:nonLTE}
\begin{tabular}{c|cc|c|c}
\\[-4.5ex]
\hline
species & \hspace*{-1mm}\#levels\hspace*{-1mm} 
        & \hspace*{-3mm}\#lines\hspace*{-2mm} 
        & coll.\,partners & references\\
\hline
\hline
O\,I               &   91 & 647 & p-H$_2$,o-H$_2$,H,H$^+$,e$^-$    
                   & $\Lambda$, NIST, K, S\\
C\,I               &   59 & 117 & 
      \hspace*{-1mm}p-H$_2$,o-H$_2$,H,H$^+$,He,e$^-$\hspace*{-2mm} 
                   & $\Lambda$, NIST\\
C\,II              &   10 & 31  & p-H$_2$,o-H$_2$,H,e$^-$        
                   & $\Lambda$, NIST \\
\end{tabular}
\end{table}

%=====================================================================
\subsection{Fluorescent UV-pumping}
\label{sec:fluor}

We have replaced the small model atoms for O, C and C$^+$ \citep[as
listed in Table~4 of][]{Woitke2009a} by larger ones, see
Table~\ref{tab:nonLTE} (this paper), to account for fluorescent UV and
optical pumping of the lower levels responsible for the far IR
fine-structure lines. The new model atoms have collisional data only
among the lowest states, but much more radiative data.  The additional
energy levels and transition probabilities are taken from the {\em
National Institute of Standards and Technology} atomic spectroscopic
database \citep{NIST2009}. We selected all evaluated transitions
longward of 912\,\AA. Rovibronic transitions are permitted and their
probabilities are of the order of $(1\!-\!10^6)\rm\,s^{-1}$, much
higher than typical collision rates ($\sim$10$^{-9}$ s$^{-1}$).

Collisional data connecting the two first electronic-excited states
$^1D_2$ at 22830~K and $^1S_0$ at 48370~K of neutral oxygen and the
three spin-orbit split ground states $^3P$ exist with electrons and
hydrogen atoms as collision partners \citep{Storzer2000,Krems2006}.
For C and C$^+$, we use the rates collected in the {\em Leiden Lambda}
database \citep{Lambda2005}, whose collision rates are mostly limited
to the ground state electronic levels.

%=====================================================================
\subsection{[OI]\,6300\,\AA\,-pumping by OH photo-dissociation}
\label{sec:OHpump}

Photo-dissociation of OH by absorption of UV photons into the
$1\,^2\Sigma^-$ state, $\lambda\!\simeq\!(973\!-\!1350)\,\AA$,
produces an electronically excited oxygen atom O($^1D$) as
\begin{equation}
  \rm OH + h\nu \longrightarrow O(^1{\it D}) + H,
\end{equation}
while absorption into the $1\,^1\Delta$ and $3\,^2\Pi$ states,
$\lambda\!\simeq\!(1100\!-\!1907)\,\AA$, leads to the ground state
$^3P$. According to \citet{Dishoeck1984}, about 55\% of the OH
photodissociations by an interstellar UV field\footnote{We note that
the branching ratio could be different for our adopted star+UV
spectrum. A maximum branching ratio of 100\% would would increase
our [OI]\,6300\,\AA\ line predictions by less than a factor of 1.8.}
result in an O-atom in the $^1D_2$ state, which happens to be the
upper level of the 6300\,\AA\ line $^1D_2\to\,^3P_2$. This chemical
pumping was suggested by \citet{Acke2005} to play a major role in the
formation of the 6300\,\AA\ line around Herbig Ae/Be stars. We take
this effect into account by introducing a quasi-collisional pumping
rate $\rm[s^{-1}]$
\begin{equation}
  C_{l\,u}^{\rm chem} = 0.55\times n_{\rm OH}\,
                        R_{\rm ph}^{\rm\,OH}\;\big{/}\;n_{\rm O}
  \label{eq:OHpump}
\end{equation} 
where $R_{\rm ph}^{\rm\,OH}$ is the OH photodissociation rate
$\rm[1/s]$.  We apply Eq.\,(\ref{eq:OHpump}) to $l=\{1,2,3\}$,
where $1\!=\!^3\!P_2$, $2\!=\!^3\!P_1$, $3\!=\!^3\!P_0$, and
$u\!=\!4\!=\!^1D_2$. We add these excitation rates to the other
collisional excitation rates $C_{l\,u}$ \citep[see Sect.\,6.1
in][]{Woitke2009a}. In writing Eq.\,(\ref{eq:OHpump}), we rely on
statistical equilibrium and assume that the majority of chemical
reaction channels forming OH start from the 3 lowest $^3P$ levels with
$n_{\rm O}\!\approx\!n_{\rm O}(1)\!+\!n_{\rm O}(2)\!+\!n_{\rm O}(3)$,
making sure that $n_{\rm O}\left(\frac{n_{\rm O}(1)}{n_{\rm
O}}C_{1\,4}^{\rm chem} \!+\! \frac{n_{\rm O}(2)}{n_{\rm
O}}C_{2\,4}^{\rm chem} \!+\! \frac{n_{\rm O}(3)}{n_{\rm
O}}C_{3\,4}^{\rm chem}\right) = 0.55\,\,n_{\rm OH}\, R_{\rm
ph}^{\rm\,OH}$, where $\frac{n_{\rm O}(l)}{n_{\rm O}}$ are the
fractional populations.

%In application to ET\,Cha we observe that the inclusion of the OH
%photo-dissociation pumping increases the [OI]\,6300\,\AA\ line flux by
%about a factor of $2\!-\!3$.

%=====================================================================
\subsection{H$_2$-pumping by its formation on dust surfaces}
\label{sec:H2pump}

The formation of H$_2$ on grain surfaces liberates the binding energy, 
causing the newly formed H$_2$ molecules to leave the surface in 
a vibrationally highly excited state. \citet{Duley1986}
estimated that the H$_2$ molecules will be released in vibrational
states as high as $\rm v\!\la\!7$, but in the rotational ground state
$J\!=\!0$ for p-H$_2$ and $J\!=\!1$ for o-H$_2$.
\begin{equation}
  \rm H + H + dust \longrightarrow H_2(v\!\le\!7, J\!=\!0/1) + dust
\end{equation}
We treat this formation-pumping by 
\begin{equation}
  C_{l\,u}^{\rm chem} = \alpha*n_{\rm H}\,R_{\rm dust}^{\rm\,H_2}
                        \;\big{/}\;n_{\rm tot}
  \label{eq:H2pump}
\end{equation} 
where $R_{\rm dust}^{\rm\,H_2}$ is the H$_2$ formation rate on dust
surfaces $\rm[1/s]$, $u$ is the index of the highest state with
$\rm v\!\le\!7$ and $J\!\le\!1$ that still has collisional
de-excitation rates in the database, and Eq.~(\ref{eq:H2pump}) is
applied to all states $l\!<\!u$, \ie to all states that have lower
excitation energy than $u$. $n_{\rm tot}$ is the total ortho or para
H$_2$ particle density, respectively, and $\alpha\!=\!n_{\rm
o/p-H_2}/n_{\rm H_2}$ is the prescribed ortho/para-H$_2$ fraction.

Our formulation of the H$_2$-pumping, therefore, leads to a constant
de-population of all states $l<u$ with a certain timescale
$C_{l\,u}^{\rm chem}\rm\,[s^{-1}]$ (simulating the disappearance of
H$_2$ in all states due to other chemical channels that are feeding the
H$_2$ formation) and an equally large population $\rm[cm^{-3}s^{-1}]$
of the state $u$ due to the formation of H$_2$ on grains.

The condition of state $u$ being collisionally connected avoids
artifacts at very low densities, where collisions are rare and the
pumping would lead to an almost complete de-population of all
low-lying states. According to our current collection of collisional
H$_2$ de-excitation rates, this condition implies $u\!=\!({\rm
v}\!=\!3,J\!=\!0/1)$.  We checked that our modelling of the
o-H$_2$v=1$\to$0\,S(1) line at 2.122\,$\mu$m does not depend much on
the choice of this upper level $u$ as long as $\rm v\!>\!1$.

%=====================================================================
\subsection{Line transfer}
\label{sec:linetrans}

During solving the chemistry and energy balance of the gas, ProDiMo
calculates the various level populations of atoms, ions and molecules
by means of an escape probability method \citep[see Sect.\,6.1.1
in][]{Woitke2009a}. These are stored and used later to perform a
formal solution of line transfer for selected spectral lines. 

The continuum $+$ line radiative transfer equation is given by
\begin{equation}
  \frac{dI_\nu}{ds} = \kappa_\nu^{\rm ext} \big(S_\nu-I_\nu\big) \ ,
  \label{eq:line_transfer}
\end{equation}
where $I_\nu$ is the spectral intensity and $s$ the distance on a
ray. The source function and extinction coefficient are given by
\begin{eqnarray}
  S_\nu &=& \frac{\epsilon_\nu^{\rm D} + \phi_\nu \epsilon_\nu^{\rm L}}
                 {\kappa_\nu^{\rm ext}}\\
  \kappa_\nu^{\rm ext} &=& \kappa_\nu^{\rm D} + \phi_\nu \kappa_\nu^{\rm L}\ .
\end{eqnarray}
Assuming isotropic dust scattering, the continuum and line transfer
coefficients are given by
\begin{eqnarray}
  \epsilon_\nu^{\rm D} &=& \kappa^{\rm abs}_\nu B_\nu(T_{\rm dust})
                            + \kappa^{\rm sca}_\nu J_\nu\\
  \kappa_\nu^{\rm D}   &=& \kappa^{\rm abs}_\nu + \kappa^{\rm sca}_\nu\\
  \epsilon_\nu^{\rm L} &=& \frac{h\nu}{4\pi} n_u A_{ul}\\
  \kappa_\nu^{\rm L}   &=& \frac{h\nu}{4\pi} \big(n_l B_{lu}-n_u B_{ul}\big)\ ,
\end{eqnarray}
where $\kappa^{\rm abs}_\nu$ and $\kappa^{\rm sca}_\nu\rm\,[cm^{-1}]$
are the local dust absorption and scattering coefficients, $B_\nu$ is
the Planck function, $J_\nu$ is the local mean intensity (taken from
the results of the continuum radiative transfer), $\nu$ is the line
centre frequency, $n_u$ and $n_l$ are the level populations
$\rm[cm^{-3}]$ in the upper and lower state, respectively,
and $A_{ul}$ and $B_{ul},B_{lu}$ are the Einstein coefficients.

The profile function $\phi_\nu\rm\,[Hz^{-1}]$ is assumed to be
given by a Gaussian with thermal $+$ turbulent broadening
\begin{eqnarray}
  x &=& {\rm v'} + \vec{n}\cdot{\vec{\rm v}}\\
  \phi_\nu &=& \frac{c}{\nu\sqrt{\pi}\Delta{\rm v}}
               \exp\left(-\frac{x^2}{\Delta{\rm v}^2}\right) \ ,
\end{eqnarray}
where ${\rm v'}$ is the observers velocity with respect to the star
along backward ray direction $\vec{n}$, $\vec{\rm v}$ is the
3D-velocity of the emitting gas with respect to the star (assumed to
be given by Keplerian orbits), and $x$ the local velocity shift. The
velocity width is $\rm\Delta{\rm v}^2 = v^2_{th}+v^2_{turb}$ and
the thermal velocity ${\rm v}_{\rm th}^2\!=\!2k\Tg/m$ where $m$ is the mass
of the line emitting species.

We use the same setup of parallel rays as described in \citet[][see
their Sect.~2.3]{Thi2010}, organised in about 150 log-equidistant
concentric rings in the image plane, each subdivided into 72 angular segments.
Equation~(\ref{eq:line_transfer}) is solved numerically on each of
these rays in the observer's frame on 151 velocity grid
points to sample v', where the transport coefficients are
pre-calculated on the grid points and later interpolated along the
rays, whereas the profile function is always calculated from
scratch. The numerical scheme features a variable step size which is
controlled by comparing the results after two consecutive steps with
the results obtained after one step of double size.

%=====================================================================
\subsection{Chemical heating}
\label{sec:chemheat}

By definition, exothermic chemical reactions convert chemical
potential energies into heat, whereas endothermic reactions consume
internal kinetic energy and actually cool the gas. We calculate
this chemical heating/cooling rate $\rm[erg/cm^3\!/s]$ as 
\begin{equation}
  \Gamma_{\rm chem} = \sum\limits_r R(r)\,\gamma^{\rm chem}_r\,\Delta H_r
  \label{eq:chemheat}
\end{equation}
where $r$ is an reaction index, $R(r)$ is the reaction rate
$\rm[1/cm^3/s]$ and $\Delta H_r$ $\rm[erg]$ is the reaction enthalpy,
which is positive for exothermic reactions and negative for
endothermic reactions
\begin{equation}
  \Delta H_r = \sum\limits_{\rm pr} \Delta H_f^0({\rm pr})
             - \sum\limits_{\rm ed} \Delta H_f^0({\rm ed}) \ .
\end{equation}
$\Delta H_f^0$ [erg] is the heat of formation of the chemical
species involved in the reactions (pr means products, ed means
educts). By simplification, we neglect the temperature-dependence of 
$\Delta H_f$, and take the values for the heat of formation
for all species from \citep{Millar1997}, who list $\Delta H_f^0$
[kJ/mol] at 0\,K in their Table~2.

The details of exothermic reactions are quite difficult and often not
precisely known. The excess chemical binding energy is seldomly
released in form of kinetic energy directly (although there are some
exceptions like, for example, dielectronic recombination which is
radiationless). But in the vast majority of cases, the reaction will
create products in some kind of electronic, vibrational and rotational 
excited states. 

%For example, charge exchange reactions like $\rm C^+\!\!+\!Fe \to
%Fe^+\!\!+\!C$ [$\Delta H_r\!=\!+3.39\,$eV] can be expected to produce
%ions and atoms in electronically excited states, whereas only a tiny
%part of $\Delta H_r$ will be released directly in form of kinetic
%energy.  If molecules are produced, they will probably be released in
%ro-vibrationally excited states.

The energy temporarily stored in these excitational states can then be
either radiated away (in which case the net heating is small),
or subsequent collisional processes can thermalise the excess
energy. The ratio between these two competing processes depends on the
critical density for collisional de-excitation, which depends largely
on type and amount of excitation, for example
$\sim\!10^{15}\rm\,cm^{-3}$ for permitted electronic transitions down
to $\sim\!10^{3}\rm\,cm^{-3}$ for rotational transitions. Further
complications arise in possibly large optical depth where created
photons cannot escape directly, scatter around and drive secondary
processes, with a higher probability to get (partly) thermalised. To
avoid all these complications, we have simply introduced an efficiency
$\gamma^{\rm chem}_r$ in Eq.~(\ref{eq:chemheat}). We also exclude
certain types of reactions from Eq.~(\ref{eq:chemheat}): reactions
which are known to produce or absorb photons, cosmic ray and cosmic
ray-induced reactions, X-ray primary and secondary reactions,
reactions on grain surfaces, and reactions which are energetically
treated in larger detail elsewhere. Some of the most important
reactions are found to be
\begin{eqnarray}
  \rm H^- + H   \to H_2 + e^-        && \quad\Delta H_r\!=\rm\!+3.72\,eV\\
  \rm H_2 + O \leftrightarrow OH + H && \quad\Delta H_r\!=\rm\!\pm0.079\,eV\\
  \rm H_2 + H   \to H + H + H        && \quad\Delta H_r\!=\rm\!-4.48\,eV\\
  \rm H_2 + H_2 \to H_2 + H + H      && \quad\Delta H_r\!=\rm\!-4.48\,eV\ ,
\end{eqnarray}
where we put $\gamma^{\rm chem}_r\!=\!1$ for the last two ``collider''
reactions, assuming that these highly endothermic reactions, which
only occur in hot gases, are in fact driven by the kinetic
energies of the colliding atoms and molecules.

\begin{figure}
\centering
\includegraphics[width=48mm]{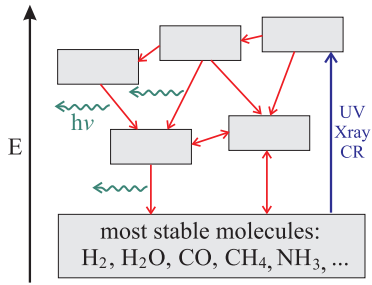}\\
\caption{Sketch of energetics in astrochemical reaction cycles.}
\label{fig:ChemHeat}
\end{figure}

In kinetic chemical equilibrium, as assumed in this paper, there is no
net formation/destruction of any chemical species, but the chemistry
is organised in complicated reaction cycles as sketched in
Fig.~\ref{fig:ChemHeat}. In the absence of cosmic rays, UV photons and
X-ray irradiation, only the energetically most favourable chemical
configurations, like $\rm H_2, H_2O, CO, CH_4, NH_3$ etc., are
abundant. However, due to the impact of these high-energy photons and
particles, much less stable molecules, atoms and ions are continuously
created which, via some complicated chemical paths, come cascading
down in energy again, eventually re-forming the abundant stable
molecules. Along these paths, some of the chemical potential energy
can be lost in form of secondary photons. The net effect of this cycle
is typically a heating, because we have -- by definition -- excluded
the primary UV, X-ray and CR reactions from
Eq.~(\ref{eq:chemheat}). The chemical heating is hence yet another way
to partly thermalise the energy of incoming high-energy photons and
particles through secondary exothermic reactions.

To our surprise, the chemical heating, even with low efficiency
$\gamma^{\rm chem}_r\!=\!0.1$, results to be an important heating
process in protoplanetary disks, in particular at the bottom of the
warm molecular layer where many of the near-mid IR spectral lines are
formed, and densities are of order $10^{9}-10^{11}\rm\,cm^{-3}$,
which, as far we are aware of, has not been noticed so far in other
disk modelling papers.

%=====================================================================
\section{The Outflow Model}
\label{app:outflow}

We have used the HVC [OI]\,6300\,\AA\ and [SII]\,6731\,\AA\ line
luminosities to estimate the outflow mass loss rate in two
independent ways as suggested by \citet[][see their Eqs.\,(A8) and
(A10)]{Hartigan1995}.  Assuming the same values for electron density
(only necessary for the OI line) and the projected velocity of the jet
as used by in Hartigan et al., we obtain an outflow mass loss rate of
$\dot{M}_{\rm outflow}\!=\!(0.9\!-\!2)\times10^{-9}\rm\,M_\odot/yr$,
similar to the mass accretion rate as estimated by \citep{Lawson2004}. 

\citet{Hollenbach1985} showed that it is possible to relate the mass
outflow rate to the [OI]\,63.2\,$\mu$m line luminosity by assuming
that each atom in the outflow passes through a single shock wave and
that the gas with $T_{\rm g}\!<\!5000$\,K cools by radiating only in
the [OI]\,63.2\,$\mu$m line. This procedure is likely to overestimate
the resulting line flux since other cooling processes can be active as
well, and since the jet material typically passes through several
shock waves close to the star. Nevertheless, we use the formula
derived by \citet{Hollenbach1985} to obtain a rough estimate of the
[OI]\,63.2\,$\mu$m line luminosity as would be expected for an
$\dot{M}_{\rm outflow}\!=\!(0.9\!-\!2)\times10^{-9}\rm\,M_\odot/yr$ outflow.
The result is a [OI]\,63\,$\mu$m line luminosity equal to $(0.6\!-\!1.5)$
times the observed value.  This estimate suggests that a substantial
part of the [OI]\,63.2\,$\mu$m line as seen by {\sc Herschel} might
originate in the outflow rather than in the disk.  We note, however,
that the peak of the HVC is shifted only by about $-42\,$km/s with
respect to the stellar velocity, and the formulae used above may not
be appropriate to such unusually low outflow/shock velocities.

Another way to estimate the [OI]\,63.2\,$\mu$m emission line flux from
an outflow is to use the shock models of \citep{Hartigan2004}, which
predict the line ratio [OI]\,6300\,\AA/[OI]\,63.2\,$\mu$m for a
variety of outflow shock parameters. The authors find 6300/63.2 line
ratios of about $(0.8\!-\!2)$. For ET\,Cha, the measured HVC
[OI]\,6300\,\AA\ line flux is $(37\!-\!87)\times 10^{-18}\rm\,W/m^2$,
hence the predicted [OI]\,63.2\,$\mu$m outflow line flux should be
$(30\!-\!175)\times 10^{-18}\rm\,W/m^2$, which is just consistent with
the line flux as observed with {\sc Herschel}, $(30.5\pm3.2)\times
10^{-18}\rm\,W/m^2$. If we include the LVC and work with the total
[OI]\,6300\,\AA\ emission line flux, the results become inconsistent,
\ie the outflow model alone would already over-predict the measured flux. 

%=====================================================================
\section{Details of the best-fitting disk model}
\label{sec:bestmodel}

\begin{figure}
  \centering
  \includegraphics[width=88mm]{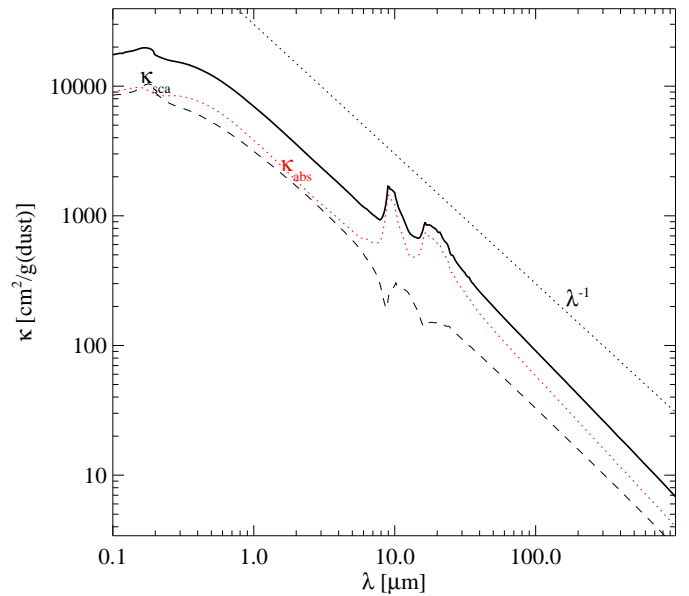} 
  \caption{Dust opacities assumed in the best-fitting disk model,
  calculated with effective mixing and Mie theory according to the
  parameters listed in Table~\ref{tab:Parameter}. Note that the
  absorption coefficient (red dotted) is larger than the
  scattering (black dashed) also at optical and near IR wavelengths, 
  which is untypical for pure silicates.}
\label{fig:dustopac}
\end{figure}

Figure~\ref{fig:dustopac} shows the dust opacities assumed in the
best-fitting model, which results to be absorption-dominated and to
scale roughly like $\/\lambda$, except for the 10\,$\mu$m and
20\,$\mu$m silicate features.

\begin{figure*}
\begin{tabular}{cc}
  \includegraphics[width=80mm]{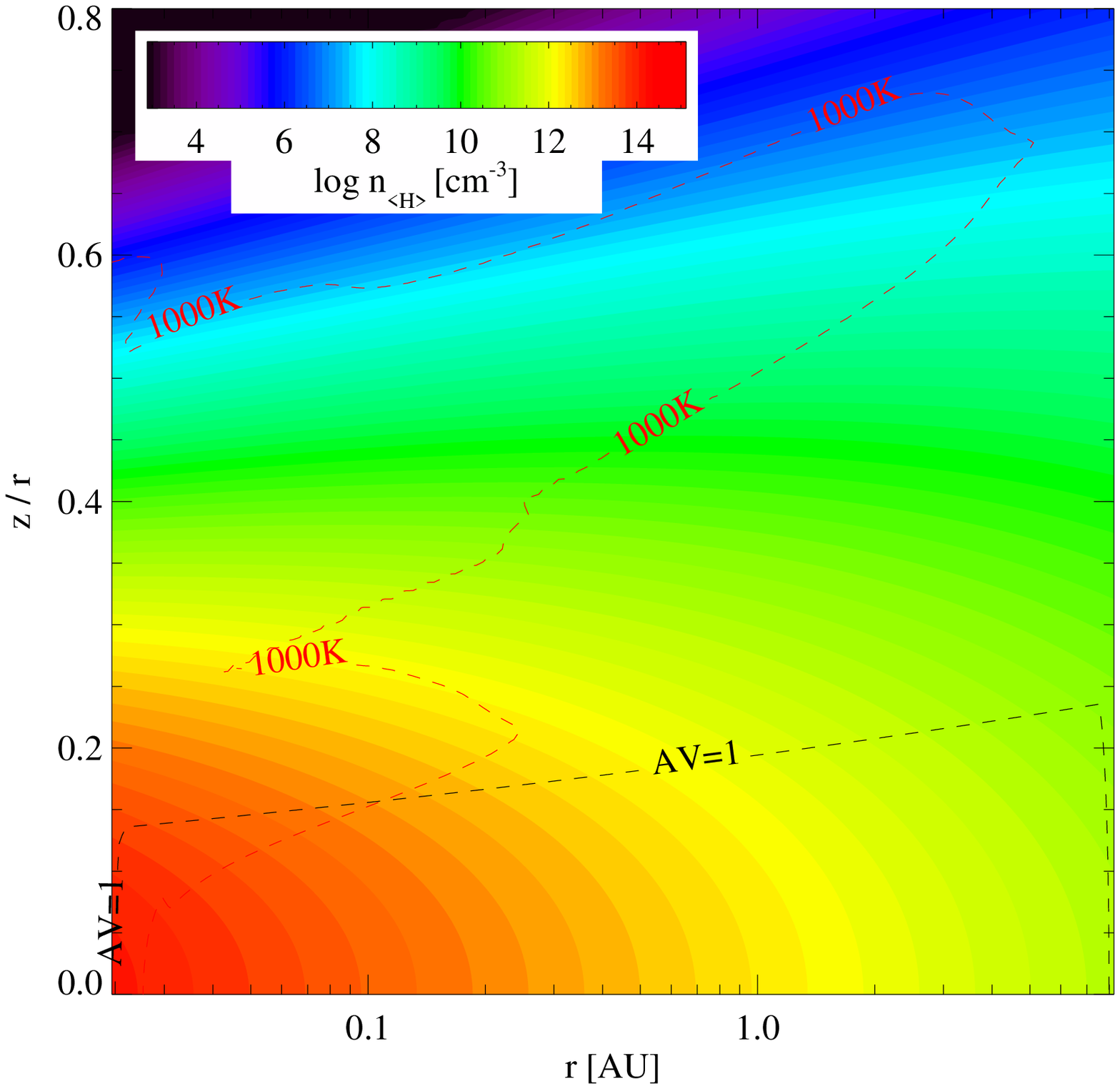} &
  \includegraphics[width=80mm]{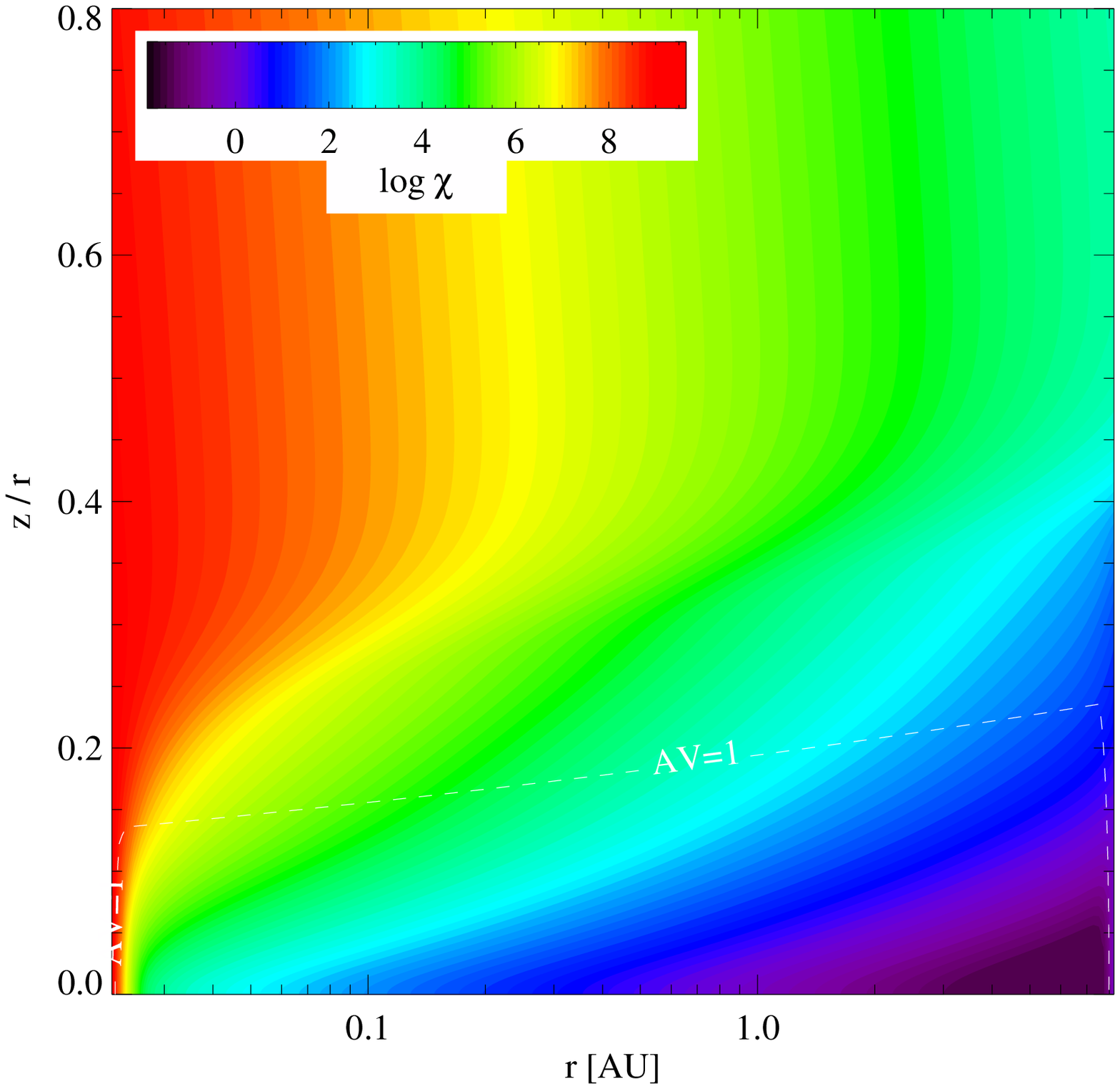} \\[-2mm]
  \includegraphics[width=80mm]{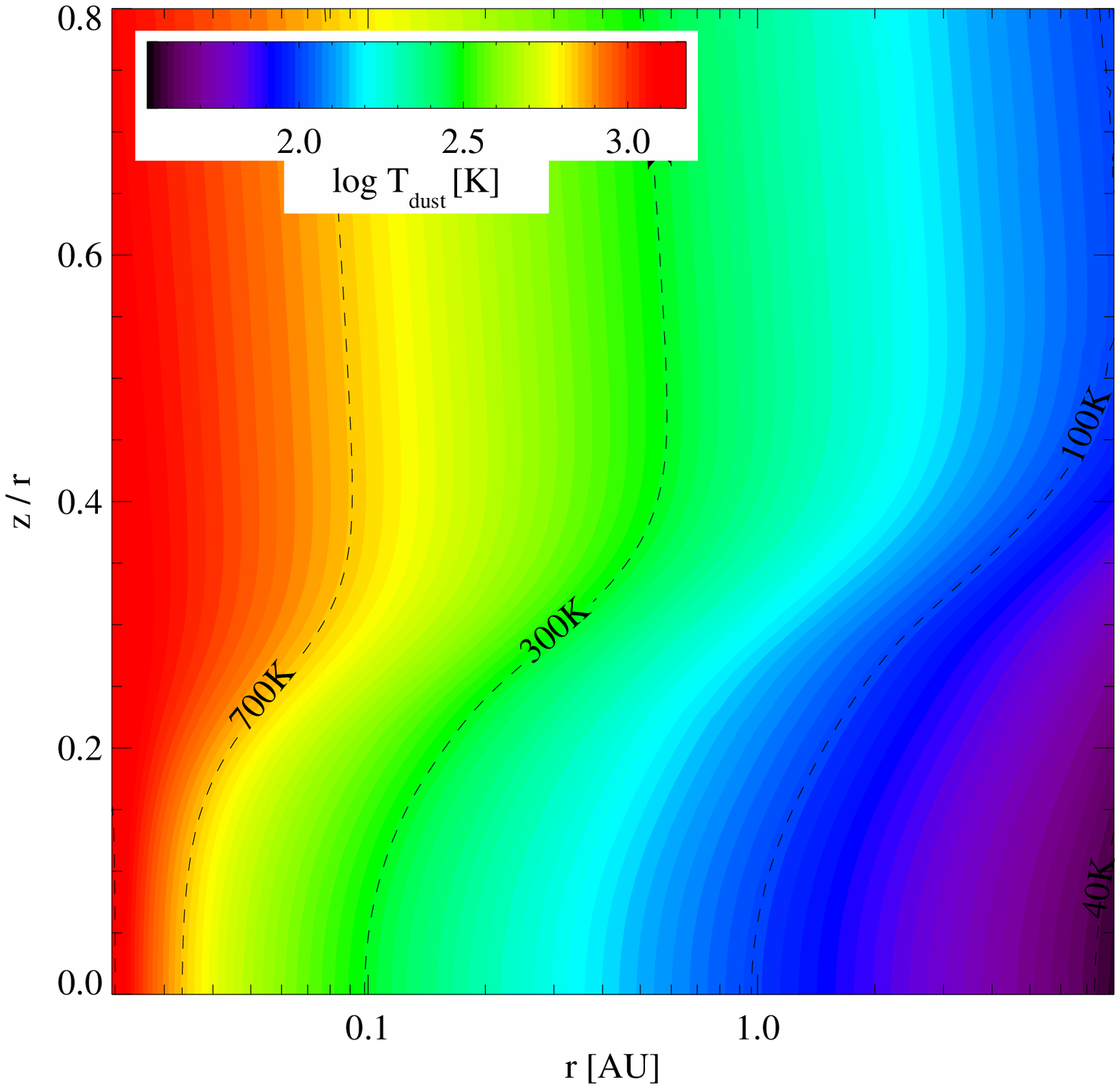} &
  \includegraphics[width=80mm]{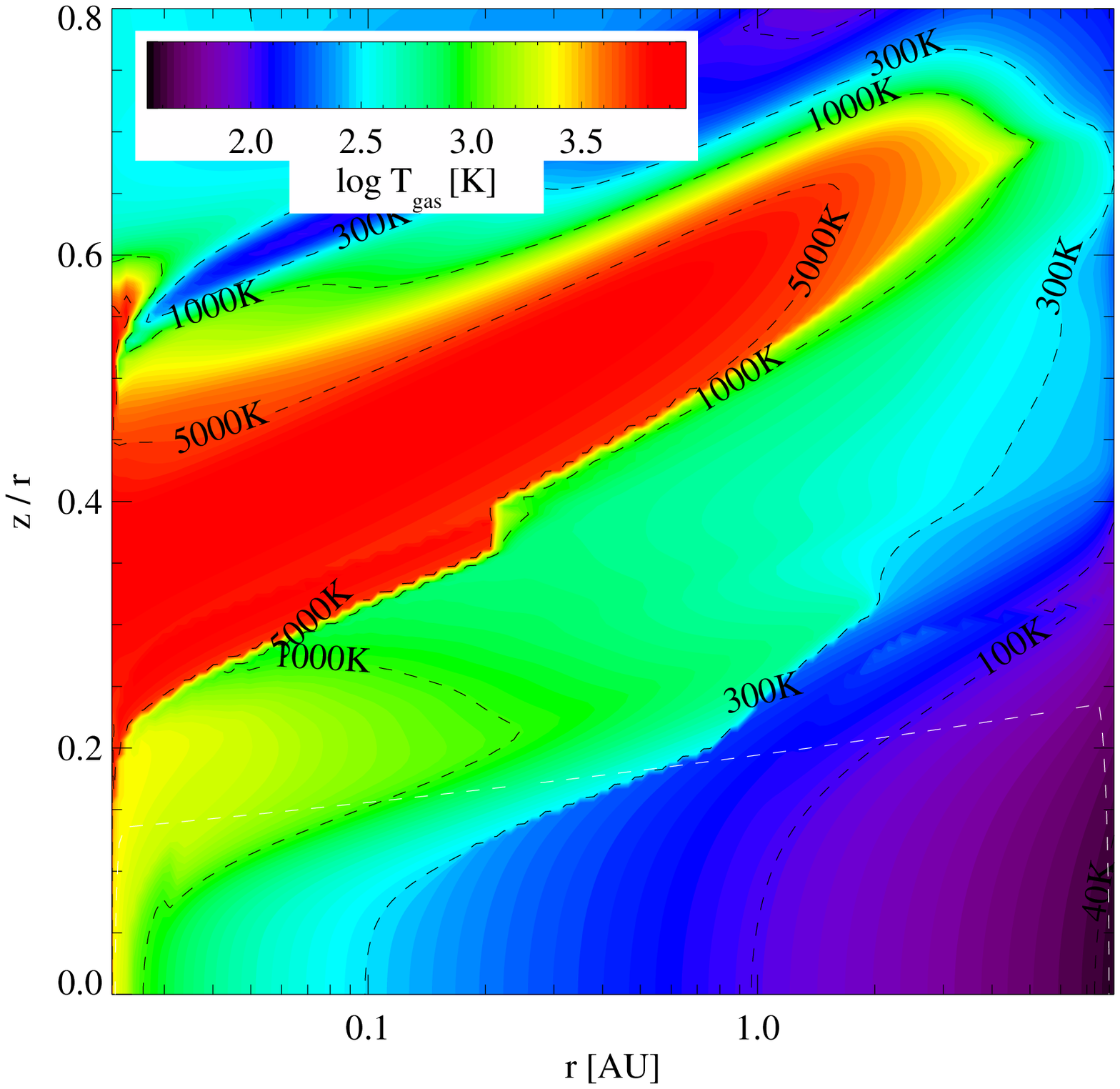} \\[-2mm]
  \includegraphics[width=90mm,height=100mm]{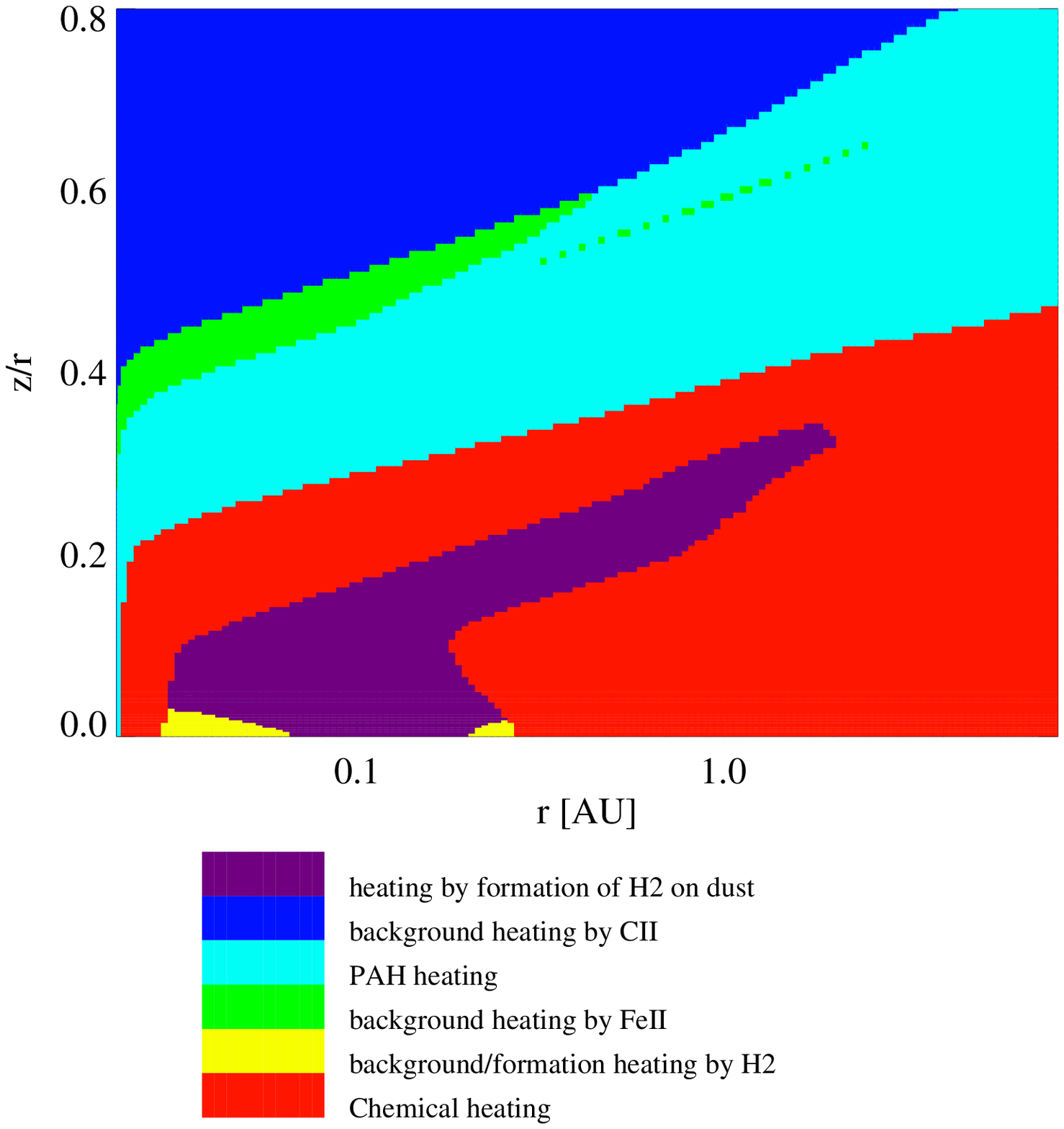} &
  \includegraphics[width=90mm,height=100mm]{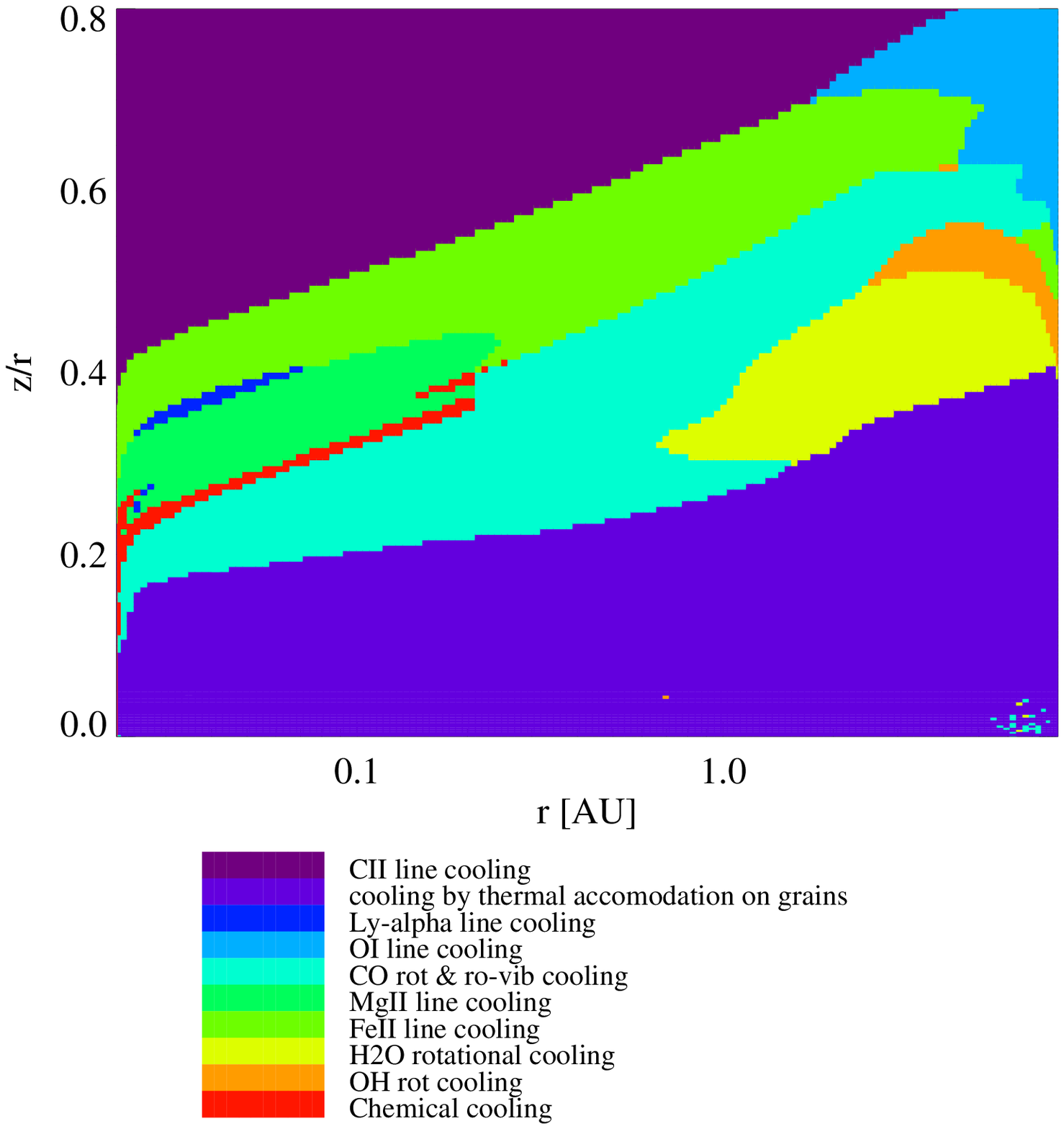} \\[-2mm]
\end{tabular}
\caption{Physical details of the best-fitting disk model. {\bf Upper row:}
  total hydrogen nuclei density $\nH\rm\,[cm^{-3}]$, with overplotted
  contours for visual extinction $A_V\!=\!1$, and gas
  temperature $T_{\rm gas}\!=\!1000\,$K, and strength of UV radiation
  field with respect to interstellar standard $\chi$. {\bf Second
  row:} dust and gas temperature structures, $T_{\rm dust}$ and
  $T_{\rm gas}$ (note the different scaling). {\bf Lower row:} Most
  important heating and cooling processes.}
\label{fig:diskmodel1}
\end{figure*}

\begin{figure*}
\begin{tabular}{cc}
  \includegraphics[width=80mm]{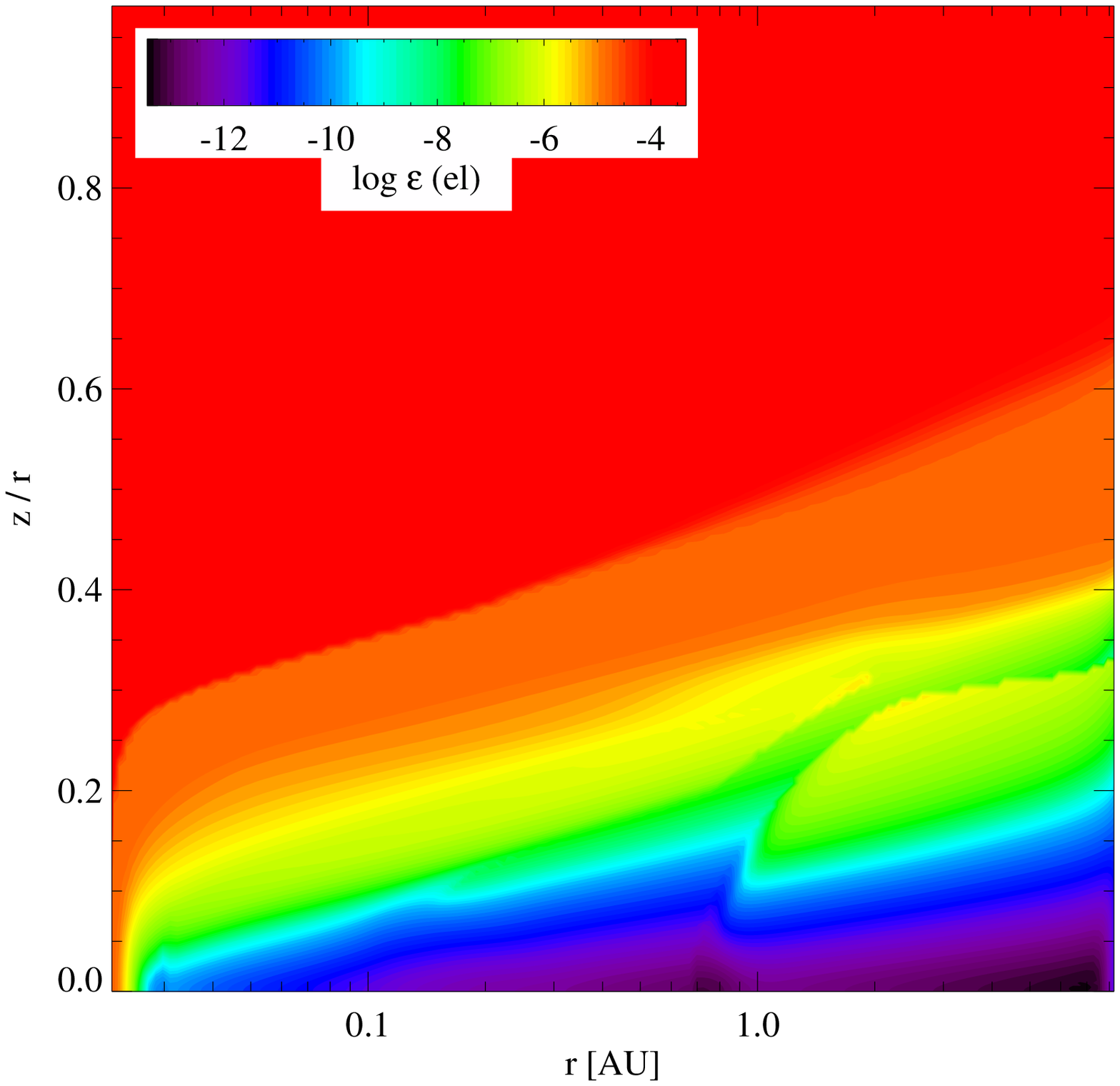} &
  \includegraphics[width=80mm]{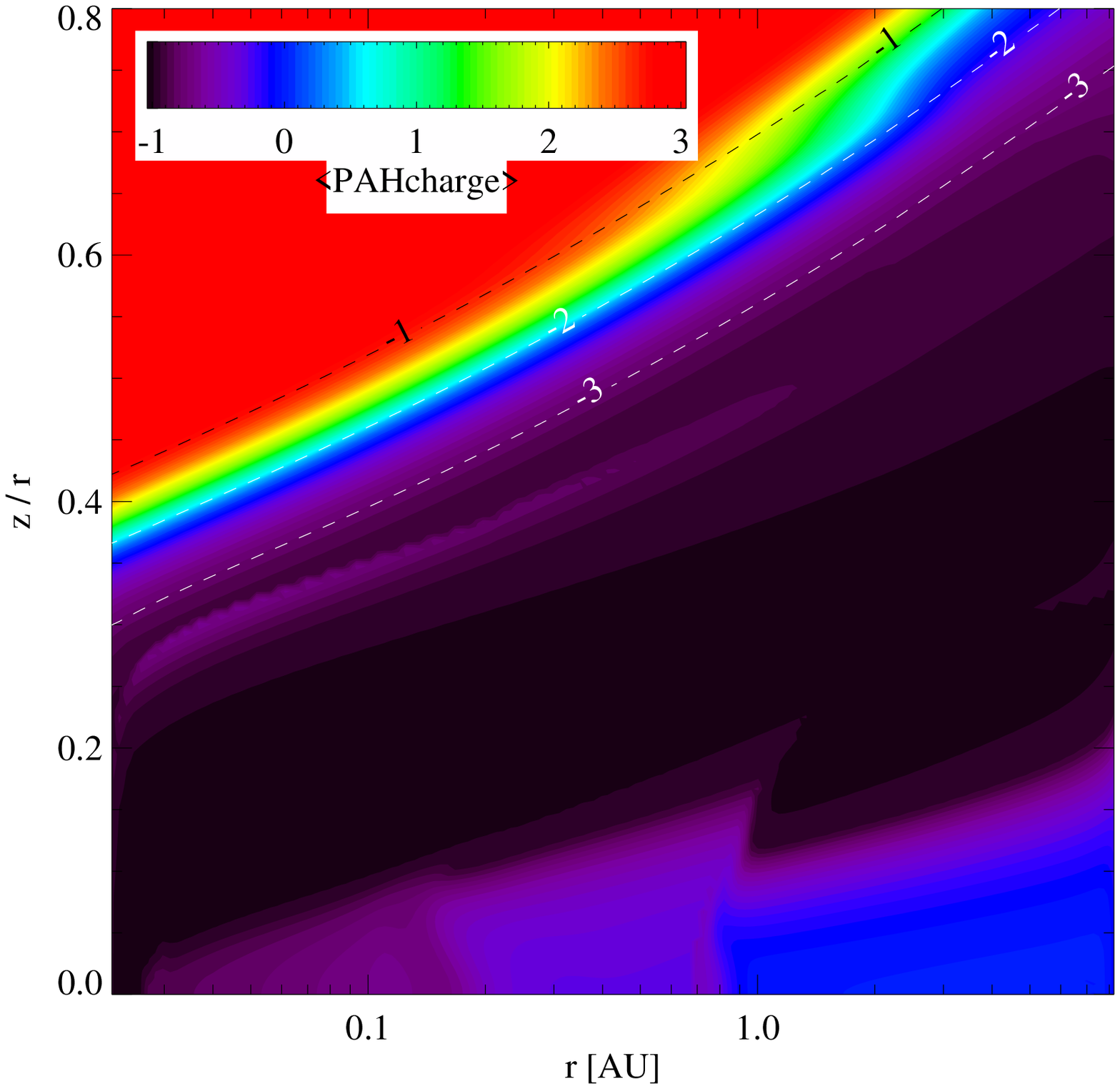} \\[-2mm]
  \includegraphics[width=80mm]{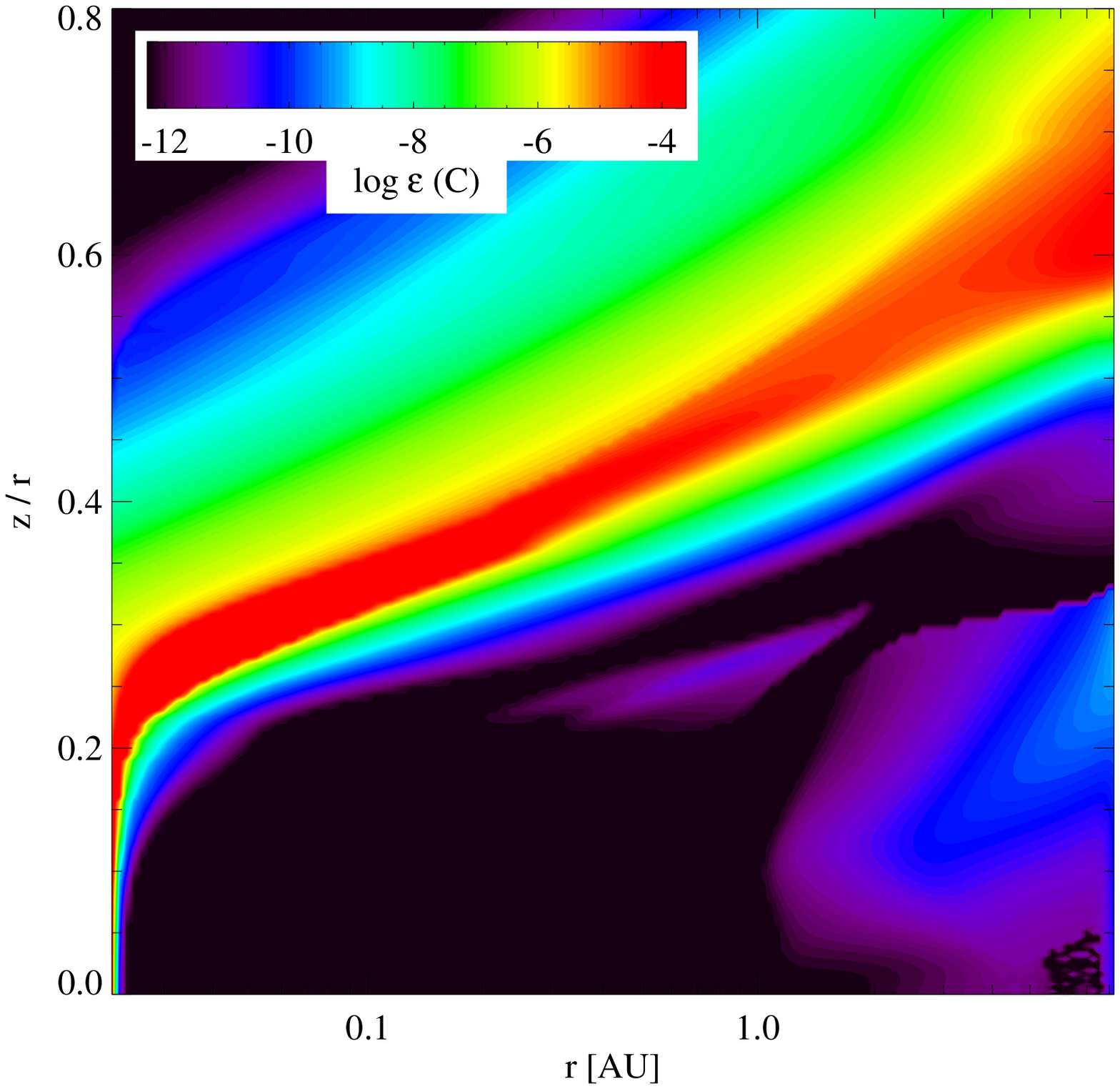} &
  \includegraphics[width=80mm]{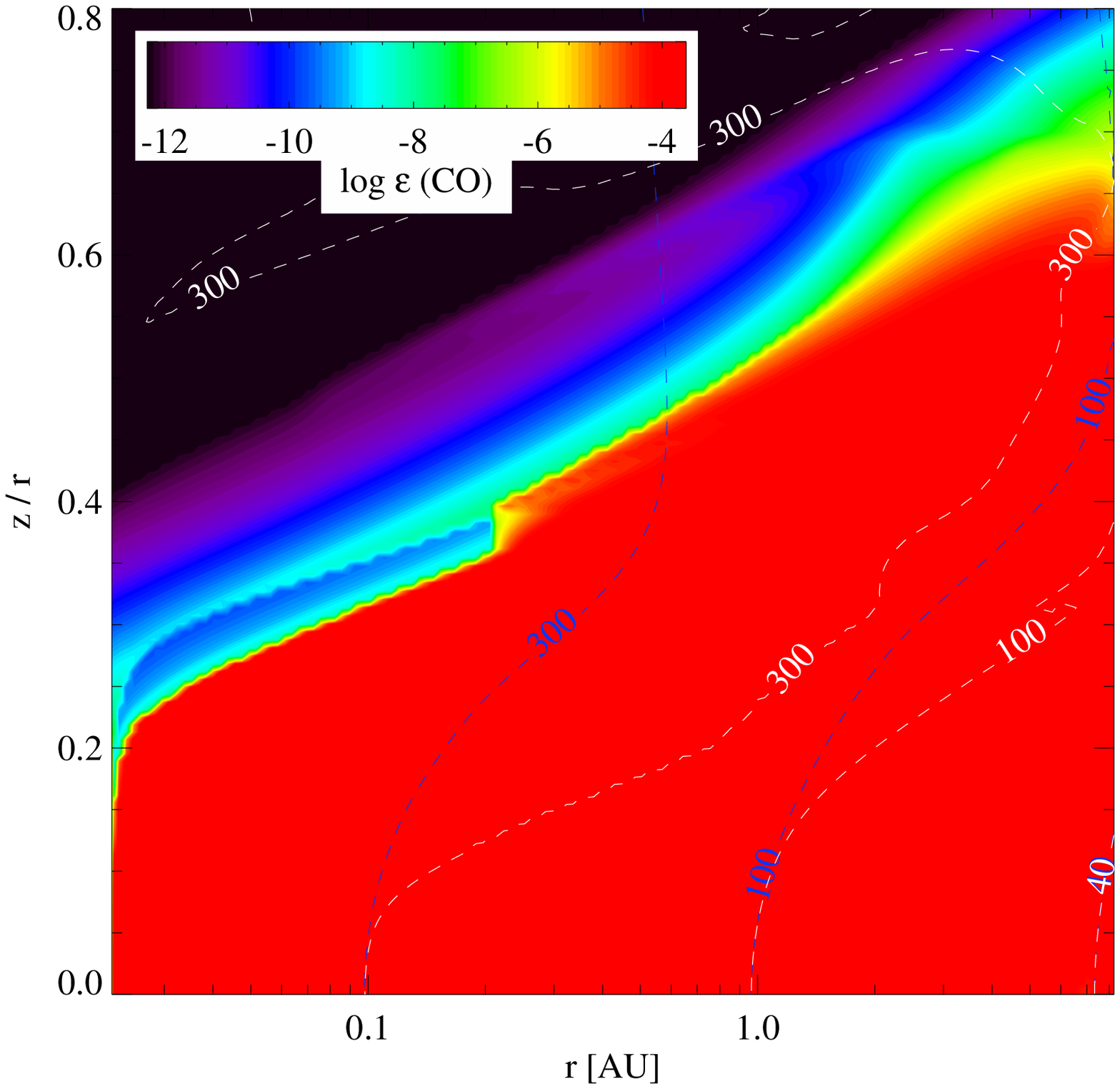} \\[-2mm]
  \includegraphics[width=80mm]{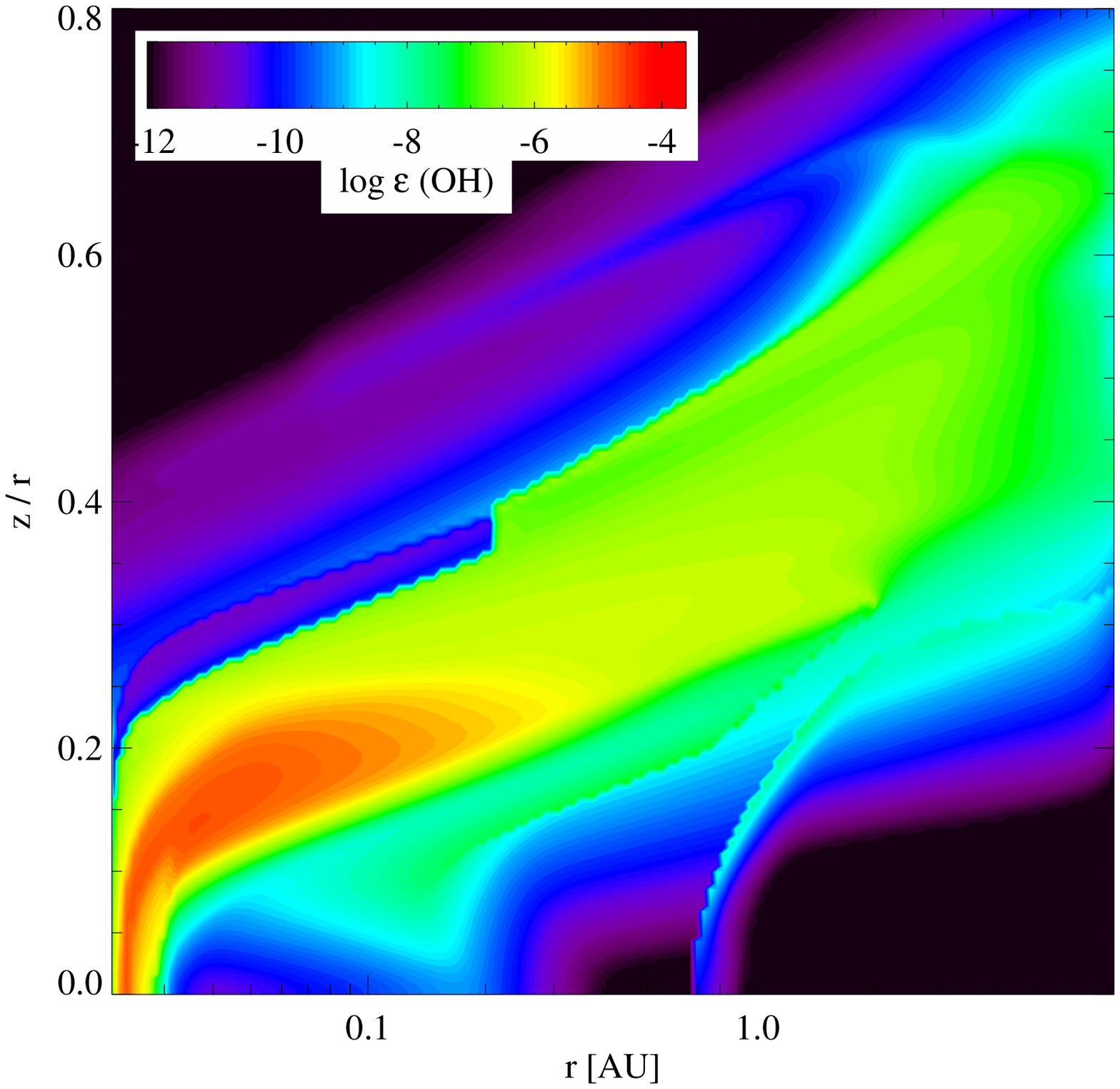} &
  \includegraphics[width=80mm]{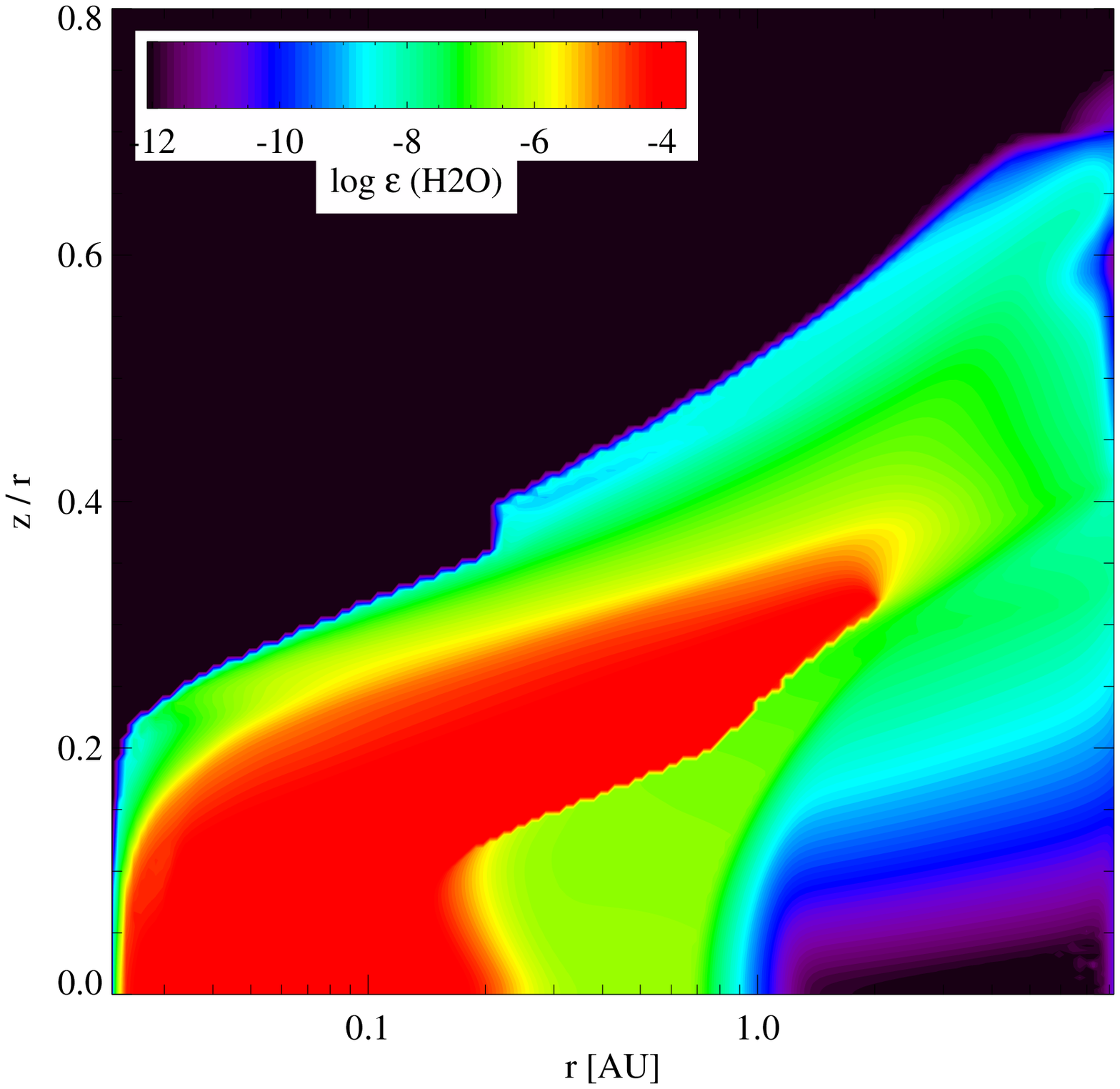} \\[-2mm]
\end{tabular}
\caption{Chemical details of the best-fitting disk model. {\bf Upper
  row:} electron concentration $n_{\rm el}/\nH$ and mean PAH-charge,
  with overplotted contours for ionisation parameter
  $\log(\chi/\nH)$. {\bf Second row:} neutral carbon and
  CO-concentration, the CO-plot includes contours for dust temperature
  (blue) and gas temperature structure (white).  {\bf Lower row:} OH
  and H$_2$O concentration.}
\label{fig:diskmodel2}
\end{figure*}

Figure~\ref{fig:diskmodel1} visualises the densities, UV radiation
field strength, and resulting gas and dust temperatures of the
best-fitting model.  We obtain the typical $T_{\rm dust}$-pattern for
passive disks \citep[see \eg][]{Pinte2009} with a shadow casted by the
innermost disk regions and dust temperatures of the order of 1450\,K
at the inner rim at 0.022\,AU, and $\sim\!35\,$K at the outer disk
edge located at 8.2\,AU. The gas temperature structure $T_{\rm
gas}(r,z)$ follows the dust temperature structure in the midplane
(where thermal accommodation dominates) but shows substantial
deviations at higher altitudes where the disk starts to become
optically thin to UV radiation.

In particular, there is a ``hot finger'' at relative heights
$z/r\!\approx\!0.3\!-\!0.6$ in this model, stretching out from the
inner rim to about 4\,AU in radius. Here, gas temperatures of the
order of 5000\,K are achieved in this model due to PAH heating versus
FeII and MgII line cooling. Just below the hot atomic layer, there is
a warm molecular layer with temperatures $300\!-\!1500$K, heated by
PAHs and chemical heating, balanced by CO ro-vibrational cooling.
The outermost layers beyond 4\,AU are featured by an equilibrium 
between PAH and chemical heating, versus OI, OH, H$_2$O and CO 
rotational line cooling.

Figure~\ref{fig:diskmodel2} shows some details about the chemical
structure of the disk. There is a quite sudden transition of the
charge of the PAHs from 3 to -1 where the ionisation parameter
$\chi/\nH$ is about 100. The PAHs are mostly negatively charged then,
except for the outer midplane where the gas almost completely
neutralises. The neutral carbon, CO, OH and H$_2$O concentrations
are typical of a photon dominated region (PDR-structure).

\begin{figure*}
  \begin{tabular}{cc}
  {\large prescribed disk structure} & {\large calculated disk structure}\\
  \hspace*{1mm}\includegraphics[width=8.8cm,height=75mm]
                                 {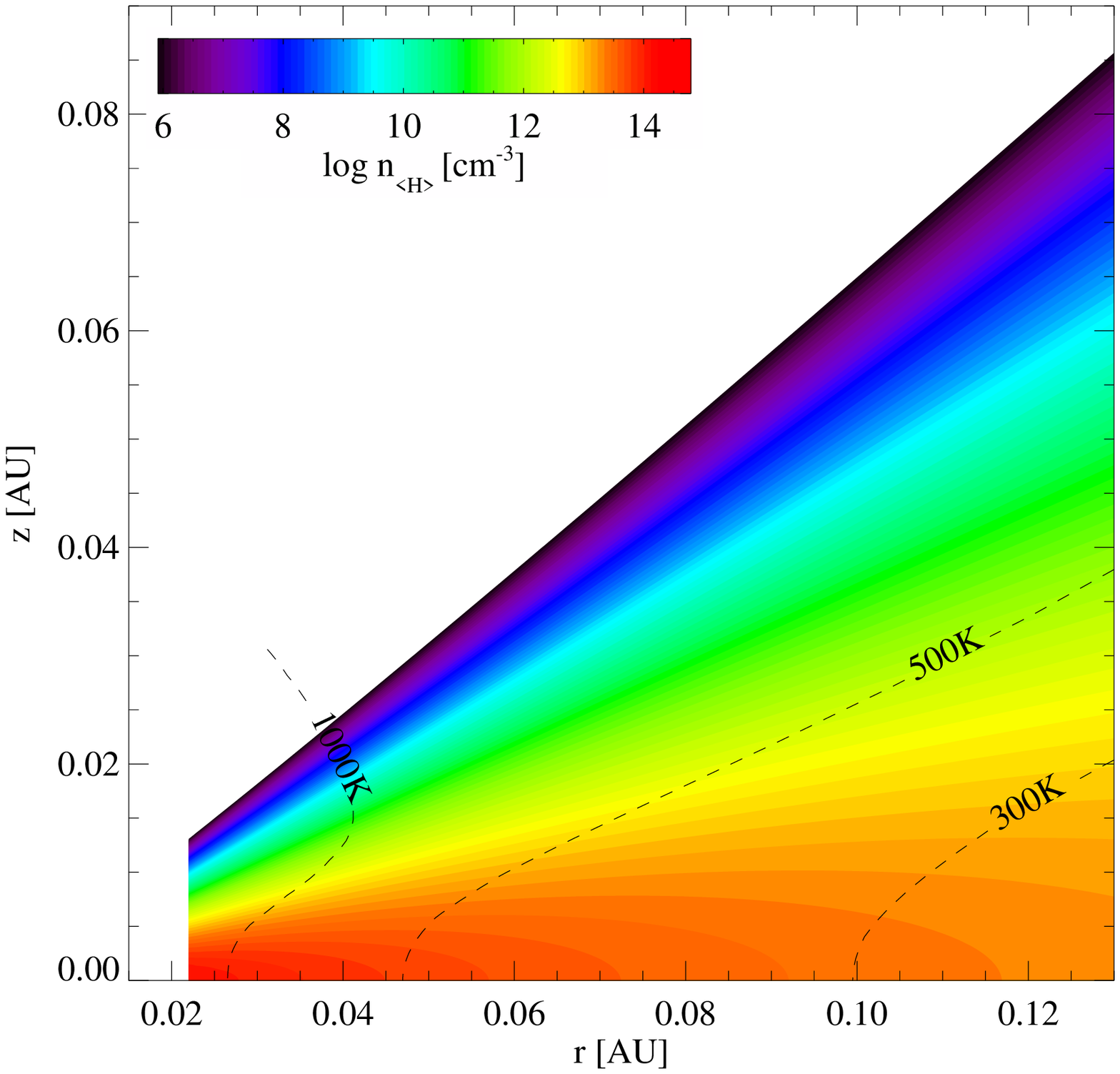} &
  \hspace*{-8mm}\includegraphics[width=8.8cm,height=75mm]
                                 {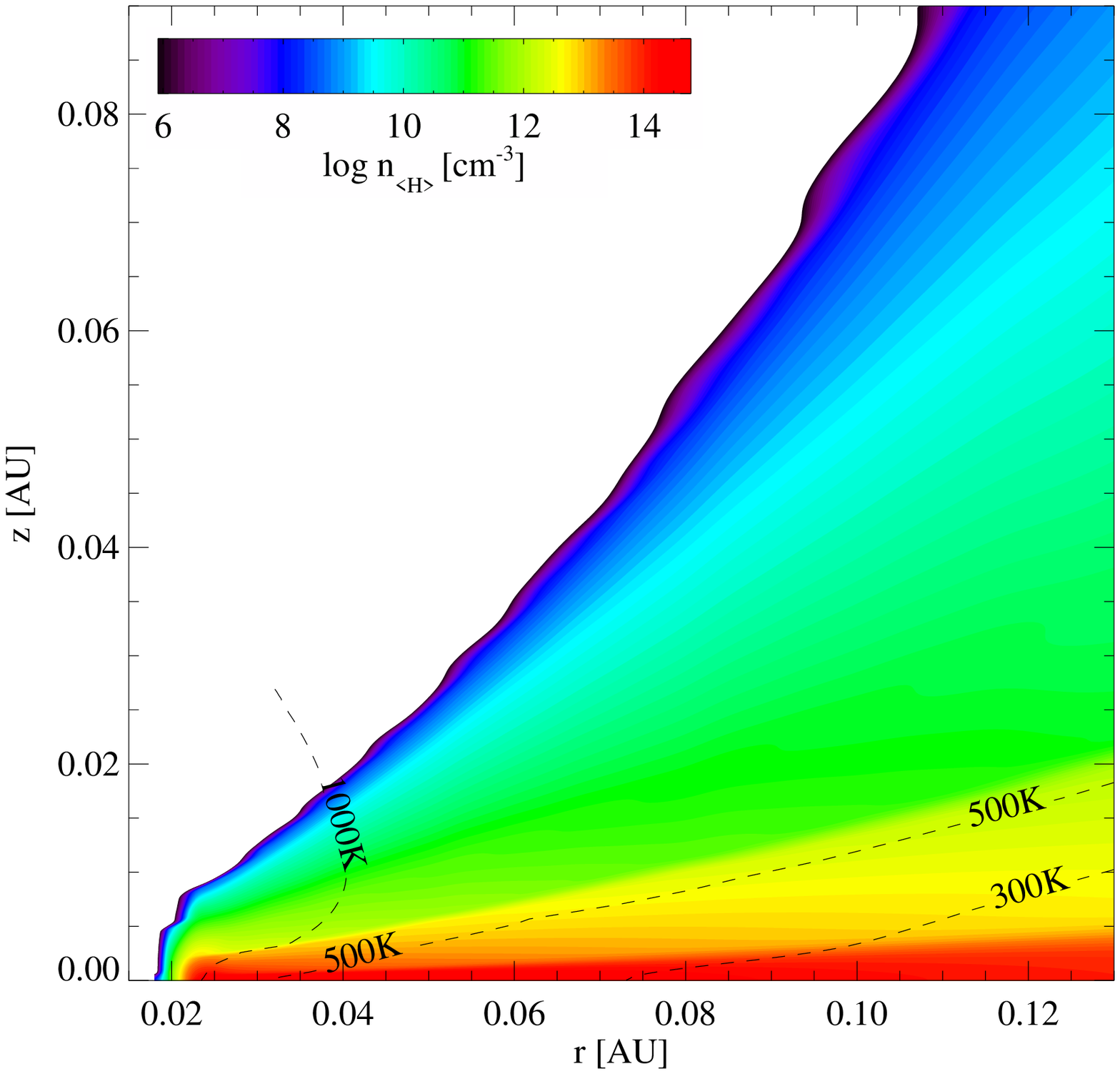} \\*[-7mm]
  \hspace*{-2mm}\begin{minipage}{8.9cm}
               {\ }\\*[3mm]
               \includegraphics[width=8.9cm,height=7.4cm]{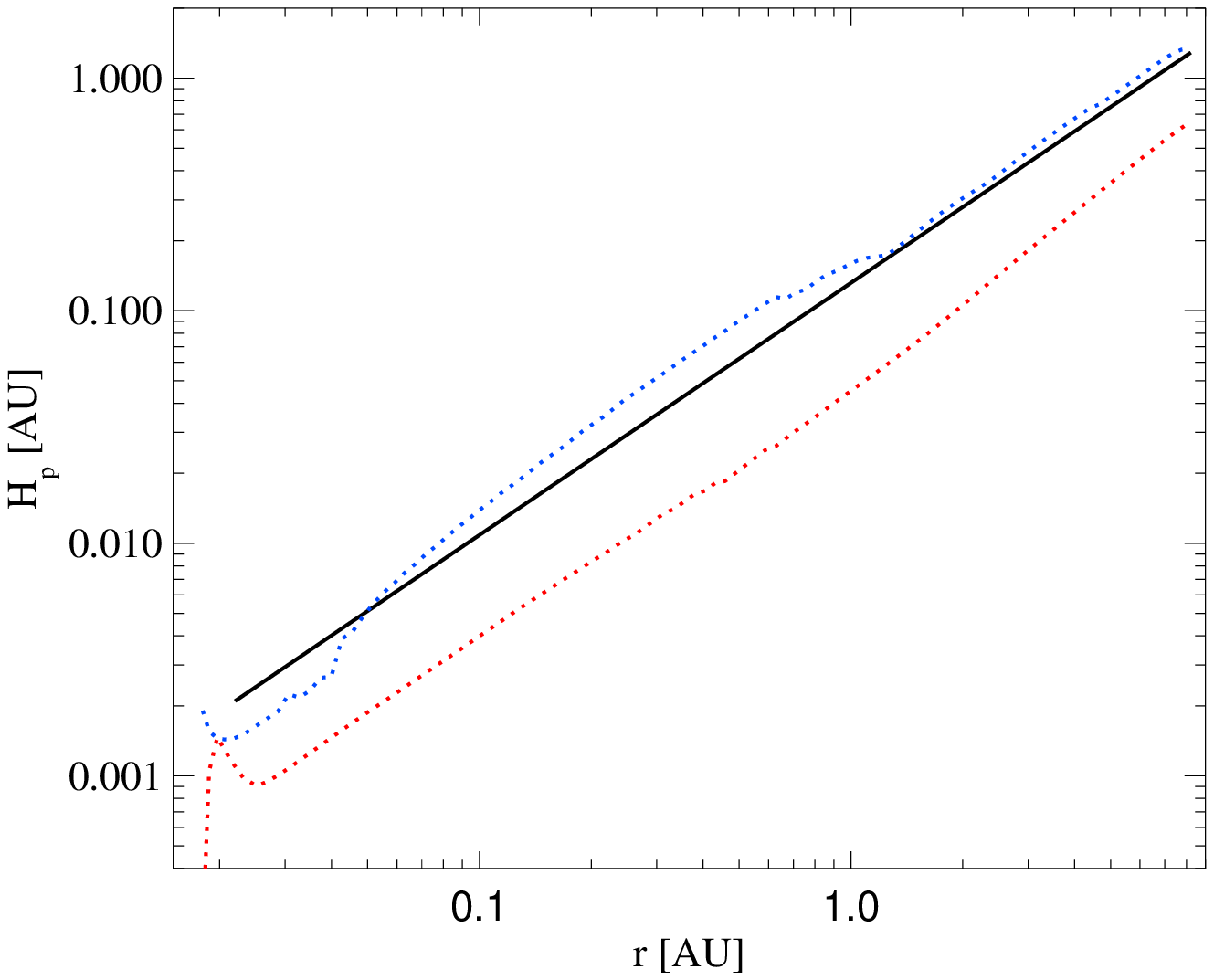}
               \end{minipage} &
  \hspace*{-7mm}\begin{minipage}{8.9cm}
                \includegraphics[width=9.0cm]{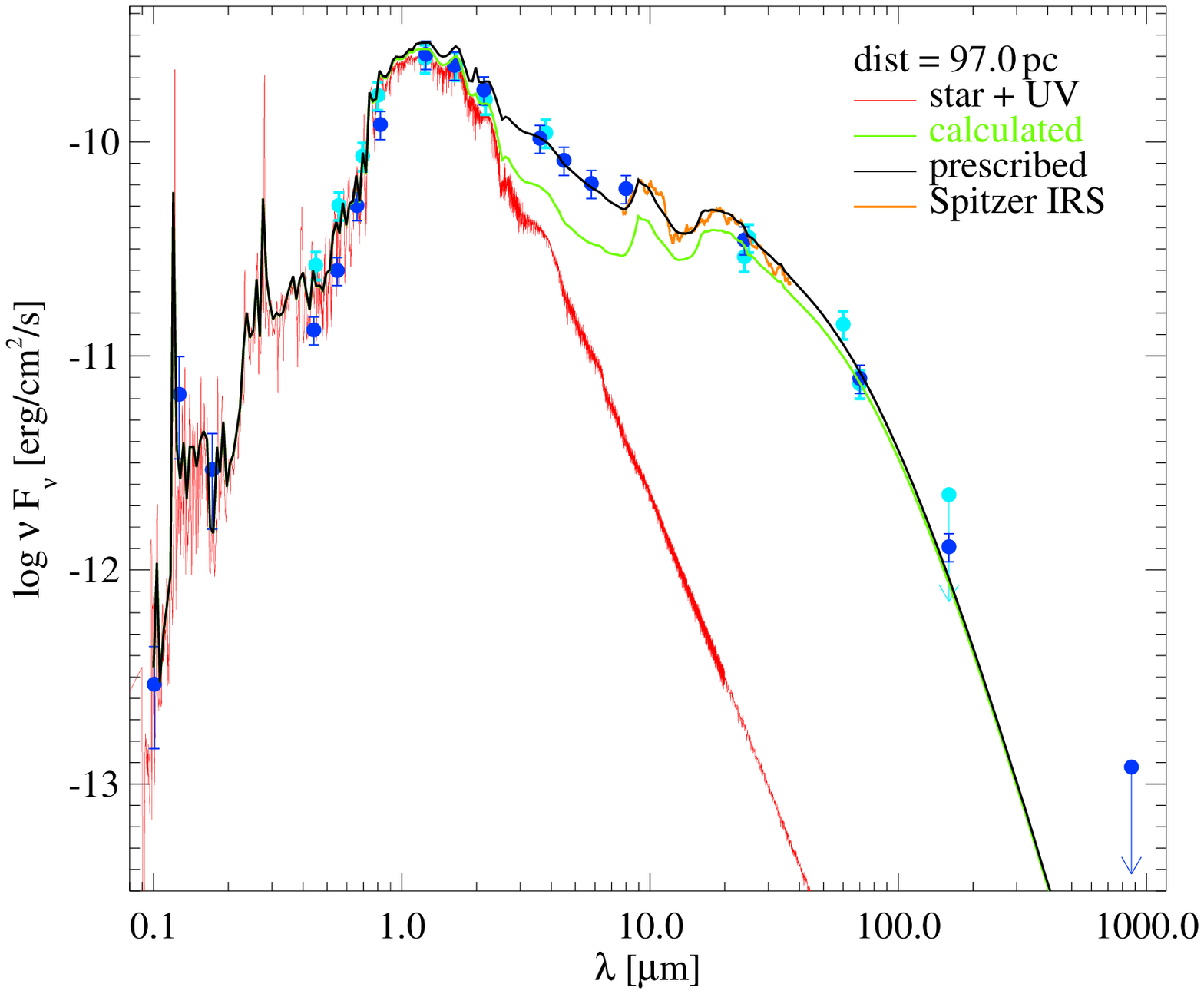} 
                \end{minipage}\\*[-5mm]
  \end{tabular}
  \caption{Density structure of inner rim and impact on SED. The {\bf
     upper left plot} shows the prescribed density structure of our
     best-fitting model. Dashed contour lines in the upper panel refer
     to the calculated dust temperatures. The {\bf upper right plot}
     represents a model with identical parameters where, however, the
     vertical disk structure is calculated consistently with
     the resultant gas temperatures and mean molecular
     weights, which results in a flatter midplane close to the star
     and too little near-mid IR excess ({\bf lower right plot}). Note
     the ``soft inner edge'' \citep[see][]{Woitke2009a}. The {\bf
     lower left plot} compares the prescribed scale height of the
     best-fitting model (full line) with the scale heights resulting
     from the model with calculated vertical disk stratification. The
     red dotted line shows these results at small relative height
     $z/r\!=\!0.05$, and the blue dotted line at $z/r\!=\!0.5$.}
  \label{fig:struc}
\end{figure*}

%=====================================================================
\section{Variation of input physics}
\label{sec:inputphysics}

\subsection{Difference between dust and gas temperature}
\label{sec:TgTd}

A control model where the gas temperature is assumed to be equal to
the dust temperature resulted in the line fluxes shown in
Table~\ref{tab:TgTd}. We conclude that modelling the gas energy
balance, mostly leading to $T_{\rm gas}\!>\!T_{\rm dust}$ in the
line forming regions, is absolutely essential to understand the
gas emission lines from protoplanetary disks.

\begin{table}[!b]
\centering
\caption{Computed line fluxes $\rm[10^{-18}\,W/m^2]$ from a model
where $T_{\rm gas}\!=\!T_{\rm dust}$ is assumed in comparison to the
best-fitting disk model.}
\label{tab:TgTd}
\vspace*{-1mm}
\begin{tabular}{lc|c|cc}
line    & $\rm\lambda\,[\mu m]$    
        & $T_{\rm gas}\!=\!T_{\rm dust}$ & best model\\
\hline
$\rm[OI]\ ^3P_1\to\, ^3P_2$        & 63.18    & 5.2    & 34.5   \\
$\rm[OI]\ ^1D_2\to\, ^3P_2$ (LVC)  & 0.6300   & 2.3    & 69.6   \\
CO\,$J\!=\!3\!\to\!2$              & 866.96   & 0.0053 & 0.014  \\
o-H$_2\rm\ v\!=\!1\!\to\!0\ S(1)$  & 2.122    & 0.0073 & 2.4    \\ 
\end{tabular}
\end{table}

\subsection{Influence of X-rays}
\label{sec:Xray}

We have run a comparison disk model with X-ray heating and chemistry
included, as has recently been implemented by \citet{Aresu2010}, with
X-ray luminosity $L_X\!=\!6\times10^{28}\,$erg/s \citep[{\sc Xmm}
observations by][]{Lopez2010}.  We assumed an X-ray emission
temperature of $T_X\!=\!10^7\,$K and a minimum energy of X-ray photons
of 0.1\,keV. This model does not result in any observable changes in
the calculated line fluxes. The modification by X-rays are only
-0.5\%, 1.1\%, 1.4\% and 0.05\% for CO$\,J\!=\!3\!\to\!2$,
[OI]\,63\,$\mu$m, [OI]\,6300\,\AA\ (LVC) and o-H$_2$\,2.122\,$\mu$m,
respectively.  Since the X-rays are attenuated by gas in the model,
but the UV photons by dust, and our best-fitting model is extremely
gas-rich (assumed gas/dust mass ratio $\sim\!23000$, see
Table~\ref{tab:Parameter}), the X-rays do not penetrate deep enough to
change the energy balance in the line emitting regions.

\subsection{Influence of viscous heating}
\label{sec:viscous}

We have run a comparison disk model with viscous (gas) heating included,
via the formula of \citet{Frank1992}
\begin{equation}
  \Gamma_{\rm vis} = \frac{9}{4}\,\rho\,\nu_{\rm kin}\,\Omega_{\rm kep}^2
                  \ ,
  \label{eq:gamma_vis}
\end{equation}
where $\rho$ is the gas mass density, $\nu_{\rm kin}=\alpha\,c_T\,H_p$
the viscosity, $\alpha$ the viscosity parameter,
$c_T\!=\!\sqrt{p/\rho}$ the isothermal sound speed,
$H_p\!=\!c_T/\Omega_{\rm kep}$ the pressure scale height, and
$\Omega_{\rm kep}\!=\sqrt{G M_\star/r^3}$ the Keplerian angular
velocity. Putting the viscosity parameter to $\alpha\!=\!0.01$, we
find no significant changes in the far-IR and (sub-)mm lines, but
modest changes in the calculated line fluxes in the optical and
near-IR. The enhancement by viscous heating is 1.0\%, 2.4\%, 25\% and
16\% for CO$\,J\!=\!3\!\to\!2$, [OI]\,63\,$\mu$m, [OI]\,6300\,\AA\
(LVC) and o-H$_2$\,2.122\,$\mu$m, respectively. Since the viscous
heating scales as $\Gamma_{\rm vis}\!\propto\!\rho$, but most cooling
processes scale as $\Lambda\!\propto\!\rho^2$, the effect of the
viscous heating is actually strongest in the low density uppermost
disk regions, which is counter-intuitive. It renders the gas
temperature in these layers unbound (artificially limited by
20000\,K), as no implemented cooling process is able to balance the
viscous heating according to Eq.~(\ref{eq:gamma_vis}) in a thin gas.
We therefore refrain from discussing the effects of viscous heating
any further in this paper.

\subsection{Self-consistent disk structure}
\label{sec:struc}

Figure~\ref{fig:struc} visualises the density structure as resultant
from a self-consistent disk model where the vertical disk
stratification is a result of radiative transfer, chemistry, and gas
energy balance, assuming vertical hydrostatic equilibrium. We find a
good match concerning the flaring angle (the slope in the lower left
plot), but a generally flatter disk structure, more condensed toward
the midplane. The $z$-dependent scale heights from the
self-consistently calculated disk model are calculated as
\begin{equation}
  H_p^2(z) = \frac{z^2}{2\,\log\,[p(r,z)/p(r,0)]} \ ,
  \label{eq:Hscale}
\end{equation}
where $p$ denotes the gas pressure. In the lower left plot of
Fig.~\ref{fig:struc} we have plotted two scale heights from the
self-consistent disk model, one measured close to the midplane (at
relative height $z/r\!=\!0.05$, red dotted line) and one measured high
above the midplane (at $z/r\!=\!0.5$, blue dotted line). The
difference between these two scale heights is a natural consequence of
our calculated gas temperature structure, with cold conditions in the
midplane and a warm/hot disk surface. 

We observe a fair match of the scale heights at $z/r\!=\!0.5$ with the
prescribed scale-heights from our best-fitting model, but in the
midplane the scale heights of the best-fitting model are about a
factor of 2-3 too large. Figure~\ref{fig:struc} also demonstrate that,
consequently, we loose our SED-fit when using the self-consistent
model.  The self-consistent model intercepts less star light,
resulting in a significant cooling and flux deficit in the near and 
mid IR as compared to the observations.

\subsection{Influence of non-radiative dust heating}
\label{sec:dustnonRE}

\begin{figure}
\vspace*{-7mm}
\hspace*{-2mm}\includegraphics[width=9.6cm]{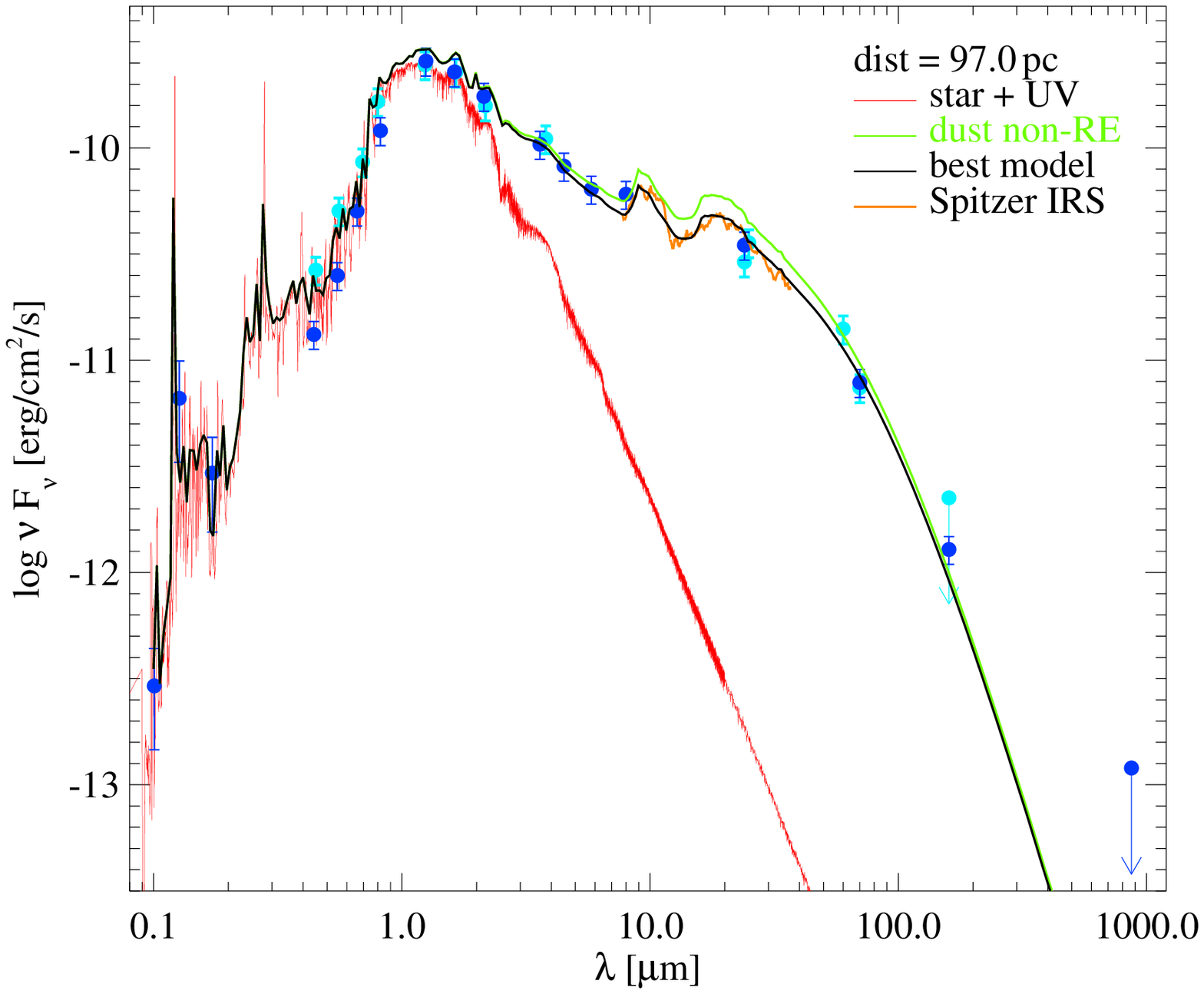}
\vspace*{-9mm}
\caption{Comparison between predicted SEDs of two models with and
  without non-radiative dust heating. The black model shows once more
  the SED of our best fitting model, without non-radiative dust
  heating. In the green model, non-radiative dust heating through
  thermal accommodation is included.  All model parameters are
  identical otherwise.}
\label{fig:SED_dustnonRE}
\end{figure}

In our best-fitting model, we have ignored non-radiative
heating/cooling of the dust grains when determining the dust
temperature structure $T_{\rm dust}(r,z)$.  However, since the gas is
typically warmer than the dust, inelastic gas-grain collisions
(thermal accommodation) lead to a collisional, \ie non-radiative,
heating of the dust.  If we include this effect \citep[see
Eqs.~(14) and (108) in][]{Woitke2009a}, we do not observe much of an
effect on the calculated line fluxes, but the dust temperatures result
to be slightly higher, with noticeable effects on the SED, see
Fig.~\ref{fig:SED_dustnonRE}.

According to this model with the dust in non-radiative equilibrium,
the temperature contrast between gas and dust is mainly driven by
exothermic chemical reactions which are active even in quite deep and
dense layers (see Fig.~\ref{fig:diskmodel1}), causing an overall heat
transfer from gas to dust in the disk of
$\sim\!4.5\times10^{-3}\rm\,L_\odot$, \ie about 5\% of the stellar
luminosity. It is this additional energy input that leaves the disk in
form of additional mid IR continuum photons, causing the depicted 
variations in Fig.~\ref{fig:SED_dustnonRE}.

The effect of non-radiative dust heating on the SED is similar to
increasing the scale heights. We have performed an additional run 
of the evolutionary strategy with enabled non-radiative dust heating.
This run did not entirely converge. The final parameter set had a gas mass
of $1.2\times10^{-3}\rm\,M_\odot$, a scale height of only $0.007\,$AU, 
\ie a reduction of $\sim\!35\%$ with respect to our best-fitting model. 
Unfortunately this is a slow and unstable option, because there is an 
additional outer iteration necessary between gas and dust temperature 
determination, to achieve consistent results, which requires about 
3 times more computational time.

\subsection{Influence of treatment of H$_2$-formation}
\label{sec:H2form}

The formation of H$_2$ on dust grain surfaces is one of the most
important first steps to initiate a rich molecular chemistry. It has
profound effects also on other abundances, for instance the C$^+$/C/CO
transition (because of the mutual H$_2$/C shielding) and on the
formation of OH and H$_2$O. Yet its rate is still rather uncertain.
Our default choice is to calculate this key chemical process according
to \citep[][called ``model~B'' in
Table~\ref{tab:H2form}]{Cazaux2004}. We have run two comparison
models with two other formulations, one with a typically smaller
H$_2$-formation rate according to \citep{Sternberg1995}. And one
with a typically larger rate according to \citep{Cazaux2010}.

Concerning the formulation of \citet{Sternberg1995}, which is valid for
standard ISM size distribution and dust/gas ratio only, we add a
scaling factor to account for deviations of the total dust surface 
per hydrogen nucleus in disks as 
\begin{equation}
  R_{\rm H2}^{\rm\,form} = 3\times10^{-18} \nH \sqrt{T_{\rm gas}}\;\;
           \frac{\langle a^2\rangle\,n_{\rm dust}/\nH} 
                {5.899\times10^{-22}{\,\rm cm^2}} \ .
  \label{eq:H2Jura}
\end{equation}
The H$_2$-formation rate coefficient $R_{\rm
H2}^{\rm\,form}\,\rm[1/s]$ needs to be multiplied by the neutral
hydrogen atom density $n_{\rm H}$ to get the H$_2$-formation rate in
$\rm[cm^{-3}\,s^{-1}]$. The normalisation factor in
Eq.~(\ref{eq:H2Jura}) results from $\langle a^2\rangle_{\rm
ISM}\,[n_{\rm dust}/\nH]_{\rm ISM}\!=\!5.899\times10^{-22}{\rm cm^2}$
under interstellar conditions, \ie for $a_{\rm min}\!=\!0.005\,\mu$m,
$a_{\rm min}\!=\!0.25\,\mu$m, $p\!=\!3.5$, $\rho_{\rm
gr}\!=\!3\rm\,g/cm^3$ and $\rho_{\rm dust}/\rho_{\rm gas}=0.01$.

%     stick = 1.0/(1.0+0.4*sqrt(0.01*(Tg+Td))+0.2*(0.01*Tg)
%                     +0.08*(0.01*Tg)**2)

Table~\ref{tab:H2form} shows a strong dependence of the predicted
o-H$_2$\,2.122\,$\mu$m line on the assumed H$_2$-formation rate on
grains. The \citet{Sternberg1995}-formalism results in a factor of
0.77 smaller and the \citet{Cazaux2010}-formalism in a factor of 4.8
larger line flux. This is a daunting example of hidden uncertainties
in astrochemical modelling. We note that the
\citet{Sternberg1995}-formalism gives a smaller $\rm
FWHM\!\approx\!30\,km/s$ that brings this model closer to the
observations.  All other calculated gas emission lines (including
those of H$_2$O) are less affected.

\begin{table}[!b]
\def\z{\hspace*{-1mm}}
\caption{Calculated o-H$_2$ and o-H$_2$O line fluxes
  $\rm[10^{-18}\,W/m^2]$ and FWHM $\rm[km/s]$ of models with different
  treatment of the H$_2$-formation on grain surfaces.}
\vspace*{-2mm}
\label{tab:H2form}
\begin{tabular}{c|cc|cc|cc}
        & \multicolumn{2}{c|}{o-H$_2$\ 2.122\,$\mu$m} 
        & \multicolumn{2}{c|}{\z o-H$_2$O\ 179.53\,$\mu$m\z} 
        & \multicolumn{2}{c}{o-H$_2$O\ 78.74\,$\mu$m}\\
        & flux & FWHM & flux & FWHM & flux & FWHM \\
\hline
\z model A  & 1.81 & 30 & 1.41 & 8.2 & 11.2 & 8.5 \\
\z model B  & 2.35 & 37 & 1.40 & 8.2 & 11.1 & 8.5 \\
\z model C  & 11.3 & 41 & 1.51 & 8.3 & 12.2 & 8.6 \\
\z observed & \z$2.5\!\pm\!0.1$\z\z & \z\z$18\!\pm\!1.2$\z\z 
            & $<5.0$ & $-$ & $<30$ & $-$\\
\end{tabular}\\[2mm]
{\footnotesize model A $=$
  \citet{Sternberg1995}, model B (best-fitting model) $=$
  \citet{Cazaux2004}, model C $=$ \citet{Cazaux2010}. The 
  other calculated gas emission lines are less affected.}
\end{table}

\end{appendix}

\end{document}